\documentclass[preprint]{aastex}
\usepackage[english]{babel}

\usepackage{epsfig}
\usepackage{graphicx,natbib}
\usepackage{amsmath,amssymb}
\usepackage{hyperref}
\hypersetup{
    colorlinks=true,
    citecolor=blue,
    linkcolor=red,
    urlcolor=black
    }  

\def\ga{\,\hbox{\hbox{$ > $}\kern -0.8em \lower 1.0ex\hbox{$\sim$}}\,}
\def\la{\,\hbox{\hbox{$ < $}\kern -0.8em \lower 1.0ex\hbox{$\sim$}}\,}
\def\aap{A\&A }
\def\mnras{MNRAS }

\newcommand{\rev}[1]{#1}
\newcommand{\revv}[1]{#1}
\newcommand{\m}[1]{#1}
\newcommand{\noprint}[1]{{}}

\begin{document}
\def\apj{Astrophys. J. }
\def\apjs{Astrophys. J., Suppl. Ser. }
\def\apjl{Astrophys. J., Lett. }
%\begin{frontmatter}

\title{Incorporating Ambipolar and Ohmic Diffusion in the AMR MHD code {\ttfamily RAMSES}}

\author{J. Masson}
\affil{\'Ecole normale sup\'erieure de Lyon, CRAL, UMR CNRS 5574, 69364 Lyon Cedex 07,  France}
\affil{Laboratoire de radioastronomie, UMR CNRS 8112, \'Ecole normale sup\'erieure et Observatoire de Paris, 24 rue Lhomond, 75231 Paris cedex 05, France}
\author{R. Teyssier}
\affil{Laboratoire AIM, CEA/DSM, CNRS, Universit\'e Paris Diderot, IRFU/SAp, 91191 Gif--sur--Yvette, France}
\affil{Institute of Theoretical Physics, University of Z\"urich, Winterthurerstrasse 190, CH-8057 Z\"urich, Switzerland}
\author{C. Mulet-Marquis}
\affil{\'Ecole normale sup\'erieure de Lyon, CRAL, UMR CNRS 5574, 69364 Lyon Cedex 07,  France}
\author{P. Hennebelle}
\affil{Laboratoire de radioastronomie, UMR CNRS 8112, \'Ecole normale sup\'erieure et Observatoire de Paris, 24 rue Lhomond, 75231 Paris cedex 05, France}
\author{G. Chabrier}
\affil{\'Ecole normale sup\'erieure de Lyon, CRAL, UMR CNRS 5574, 69364 Lyon Cedex 07,  France}
\affil{School of Physics, University of Exeter, Exeter, EX4 4QL, UK}

\date{}

\begin{abstract}

We have implemented non-ideal Magneto-Hydrodynamics (MHD) effects in the Adaptive Mesh Refinement (AMR) code RAMSES,
namely ambipolar diffusion and Ohmic dissipation, as additional source terms in the ideal MHD equations.
We describe in details how we have discretized these terms using \m{the} adaptive Cartesian mesh, and how the time step is \m{diminished} 
with respect to the ideal case, in order to perform a stable time integration. We have performed a large suite of test runs,
featuring the Barenblatt diffusion test, the Ohmic diffusion test, the C-shock test and the Alfven wave test. For the latter, we have performed a careful
truncation error analysis to estimate the magnitude of the numerical diffusion induced by our Godunov scheme, \m{allowing us} to estimate the spatial resolution that is required to address non-ideal MHD effects reliably. We show that our scheme is second-order accurate,
and is therefore ideally suited to study non-ideal MHD effects in the context of star formation and molecular cloud dynamics.

\end{abstract}


%\end{frontmatter}

\maketitle

\section{Introduction}

The impact of magnetic fields on various objects in astrophysics is now well established. They play a major role \m{on} a wide range of scales, from the study of the early universe, the stellar and intergalactic medium to the formation and interiors of stars or the accretion flows around stellar objects. They are difficult to study both \m{from an observational} and a theoretical (and numerical) point of view. Several implementations of ideal MHD have been performed since the last decade (\citet{fromang}, \citet{StoneNorman}, \citet{Machida2005} among others), and numerical issues concerning the divergence free condition \m{have now been resolved}. However, ideal magnetohydrodynamics (MHD) is in many circumstances a poor approximation and non-ideal MHD effects need to be thoroughly considered.

Ambipolar diffusion is expected to play a major role in star formation (\citet{MestelSpitzer56}), at the scale of molecular clouds by enabling the collapse of otherwise magnetically supported clouds (\citet{BasuCiolek04}) and at the scale of the first Larson's core with the formation of a centrifugally supported disk and the well-known {\it fragmentation crisis} (\citet{HennebelleTeyssier2008}). \m{Ambipolar diffusion is also important in protoplanetary disks} as they are in general only partially ionised. The microscopic and entropic heating resulting from the drift and collision between neutral and charged species is another very important \m{and relatively unknown} aspect which is \m{crucial} as soon as cooling or heating of the gas (thus radiative transfert) is taken into account (in contrast it is not relevant \m{when using} a barotropic equation of state).

Magnetic resistivity effects range from prohibiting long-term MHD turbulence in molecular clouds (\citet{BasuDapp2010}) to preventing the {\it magnetic braking catastrophe} on small scales (\citet{DappBasu2010}). Its importance is also crucial in order to study disk formation around protostellar objects (\citet{Krasnopolsky}) and the physics of binary formation and brown dwarfs.

Thus, it appears necessary to introduce the ambipolar and Ohmic diffusion in a 3D MHD code. Before exploring the astrophysical impact of such a study, however, the accuracy of the treatment of the complete MHD set of equations must be unambiguously assessed. This is the very aim of the present paper, in which we describe a prescription to incorporate ambipolar and Ohmic diffusion in the multi-dimensional MHD AMR code {\ttfamily RAMSES} (\citet{teyssier}), extending the ideal MHD version presented in \citet{teyssierMHD} and \citet{fromang}.

Several numerical treatments have been derived from ideal MHD models, following different aims and thus \m{using} different methods. \label{rev3}\rev{The first attempt to implement ambipolar diffusion in a code was made by \citet{BlackScott1982} using an iterative approximation in an implicit first-order code. \citet{toth1994} used a semi-explicit method in a two-dimensional code to investigate instabilities in C-schocks. \citet{McLow} presented a widely used explicit method (\citet{Choietal}, \citet{MellonLi2009}, \citet{LiKrasnopolskyShang}) to implement single-fluid ambipolar diffusion in the strong coupling limit, and then developed a two-fluid model in order to capture shock instabilities. \citet{TilleyBalsara2008} and more recently \citet{TilleyBalsara2011} presented a semi-implicit scheme for solving two-fluid ambipolar diffusion, arguing that the single fluid approximation does not carry the full set of MHD waves that can propagate in a poorly ionized system. Multi-fluid approaches including ambipolar diffusion and Ohmic diffusion have been suggested by \cite{Falle2003}, or \cite{Osullivan2006} and then investigated by e.g. \citet{KunzMouschovias2009}. Recently, \citet{LiKrasnopolskyShang} used the single-fluid approach including more realistic resistivities based on a multi-fluid approach for ambipolar diffusion, Ohmic diffusion and Hall effect in two-dimensional (axi-symmetric) calculations. Another approach has been used by \citet{Machida_etal06} was to describe both ambipolar diffusion and Ohmic diffusion in one single Laplace operator $\eta \Delta B$, with $\eta$ taking into account every diffusive process at stake.} \m{These numerous studies have also given rise to several numerical tests, a number of which we will either perform directly or slightly modify to assess the accuracy of our treatment.}

Our current study focusses on the one-fluid approximation (\citet{Shu_etal87}), as in previous calculations by \citet{McLow} and \citet{DuffinPudritz}. \label{revv1}\revv{We used a direct explicit method to implement non-ideal MHD terms in both the induction and energy equations (\citet{McLow}) in an AMR framework. We did not choose to account for non-ideal effects by adding ambipolar diffusion and Ohmic dissipation in a single Laplace operator as \citet{Machida_etal06}. Instead we kept the full expressions and proceeded separately for each non-ideal effect.}

The paper is organized as follows. In \S~\ref{amb}, we first derive the equations for ambipolar diffusion in the single fluid approximation. We then describe the various tests we have performed, first without the hydrodynamics and then in a complete MHD situation, exploring in particular the propagation of Alfv\'en waves. Comparisons with existing analytical or benchmark solutions are presented in details, demonstrating the validity and the accuracy of our scheme. \S~\ref{mag} addresses the case of Ohmic diffusion, following the same procedure as for ambipolar diffusion, while \S~\ref{concl} is devoted to the conclusion.

\section{Ambipolar diffusion}
\label{amb}

\subsection{Equations}

When the ions pressure and momentum are negligible compared to those of neutral species (as is the case for example in molecular clouds), the Lorentz force exerted on the ions is in equilibrium with the drag force exerted by the neutrals, which corresponds to a situation of strong coupling between the neutral fluid and the field lines. In such a situation, the plasma can be adequately described by a single fluid (\citet{Shu_etal87} and \citet{Choietal}) of mass density $\rho$, and neutrals and ions mass densities $\rho_n\approx \rho$ \m{and} $\rho_i\ll \rho_n$ respectively. Interestingly, in the case of the one-fluid approximation the results can be directly compared with the ones obtained with ideal MHD giving clear insights about MHD wave propagation properties in the non-ideal case (\citet{Balsara}).The present study \m{is} devoted to the technical resolution of the resistive MHD equations \m{and} we will also ignore gravity (and thus the Poisson equation). The MHD equations are given by the usual continuity, momentum, energy and induction equations, completed by the magnetic field divergence-free condition (in rational units, $\mathbf{B}_{rat} = \mathbf{B}_{cgs}/ \sqrt{4 \pi} $):

\begin{align}
\frac{\partial \rho}{\partial t} + {\nabla} . (\rho \mathbf{v}) &= 0 \label{eqcont} \\
\rho \frac{\partial \mathbf{v}}{\partial t} + \rho \,(\mathbf{v}.{\nabla})  \mathbf{v} + {\nabla} P - \mathbf{F}_{L} &= 0 \label{eqqtmvt}\\
\frac{\partial E_{tot}}{\partial t} + {\nabla}.\big( (E_{tot}+P_{tot})\mathbf{v}  - (\mathbf{v}.\mathbf{B})\mathbf{B} - \mathbf{E}_{AD} \times \mathbf{B} \big) &= 0 \label{eqenergDA} \\
\frac{\partial \mathbf{B}}{\partial t} - {\nabla} \times ( \mathbf{v} \times \mathbf{B}) - {\nabla} \times \mathbf{E}_{AD} &= 0 \label{eqdbdt} \\
{\nabla}\cdot {\mathbf{B}} &= 0 \label{eqdivb}.
\end{align}
  
$E_{tot}$ denotes the total energy
\begin{equation}
E_{tot} = \rho \epsilon + \frac{1}{2}\rho v^2 + \frac{1}{2}B^2,
\label{energtot}
\end{equation}
\m{where} $\epsilon$ is the specific internal energy.

$P_{tot}$ is the total pressure
\begin{equation}
P_{tot} = (\gamma-1) \rho \epsilon + \frac{1}{2}B^2,
\label{pressuretot}
\end{equation}
where $\gamma$ is the adiabatic index.

$\mathbf{F}_{L}$ denotes the Lorentz force
\begin{equation}
\mathbf{F}_{L}= ({\nabla} \times  \mathbf{B}) \times \mathbf{B},
\label{lorentz}
\end{equation}
with
\begin{equation}
\mathbf{v_i}-\mathbf{v_n} = \frac{1}{\gamma_{AD} \rho_i \rho} \mathbf{F_L}.
\end{equation}

The ambipolar electromagnetic force (EMF) is given by
\begin{equation}
\mathbf{E}_{AD} =({\mathbf{v}_i}-{\mathbf{v}_n})\times  \mathbf{B} = \frac{1}{\gamma_{AD} \rho_i \rho}\mathbf{F}_{L} \times \mathbf{B},
\label{Fad}
\end{equation}
where $\mathbf{v}_i$ and $\mathbf{v}_n$ denote respectively the ions and neutrals velocities, and $\gamma_{AD}$ is the drift coefficient between ions and neutrals due to ambipolar diffusion. The last equality in Equation~(\ref{Fad}) illustrates the balance between magnetic and drag forces in the ion fluid, while Equation~(\ref{eqenergDA}) is accounting for ambipolar heating of neutrals by ions (\citet{Shu_1992}). In order to write both Equation~(\ref{eqenergDA}) and Equation~(\ref{Fad}) we need to assume that the velocity drift between ions and neutrals and the one between electrons and neutrals are not too different (by a factor $\simeq \frac{m_i}{m_e}$). Therefore, as long as the Hall effect is negligible, these equations remain valid (see \citet{PintoGalli1} \m{for a more detailed study}). 

Equation~(\ref{eqenergDA}) (the conservation of energy: $\frac{\partial E}{\partial t} + \nabla \cdot \mathbf{\mathcal{F}_{\textrm{energy}}} = 0$) is equivalent to
\begin{equation}
\rho T \frac{d s}{d t} = \frac{\| (\mathbf{\nabla} \times \mathbf{B}) \times \mathbf{B} \|^2}{\gamma_{AD} \rho \rho_i}, \label{eqn001}
\end{equation} 
where we can see that the neutrals-ions friction term heats up the gas and increases the entropy.

\subsection{Computing the ambipolar diffusion terms}

In this section, we describe the numerical implementation of the previous equations, focusing on the ambipolar diffusion terms in the energy and induction equations.

\subsubsection{The ambipolar EMF} 
\label{ambterm}

The ambipolar term in the induction equation can be considered as an additional electromotive force (EMF). To update the magnetic field the values of the EMF have to be defined as time and space averages along cell edges (\citet{teyssierMHD}). For instance, the EMF in the $z$ direction is defined at $x_{i-\frac{1}{2}},y_{j-\frac{1}{2}},z_{k}$ (and the same for the other directions with circular permutations) \label{revv3}\revv{where i,j and k are the cell indices in the $x,y,z$ directions respectively}.

We focus now on the EMF in the $z$ direction, defined at $x_{i-\frac{1}{2}},y_{j-\frac{1}{2}},z_{k}$ and explain in details how it is computed. $\mathbf{E}_{AD}$ writes 
\begin{align}
\mathbf{E}_{AD} = \frac{1}{\gamma_{AD} \rho_i \rho} \left[ ({\nabla} \times  \mathbf{B}) \times \mathbf{B} \right]  \times \mathbf{B} = \mathbf{v_d} \times \mathbf{B}
\end{align} 
 with the drift velocity $\mathbf{v_d} = \mathbf{v_i}-\mathbf{v_n} = \frac{1}{\gamma_{AD} \rho_i \rho} \mathbf{F_L}$. We therefore have to evaluate  $\frac{1}{\gamma_{AD} \rho_i \rho}, \ \mathbf{F_L}$ and $\mathbf{B}$ at $x_{i-\frac{1}{2}},y_{j-\frac{1}{2}},z_{k}$, and then calculate
\begin{align}
E^{AD}_{z;i-\frac{1}{2},j-\frac{1}{2},k} &= (v_d)_{x;i-\frac{1}{2},j-\frac{1}{2},k} B_{y;i-\frac{1}{2},j-\frac{1}{2},k} - (v_d)_{y;i-\frac{1}{2},j-\frac{1}{2},k} B_{x;i-\frac{1}{2},j-\frac{1}{2},k}.
\end{align}

\begin{center}
\begin{figure}
\includegraphics[width=0.33\textwidth]{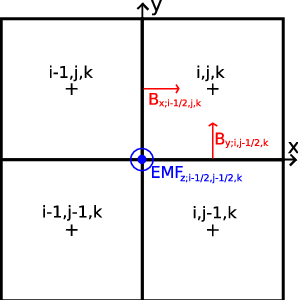}
\includegraphics[width=0.315\textwidth]{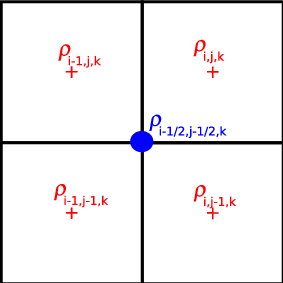}
\caption{Left: coordinates of the center of neighbouring cells, with the natural places where the magnetic field and the EMF are defined. Right: computation of the density at $x_{i-\frac{1}{2}},y_{j-\frac{1}{2}},z_{k}$ by averaging over neighbouring cells.}
\label{schemas12}
\end{figure}
\end{center}

The {\ttfamily RAMSES} code is based on the Constrained Transport scheme for the magnetic field evolution (\citet{teyssierMHD}), for which the components of the magnetic field are defined at the center of cell faces: if $x_i,y_j,z_k$ are the coordinates of a cell center, $B_x$ is defined at position $x_{i-\frac{1}{2}},y_j,z_k$, $B_y$ at $x_i , y_{j-\frac{1}{2}},z_k$ and $B_z$ at $x_i ,y_j, z_{k-\frac{1}{2}}$ (see Figure~\ref{schemas12}). Each magnetic field component is computed using the finite-surface approximation, which reads for the x component:
\begin{align}
 \left\langle B_{x;i-\frac{1}{2},j,k} \right\rangle = & \frac{1}{\Delta y} \frac{1}{\Delta z} \int_{y_{i-\frac{1}{2}}} ^{y_{i+\frac{1}{2}}} \int_{z_{i-\frac{1}{2}}} ^{z_{i+\frac{1}{2}}} B_x(x_{i-\frac{1}{2}},y,z) dy dz,
\end{align} 
while other components are defined by circular permutations.

We also need to define the drift velocity $\mathbf{v_d}$  at $x_{i-\frac{1}{2}},y_{j-\frac{1}{2}},z_{k}$ using the Lorentz force $\mathbf{F_L}$, the density $\rho$ and the ions density $\rho_i$. The two latter quantities \m{are} cell-centered quantities (in contrast to the magnetic field):
\begin{align}
 \left\langle \rho_{i,j,k} \right\rangle = & \frac{1}{\Delta x} \frac{1}{\Delta y} \frac{1}{\Delta z} \int_{x_{i-\frac{1}{2}}} ^{x_{i+\frac{1}{2}}} \int_{y_{i-\frac{1}{2}}} ^{y_{i+\frac{1}{2}}} \int_{z_{i-\frac{1}{2}}} ^{z_{i+\frac{1}{2}}} \rho(x,y,z) dx dy dz,
\end{align}
and 
\begin{align}
 \left\langle \rho_{\textrm{ions};i,j,k} \right\rangle = & \frac{1}{\Delta x} \frac{1}{\Delta y} \frac{1}{\Delta z} \int_{x_{i-\frac{1}{2}}} ^{x_{i+\frac{1}{2}}} \int_{y_{i-\frac{1}{2}}} ^{y_{i+\frac{1}{2}}} \int_{z_{i-\frac{1}{2}}} ^{z_{i+\frac{1}{2}}} \rho_{i}(x,y,z) dx dy dz,
\end{align}

We then define the edge-centered density (and the ions edge-centered density) as an arithmetic average of surrounding cells (see Figure~\ref{schemas12}, right panel):
\begin{align}
\rho_{i-\frac{1}{2},j-\frac{1}{2},k} = & \frac{1}{4} [  \rho_{i,j,k}+\rho_{i,j-1,k} + \rho_{i-1,j,k}+\rho_{i-1,j-1,k} ].
\end{align}
  
\begin{center}
\begin{figure}
\includegraphics[width=0.33\textwidth]{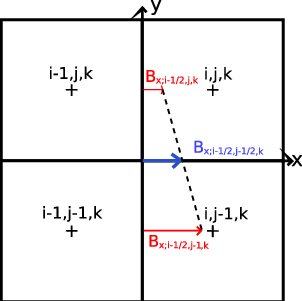}
\includegraphics[width=0.33\textwidth]{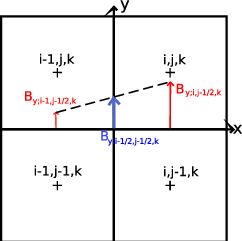}
\caption{Computation of the magnetic field $B_x$ and $B_y$ at $x_{i-\frac{1}{2}},y_{j-\frac{1}{2}},z_{k}$.}
\label{schemas32}
\end{figure}
\end{center}

This definition is adapted for the components of the magnetic field to compute them at $x_{i-\frac{1}{2}},y_{j-\frac{1}{2}},z_{k}$ (see Figure~\ref{schemas32}):  
\begin{align}
B_{x;i-\frac{1}{2},j-\frac{1}{2},k} &= \frac{1}{2} \Big[ B_{x;i-\frac{1}{2},j,k} + B_{x;i-\frac{1}{2},j-1,k} \Big]  \label{eqnBx} \\
B_{y;i-\frac{1}{2},j-\frac{1}{2},k} &= \frac{1}{2} \Big[ B_{y;i,j-\frac{1}{2},k} + B_{y;i-1,j-\frac{1}{2},k} \Big]  \label{eqnBy} .
\end{align}

Given the Lorentz force at $x_{i-\frac{1}{2}},y_{j-\frac{1}{2}},z_{k}$ (see Sections~\ref{calculexplicitlorentz} and \ref{divduflux}), it is possible to compute the first component (z direction, with unit vector $\mathbf{e_Z}$) of the ambipolar EMF: $\mathbf{E_{AD}} \cdot \mathbf{e_z} = E^{AD}_{z;i-\frac{1}{2},j-\frac{1}{2},k}$, while the two other components are obtained through circular permutations.

These ambipolar EMFs are then added to the ideal MHD EMFs calculated with a 2D Riemann solver, as described in \citet{teyssierMHD} and \citet{fromang}.

\subsubsection{The Lorentz force as the product of the current and the magnetic field}
\label{calculexplicitlorentz}

We now focus on the computation of the Lorentz force $\mathbf{F_L}$ at $x_{i-\frac{1}{2}},y_{j-\frac{1}{2}},z_{k}$. The first way to calculate this term is to explicitly compute the magnetic field components and the current $\mathbf{J} = \nabla \times \mathbf{B}$ at 
$x_{i-\frac{1}{2}},y_{j-\frac{1}{2}},z_{k}$, as:
\begin{equation}
\mathbf{F_L} = \mathbf{J} \times \mathbf{B} .
\end{equation}

$\mathbf{J_z} = \frac{\partial B_y}{\partial x}-\frac{\partial B_x}{\partial y}$ is naturally defined at $x_{i-\frac{1}{2}},y_{j-\frac{1}{2}},z_{k}$ and $\mathbf{J_x}$ and $\mathbf{J_y}$ are naturally defined respectively at $x_{i},y_{j-\frac{1}{2}},z_{k-\frac{1}{2}}$ and $x_{i-\frac{1}{2}},y_{j},z_{k-\frac{1}{2}}$. In order to have all three components of $\mathbf{J}$ at the location of the EMF ($x_{i-\frac{1}{2}},y_{j-\frac{1}{2}},z_{k}$) we need to use the magnetic field components at specific positions, as follows: 
\begin{align}
J_x = {\Delta y}^{-1} \big( &B_{z;i-\frac{1}{2},j,k} - B_{z;i-\frac{1}{2},j-1,k} \big) - {\Delta z}^{-1} \big( B_{y;i-\frac{1}{2},j-\frac{1}{2},k+\frac{1}{2}} - B_{y;i-\frac{1}{2},j-\frac{1}{2},k-\frac{1}{2}} \big) \label{eqn002} \\ 
J_y = {\Delta z}^{-1} \big( &B_{x;i-\frac{1}{2},j-\frac{1}{2},k+\frac{1}{2}} - B_{x;i-\frac{1}{2},j-\frac{1}{2},k-\frac{1}{2}} \big) - {\Delta x}^{-1} \big( B_{z;i,j-\frac{1}{2},k} - B_{z;i-1,j-\frac{1}{2},k} \big) \label{eqn003} \\
J_z = {\Delta x}^{-1} \big( &B_{y;i,j-\frac{1}{2},k} - B_{y;i-1,j-\frac{1}{2},k})\big)  - {\Delta y}^{-1} \big( B_{x;i-\frac{1}{2},j,k} - B_{x;i-\frac{1}{2},j-1,k} \big) \label{eqn004}.
\end{align}

As above, we have to express each of these terms through arithmetic averages of the naturally defined components of the magnetic field:
\begin{eqnarray}
B_{x;i-\frac{1}{2},j,k} \quad &\textrm{and}& \quad B_{y;i,j-\frac{1}{2},k} \label{direct}\\
B_{z;i,j-\frac{1}{2},k} \quad &\textrm{and}& \quad B_{z;i-\frac{1}{2},j,k} \label{mag1}\\
B_{y,i-\frac{1}{2},j-\frac{1}{2},k-\frac{1}{2}} \quad &\textrm{and}& \quad B_{x,i-\frac{1}{2},j-\frac{1}{2},k-\frac{1}{2}} \label{mag2}.
\end{eqnarray}

\m{$B_{x}$ and $B_{y}$ are naturally defined at $x_{i-\frac{1}{2}},y_{j},z_{k}$ and  $x_{i},y_{j-\frac{1}{2}},z_{k}$ respectively, and thus the two terms in Equation~(\ref{direct}) need not to be computed again.}

\begin{center}
\begin{figure}
\includegraphics[width=0.33\textwidth]{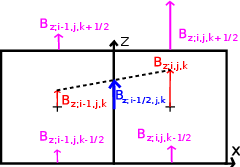}
\includegraphics[width=0.33\textwidth]{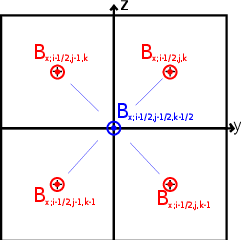}
\caption{Left: $B_{z;i-\frac{1}{2},j,k}$ as an average over surrounding cells. Right: $B_{x,i-\frac{1}{2},j-\frac{1}{2},k-\frac{1}{2}}$.}
\label{schemas45}
\end{figure}
\end{center}

The terms \m{in} Equation~(\ref{mag1}) and Equation~(\ref{mag2}) are obtained thanks to averages on four components (see Figure~\ref{schemas45}):
\begin{align}
B_{z;i-\frac{1}{2},j,k} &= \frac{1}{4}\big[ B_{z;i,j,k+\frac{1}{2}} + B_{z;i,j,k-\frac{1}{2}}  + B_{z;i-1,j,k+\frac{1}{2}} + B_{z;i-1,j,k-\frac{1}{2}} \big] \nonumber \\
      &= \frac{1}{2}\big[ B_{z;i,j,k} + B_{z;i-1,j,k} \big] \\
B_{z;i,j-\frac{1}{2},k} &= \frac{1}{2}\big[ B_{z;i,j,k} + B_{z;i,j-1,k} \big] \\
B_{x,i-\frac{1}{2},j-\frac{1}{2},k-\frac{1}{2}} &= \frac{1}{4} [B_{x;i-\frac{1}{2},j,k} + B_{x;i-\frac{1}{2},j-1,k} + B_{x;i-\frac{1}{2},j,k-1} + B_{x;i-\frac{1}{2},j-1,k-1} ] \\
B_{y,i-\frac{1}{2},j-\frac{1}{2},k-\frac{1}{2}} &= \frac{1}{4} [B_{y;i,j-\frac{1}{2},k} + B_{y;i-1,j-\frac{1}{2},k}  + B_{y;i,j-\frac{1}{2},k-1} + B_{y;i-1,j-\frac{1}{2},k-1} ] .
\end{align}

The Lorentz force for the other components of the EMF ($x$ and $y$ directions) are obtained through circular permutations.

\subsubsection{The Lorentz force as the divergence of a flux}
\label{divduflux}

Another way to compute the Lorentz force is to express it as the divergence of a well chosen flux:
\begin{equation}
\mathbf{F_L} = (\nabla \cdot \mathbf{\mathcal{F}_i}) \mathbf{e_i} ,
\end{equation}
and
\begin{equation}
\mathbf{\mathcal{F}_i} = B_i B_j \mathbf{e_j} - \delta_{ij} e_{\textrm{mag}} \mathbf{e_j},
\end{equation}
with $i,j \in [x,y,z]$ and $e_{\textrm{mag}} = \frac{1}{2} (B_x^2 + B_y^2 + B_z^2)$.

Let us focus on the $x$ component of the Lorentz force for the EMF in the $z$ direction. It reads:
\begin{align}
\mathbf{F_L} \cdot \mathbf{e_x} = \partial_x (B_x^2 - e_{\textrm{mag}}) + \partial_y (B_x B_y) + \partial_z (B_x B_z).
\end{align}

In order to compute the Lorentz force at $x_{i-\frac{1}{2}},y_{j-\frac{1}{2}},z_{k}$, we compute each term at specific positions:
\begin{align}
\partial_x (B_x^2 - e_{\textrm{mag}}) = \frac{1}{2 \Delta x} \big[ &(B_{x;i,j-\frac{1}{2},k}^2 - B_{x;i-1,j-\frac{1}{2},k}^2 \big) \nonumber \\
 - &( B_{y;i,j-\frac{1}{2},k}^2 -  B_{y;i-1,j-\frac{1}{2},k}^2) \nonumber \\
 - &( B_{z;i,j-\frac{1}{2},k}^2 - B_{z;i-1,j-\frac{1}{2},k}^2)  \big] \label{eqn005}\\ 
\partial_y (B_x B_y)                  =& \frac{1}{\Delta y} \big[  B_{x,i-\frac{1}{2},j,k} B_{y,i-\frac{1}{2},j,k}  - B_{x,i-\frac{1}{2},j-1,k} B_{y,i-\frac{1}{2},j-1,k}  \big] \\
\partial_z (B_x B_z)                  =& \frac{1}{\Delta z} \big[  B_{x,i-\frac{1}{2},j-\frac{1}{2},k+\frac{1}{2}} B_{z,i-\frac{1}{2},j-\frac{1}{2},k+\frac{1}{2}}  -  B_{x,i-\frac{1}{2},j-\frac{1}{2},k-\frac{1}{2}} B_{z,i-\frac{1}{2},j-\frac{1}{2},k-\frac{1}{2}}  \big]. 
\end{align}

We then only need to compute each component of the magnetic field at $x_{i-\frac{1}{2}},y_{j-\frac{1}{2}},z_{k}$ in order to get the EMF in the $z$ direction. 

\begin{center}
\begin{figure}
\includegraphics[width=0.5\textwidth]{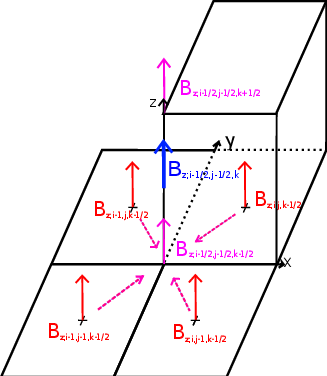}
\caption{Computation of $B_{z;i-\frac{1}{2},j-\frac{1}{2},k}$ as an average over eight naturally defined magnetic components.}
\label{schema6}
\end{figure}
\end{center}

As explained in the previous paragraph, an average over well chosen (where the magnetic field is naturally defined) surrounding cells is used (see Figures~\ref{schemas32} and~\ref{schema6}):

\begin{align}
  B_{x;i,j-\frac{1}{2},k} &= \frac{1}{2}\big[ B_{x;i,j,k} + B_{x;i,j-1,k} \big] \\
  B_{z;i,j-\frac{1}{2},k} &= \frac{1}{2}\big[ B_{z;i,j,k} + B_{z;i,j-1,k} \big] \\
  B_{y;i-\frac{1}{2},j,k} &= \frac{1}{2}\big[ B_{y;i,j,k} + B_{y;i-1,j,k} \big] \label{eqn00001}\\
  B_{x,i-\frac{1}{2},j-\frac{1}{2},k-\frac{1}{2}} &= \frac{1}{4} [B_{x;i-\frac{1}{2},j,k} + B_{x;i-\frac{1}{2},j-1,k} \nonumber \\
                        & + B_{x;i-\frac{1}{2},j,k-1} + B_{x;i-\frac{1}{2},j-1,k-1} ] \\
  B_{z,i-\frac{1}{2},j-\frac{1}{2},k-\frac{1}{2}} &= \frac{1}{4} [B_{z;i,j,k-\frac{1}{2}} + B_{z;i,j-1,k-\frac{1}{2}} \nonumber \\
                        & + B_{z;i-1,j,k-\frac{1}{2}} + B_{z;i-1,j-1,k-\frac{1}{2}} ] \label{eqn00002} ,
\end{align}
and
\begin{align}
  B_{x;i-\frac{1}{2},j-\frac{1}{2},k} &= \frac{1}{2}[B_{x;i-\frac{1}{2},j,k} + B_{x;i-\frac{1}{2},j-1,k} ] \\
  B_{y;i-\frac{1}{2},j-\frac{1}{2},k} &= \frac{1}{2}[B_{y;i,j-\frac{1}{2},k} + B_{y;i-1,j-\frac{1}{2},k} ] \\
  B_{z;i-\frac{1}{2},j-\frac{1}{2},k} &= \frac{1}{4} \Big[  \frac{1}{2}( B_{z;i,j,k-\frac{1}{2}}+B_{z;i,j-1,k-\frac{1}{2}} ) \nonumber \\
                      &   +                 \frac{1}{2}( B_{z;i-1,j,k-\frac{1}{2}}+B_{z;i-1,j-1,k-\frac{1}{2}} ) \nonumber \\
                      &   +                 \frac{1}{2}( B_{z;i,j,k+\frac{1}{2}}+B_{z;i,j-1,k+\frac{1}{2}} ) \nonumber \\
                      &   +                 \frac{1}{2}( B_{z;i-1,j,k+\frac{1}{2}}+B_{z;i-1,j-1,k+\frac{1}{2}} ) \Big] \label{eqn006} \nonumber \\
                      &= \frac{1}{4}[B_{z;i,j,k}+B_{z;i,j-1,k} \nonumber \\
                      &+ B_{z;i-1,j,k}+B_{z;i-1,j-1,k}] .
\end{align}

Again and as highlighted previously, in order to get the two other components of the EMF one only needs to perform circular permutations.

These two methods (described in Sections~\ref{calculexplicitlorentz} and \ref{divduflux}) to compute the Lorentz force are implemented in {\ttfamily RAMSES} and show similar performances. \label{revv5}\revv{This method might work better under certain conditions, for a particular setup of the magnetic field lines. Nonetheless, when counting the number of floating point operations, the computer performs using this method 4911 more additions and 8047 more multiplications for a given cell than with the previously described method.}

\subsubsection{Contribution of ambipolar diffusion to the energy flux \label{fluxenergAD}}

The ambipolar energy flux (see Equation~\ref{fluxambi}) has to be evaluated on each face of the cell, that is to say at locations  $(x_{i \pm \frac{1}{2}},y_j,z_k)$, $(x_i,y_{j \pm \frac{1}{2}},z_{k})$ and $(x_i,y_j,z_{k \pm \frac{1}{2}})$. Again, as in \S~\ref{ambterm}, the needed components are obtained thanks to averages over neighbouring cells (averages which are not detailed here). 
\begin{equation}
\mathbf{\mathcal{F}_{AD}} = - \mathbf{E_{AD}} \times \mathbf{B} = - \frac{1}{\gamma_{AD} \rho_i \rho} \left( (\mathbf{J} \times  \mathbf{B}) \times  \mathbf{B} \right)  \times  \mathbf{B} \label{fluxambi}.
\end{equation}

\subsubsection{Computation of the time step in presence of ambipolar diffusion}

The ambipolar diffusion timescale can be estimated through the drift velocity of ions. Recalling Equation~(\ref{Fad}) we get:
\begin{equation}
\begin{split}
\|\mathbf{v_{\textrm{drift}}}\| & \propto \frac{1}{\gamma_{AD} \rho \rho_i} \| \mathbf{F_L} \| \\
                              & \propto \frac{v_A^2}{\gamma_{AD} \rho_i L_{AD}},
\end{split}
\end{equation}
where $L_{AD}$ is a characteristic length for ambipolar diffusion, which can be estimated as \label{rev2} \rev{$L_{AD}^{-1} = \frac{\nabla (\|B\|)}{\|B\|}$}. We then have the timescale:
\begin{equation}
\tau_{AD} = \frac{L_{AD}}{\|\mathbf{v_{\textrm{drift}}}\|} =  \frac{\gamma_{AD} \rho_i L_{AD}^2}{v_A^2}.
\end{equation}

Written as a diffusion, $\tau_{AD} = \frac{L_{AD}^2}{D}$ with the ambipolar diffusion coefficient $D = \frac{v_A^2}{\gamma_{AD} \rho_i}$, where $\mathbf{v_A}=\frac{\mathbf{B}}{\sqrt{\rho}}$ is the Alfv\'en speed and $(\gamma_{AD} \rho_i)^{-1}$ is the characteristic collision time between ions and neutrals. \label{rev6}\rev{A Von Neumann analysis for the diffusion part of the equation can be performed for the scheme used:
\begin{align}
\frac{\partial \mathbf{B}}{\partial t} - {\nabla} \times \mathbf{E}_{AD} &= 0 .
\end{align}
It can be differenced \revv{(in one dimension)}:
\begin{align}
\frac{B_{x;i-\frac{1}{2},j,k}^{n+1} - B_{x;i-\frac{1}{2},j;k}^{n}}{\Delta t} &=  D \frac{\Delta t}{{\Delta x}^2} (B_{x;i-\frac{1}{2},j+1,k}^{n} - 2 B_{x;i-\frac{1}{2},j,k}^{n} + B_{x;i-\frac{1}{2},j-1,k}^{n}).
\end{align}
Using $B_j^n = \epsilon^n e^{ikjh}$,
\begin{align}
\epsilon=1+2 \frac{D \Delta t}{{\Delta x}^2}(\cos(kh)-1).
\label{eqnvonneumann}
\end{align}
Equation~(\ref{eqnvonneumann}) shows that the scheme is stable according to Von Neumann stability analysis provided the coefficient is lower than 0.5:
\begin{align}
| \epsilon | < 1 \Leftrightarrow \Delta t < \frac{1}{2} \frac{{\Delta x}^2}{D} = \frac{1}{2} \frac{{\Delta x}^2 \gamma_{AD} \rho_i \rho}{B^2}. 
\end{align}}
\label{revv2}\revv{For the three dimensional case, this time-step constraint is more stringent than for the one dimensional case presented above.}

Therefore, the time step used to update the solution is computed by taking the minimum of the usual MHD Courant condition (\citet{fromang}) and the ambipolar timestep defined by 
\begin{equation}
t_{AD} =  0.1 \times \min (\frac{\gamma_{AD} \,  \rho_i }{v_A^2}\, \Delta x ^ 2 )
\label{timestepAD},
\end{equation}
where the minimum is taken over all the cells of the computational grid. \rev{The coefficient $0.1 < \frac{1}{2}$ is taken to achieve better convergence. This choice is based on the various tests performed, and might not be suited \m{to all} other problems.} \label{rev4}\rev{As can be noted in equation (\ref{timestepAD}), the time-step scales as $\Delta x ^{ 2}$. Even though this is very demanding in terms of numerical resources as the grid becomes more and more refined, there are means to speed-up the calculations, as explained in the following paragraph.}

The ambipolar time step is proportional to $\rho_i$ (see Equation~(\ref{timestepAD})), which is assumed to be proportional to $\rho^k$: $\rho_i=C\sqrt{\rho}$ (see \citet{elmegreen1979}). Both the factor ($C$) and the power law ($\rho^{\frac{1}{2}}$) are very dependent on the microphysics and the geometry of the grains. This assumption is thus made for \m{the} sake of simplicity, but might not always be valid. In some cases, for example in star formation simulations, the time-step can become unphysically small in very diffuse regions where the ionisation approximated as above (Equation~(\ref{timestepAD})) is very small. Following \citet{nakamura2008}, we use a threshold in order to limit the time-step when needed. \rev{On the other hand, in very dense parts where the grid is fairly refined (where $\Delta x$ is small), the dependence of $\rho_i$ and $\gamma_{AD}$ with the density \m{prevent the time-step from becoming too becoming too small}. This situation has to be studied for each different physical problem and can't be assumed once and for all. We will address this issue in the case of \m{star-formation} in a forthcoming paper.}

%Another way to speed up calculations is to use the super time stepping (STS) method (see \citet{AlexiadesSTS}), as done by \citet{Choietal}. The super time stepping consists in demanding stability over many time-steps of well-chosen length, instead of demanding to fullfill the Courant condition at each time-step. The gain in computation time is proportional to $\sqrt{N}$ where $N$ is the number of time steps respecting the Courant condition necessary to advance from a time $t_0$ to $t_0 + \Delta t$. The results with STS simulations are in perfect agreement with those with an explicit time step, as can be seen Figure~\ref{compstspassts}. Both approaches (threshold and STS) are integrated into {\ttfamily RAMSES}.

%\begin{center}
%\begin{figure}
%%\includegraphics[width=0.33\textwidth,angle=270]{compSTSexplicit}
%\includegraphics[width=0.63\textwidth,angle=270]{compSTSexplicit}
%\caption{Comparison between the simulation with explicit time step and super time stepping for Alfv\'en propagating waves, with a fully refined grid of 32 cells.}
%\label{compstspassts}
%\end{figure}
%\end{center}

\label{rev9}
\rev{\subsection{The AMR scheme}}

\rev{The AMR algorithm used in {\ttfamily RAMSES} is described in \citet{teyssier}, and its extension to MHD is first described in \citet{teyssierMHD} and then in \citet{fromang}. We briefly recall the main features here. It is a tree-based AMR code whose data structure is a 'Fully Threaded Tree" (\citet{Khokhlov}). The grid is divided into "octs" which are groups of 8 cells with the same parent cell. The first level of refinement ($l=1$) corresponds to the unit cube, which defines the computational domain. The grid is recursively refined from the $l = 1$ to the minimum level of refinement $l_{min}$, in order to build the base Cartesian grid. Adaptive refinement then proceeds from this coarse grid up to the user-defined maximum level of refinement $l_{max}$. When $l_{max} = l_{min}$ the computational grid is a traditional Cartesian grid. Issues arise when refined cells are created, in the case where $l_{max} > l_{min}$. Concerning the non-ideal MHD, the EMF and energy fluxes are simply added to the existing ideal MHD EMF and fluxes. As a consequence, there are no more complications in refining and derefining cells than in the ideal MHD case.}

\rev{\subsubsection{Divergence-free prolongation operator: refining cells}}

\rev{The "prolongation operator" is the creation of a new "oct" of 8 cells when a cell is newly refined. Cell-centered variables and magnetic field components are needed for each refined cell. This is usually done using a conservative interpolation of the variables, yet in the case of magnetic fields, the divergence-free constraint has to be fulfilled by each of the new cells which makes things more complicated in details. A critical step has been solved by \citet{Balsara2001} and \citet{tothroe} in the constrained transport framework. The idea developed in those articles is to use slope limiters to interpolate the magnetic field components in each parent face conserving the flux, and then to perform a three dimensional (which is divergence-free inside the cell volume) reconstruction in order to compute the new magnetic field components for each children faces. The same slope limiters as the ones used in the Godunov scheme for the hydrodynamics are used in this step.}

\rev{\subsubsection{Magnetic flux corrections: derefining cells}}

\rev{The "Restriction Operator" is, in the multigrid terminology, the operation of derefining a split cell. The divergence-free constraint still needs to be satisfied, so that the magnetic field components in the coarse faces are simply the arithmetic averages of the four fine faces values. This is the parallel in MHD of the "flux correction step" for the Euler system.}

\rev{\subsubsection{EMF corrections}}

\rev{This is specific to the induction equation: for a coarse face adjacent to a refined face, the coarse EMF in the conservative update of the solution needs to be replaced by the arithmetic average of the two fine EMF vectors. This is mandatory to guarantee that the magnetic field remains divergence-free even at coarse/fine boundaries.}

\subsection{Tests for the ambipolar diffusion}

\subsubsection{The Barenblatt diffusion test}
\label{ambalone}

In this section, we first test the accuracy of the calculation of the ambipolar term alone. For sake of simplicity, we assume that the magnetic field has the form $B_y(x,z)$, with $B_x=0$ and $B_z=0$, that all the velocities remain equal to zero and that density and thermal pressure are constant. The induction equation takes the form of a diffusion equation: 
\begin{align}
\frac{\partial B_y}{\partial t} = \frac{\partial}{\partial x} \Big( \frac{B_y^2}{\gamma_{AD} \rho_i \rho} \frac{\partial}{\partial x} (B_y) \Big) 
 + \frac{\partial}{\partial z} \Big( \frac{B_y^2}{\gamma_{AD} \rho_i \rho}  \frac{\partial}{\partial z} (B_y) \Big),
 \end{align}
which can also be written in compact form: 
\begin{eqnarray}
\frac{\partial B_y}{\partial t} = {\nabla} . \Big( \frac{v_{A}^2}{\gamma_{AD} \rho_i }  {\nabla} B_y \Big) \label{eqn10}.
\label{bybaren}
\end{eqnarray}

This is a non-linear diffusion equation, since the diffusion coefficient, $\eta_{AD} = \frac{v_{A}^2}{\gamma_{AD} \rho_i } $, depends non-linearly on the magnetic field. Here, $v_{A}=B_y/{\sqrt \rho}$ denotes the y-component of the Alfv\'en velocity. The solution of this problem with a Dirac pulse as initial condition (known as the Barenblatt-Pattle solutions) has been derived by \citet{refbaren2} (See Appendix~\ref{annexebaren} for more details about the analytical solution).

The initial states in one and two dimensions are respectively:
\begin{align}
B_{y0} &= \left\{ \begin{array}{ll}
 1 \; & \textrm{if} \, \|x-x_{center}\| \leq 0.9 \, {\Delta x}_{level=3}\\
 0 \; & \textrm{elsewhere} 
\end{array} \right. \\
B_{y0} &= \left\{ \begin{array}{ll}
 1 \; & \textrm{if} \; \sqrt{(x-x_{center})^2 + (z-z_{center})^2} \leq 0.9 \, {\Delta x}_{level=3} \\
 0 \; & \textrm{elsewhere} .
\end{array} \right.
\end{align}
\label{revv4}\revv{with ${\Delta x}_{level=3}$ being the cell size at the lowest level of refinement used \revv{(in this case: 3)}. This ensures that the initial perturbation is the same for the case of an AMR grid and a fully refined grid.}

We performed the test both with a uniform grid and using the AMR with the same maximum level. \label{rev5}\rev{The level of refinement refers to the number of cells used: $2^{N D}$ cells are used for the level of refinement $N$ in a $D$-dimensional calculation.} As seen in the figures, the agreement between the numerical and the analytical curves is excellent, a few tenths of a percent on average. The results obtained on an AMR grid (with levels varying from 3 to 7, corresponding to a mesh size $\Delta x=0.5^3$ and $\Delta x=0.5^7$) are almost as good as the ones obtained on a regular grid corresponding to the highest level of refinement (level 7, with a cell size $\Delta x = 0.5^7$ and 128 cells): the maximum relative error is less than one percent except where the magnetic field equals zero. The difference between AMR and uniform grid is less than $2.10^{-4}$ for values of magnetic field of about $0.01$. 

\begin{figure*}
\begin{center}
\includegraphics[width=0.34\textwidth,angle=270]{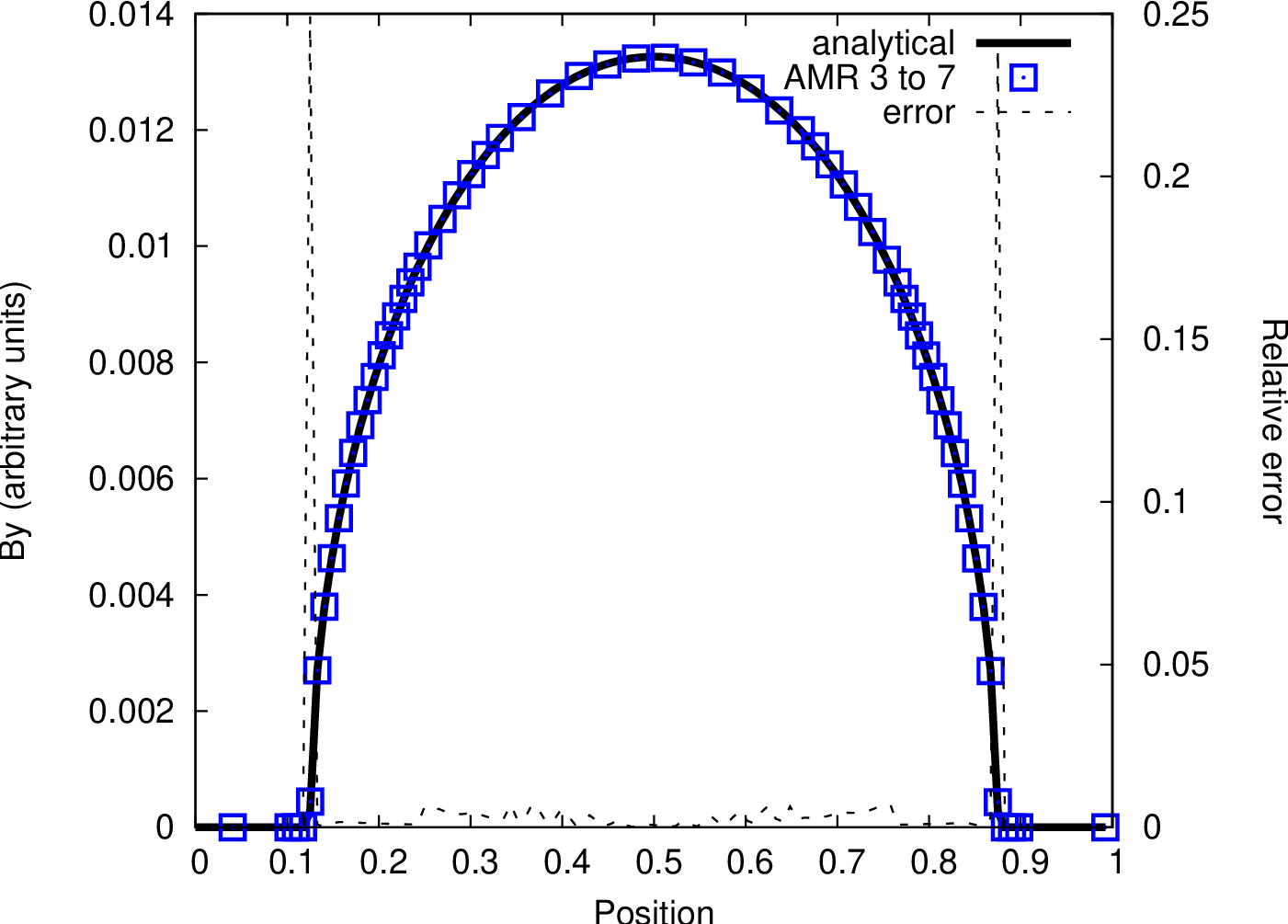}
\includegraphics[width=0.34\textwidth,angle=270]{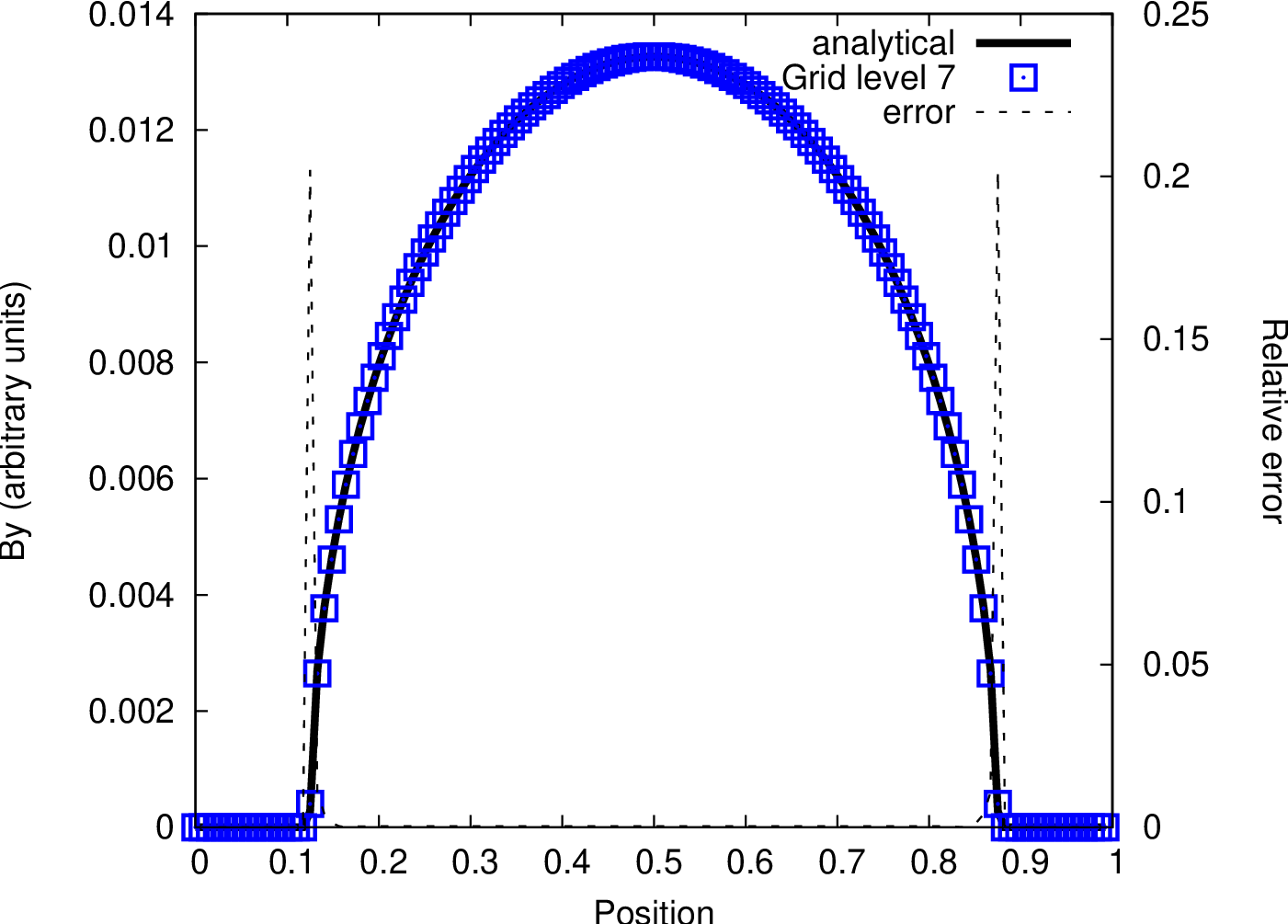}
\caption{Barenblatt diffusion test for ambipolar diffusion at $t=200$ with $B_y$ being a function of $x$ only. The left panel is a snapshot of the AMR run with levels from 3 to 7. The right panel corresponds to a fully refined Cartesian grid up to level 7.}
\label{figbaren1}
\end{center}
\end{figure*}

\begin{figure*}
\begin{center}
\includegraphics[width=0.37\textwidth,angle=270]{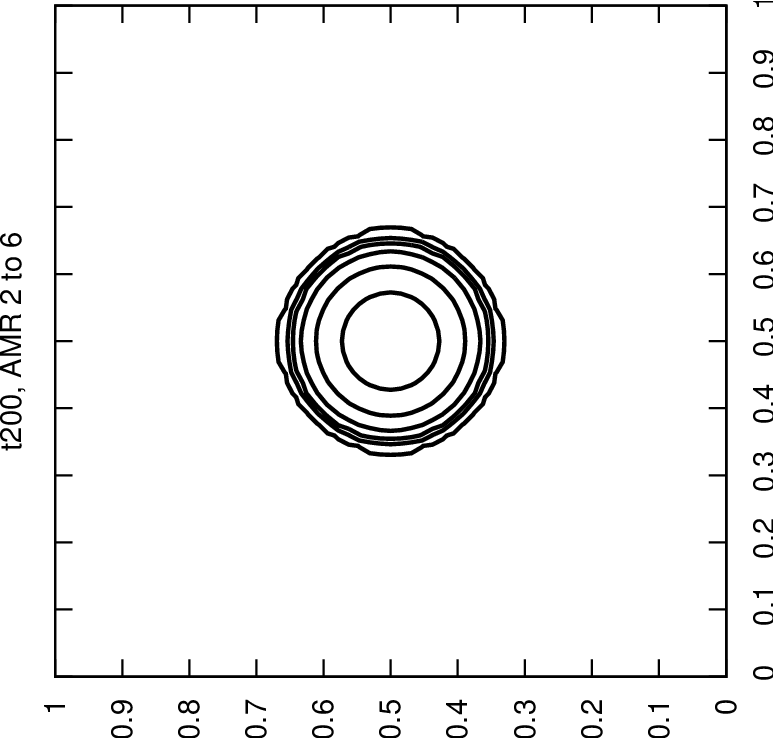}
\includegraphics[width=0.37\textwidth,angle=270]{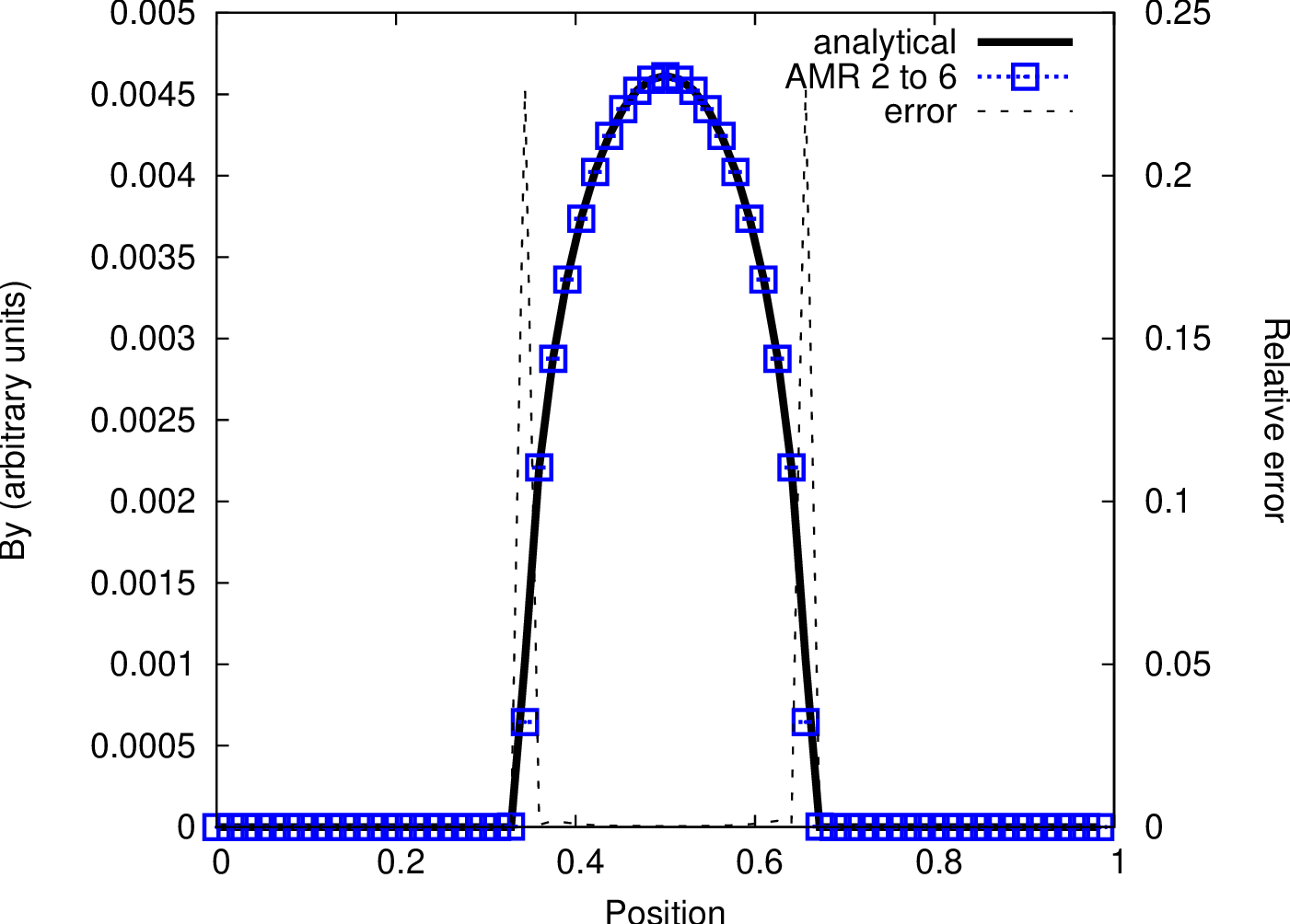}
\caption{Barenblatt diffusion test in 2D with $B_y$ depending on $x$ and $z$. Here, the calculation was performed on an AMR grid from level 2 to 6. The left snapshot is a 2D contour plot at $t=200$: the symmetry of the solution is preserved. The right snapshot (same legend as in Figure~\ref{figbaren1}) is a 1D cut across the maximum at $t=200$.}
\label{figbaren2}
\end{center}
\end{figure*}

The results for $B_y(x)$ are shown on Figure~\ref{figbaren1}, where we have taken $\gamma_{AD}=1$, $\rho_i=1$ and $\rho=1$ and on Figure~\ref{figbaren2} for $B_y(x,z)$.

\begin{figure}
\begin{center}
\includegraphics[width=0.54\textwidth,angle=270]{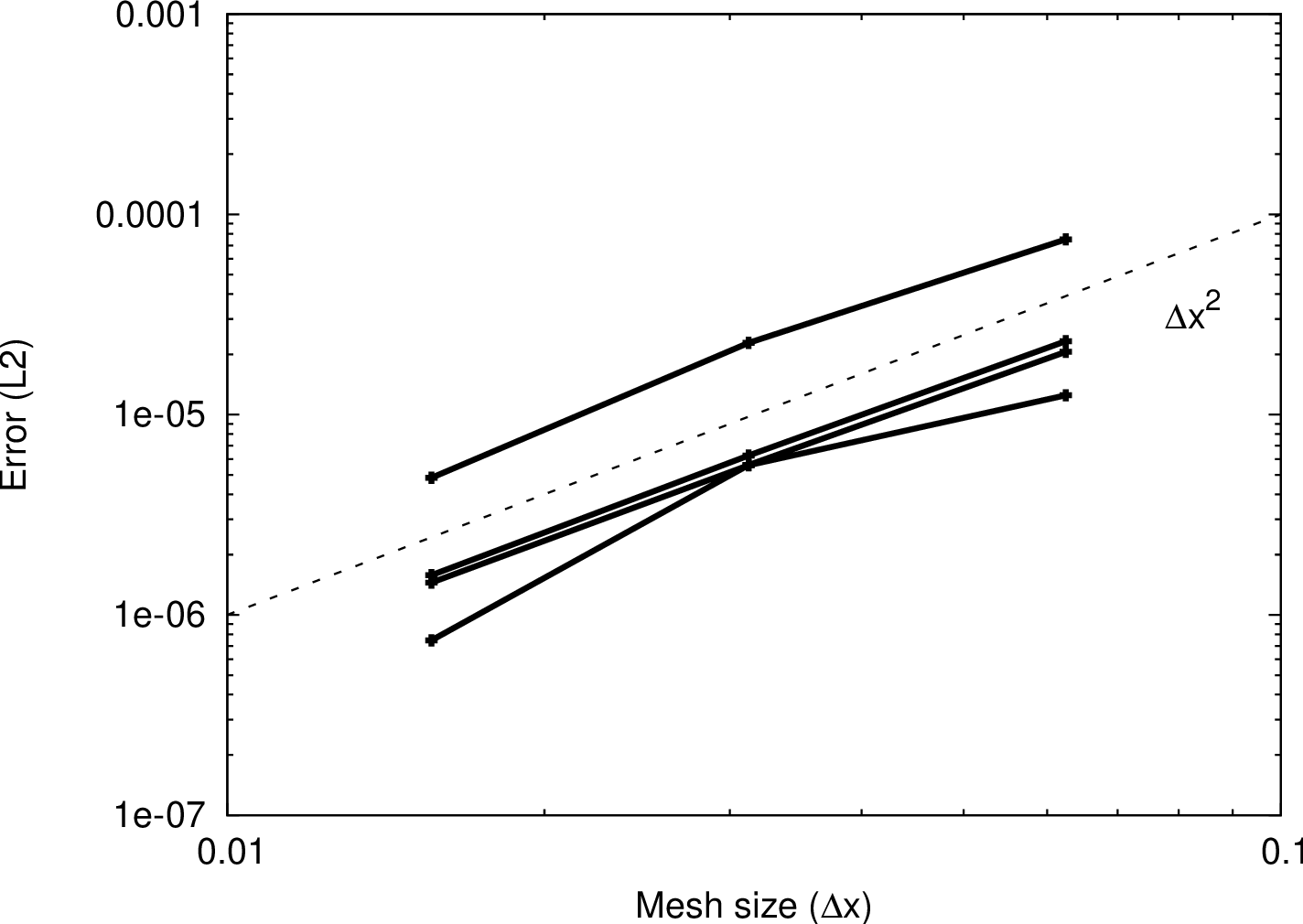}
\caption{Evolution of the error for the Barenblatt test for several times, taking the error as $\epsilon = \sqrt{\sum_{i=1}^{N} \, \frac{(By_{\textrm{numerical}}-By_{\textrm{analytical}})^2}{N}}$, with N the number of cells. The dashed line corresponds to $\epsilon \propto \Delta x^{2}$.}
\label{figordbaren}
\end{center}
\end{figure}

The grid is refined if the gradient of magnetic field is greater than 0.1 (this insures for this test that the error on the AMR grid and on the regular grid are about the same). We also checked that the same accuracy is obtained for any orientation of the magnetic field (using $Bx$ or $Bz$ instead of $By$). Figure~\ref{figordbaren} represents the evolution of the error calculated as $\epsilon = \sqrt{\sum_{i=1}^{N} \, \frac{(By_{\textrm{numerical}}-By_{\textrm{analytical}})^2}{N}}$ as function of the mesh size ($N$ being the number of cells for each level). 

\label{rev_comput}\rev{In terms of computational time for this particular test using the refinement strategy described above, the time is about the same for a regular grid at level 7 as for a grid going from levels 5 to 7, but there is a gain of about $40\%$ in the number of cells. One level further (regular grid at level 8 or AMR grid going from levels 5 to 8) the computation is $30\%$ faster in the AMR case and needs $55\%$ less cells. For a regular grid at level 9 or an AMR grid going from levels 5 to 9 the calculation is $60\%$ faster with $70\%$ less cells needed.}

\subsubsection{The C-shock test}
\label{chocoblDA}
\label{Cshock}

Following \citet{DuffinPudritz} and \citet{McLow}, we have tested our new scheme for the case of both isothermal and non-isothermal oblique C-shock including ambipolar heating as given by Equation~(\ref{eqenergDA}). We start from a steep function as initial state for the different variables, whose values are the ones taken at infinity ahead of and behind the shock. Our calculation takes place in the frame of the shock. The post- and pre-shock values are displayed in Table~\ref{tabchocoblDA}. The angle between the shock normal and the magnetic field is set to 45$^\circ$.

\begin{table*}
\begin{center}
\begin{tabular}{|c|c|c|c|c|c|c|} \hline
Variable & $\rho$ & $v_x$ & $v_y$ & $B_x$ & $B_y$ & $P$ \\
\hline
Pre-shock value & 0.5 & 5 & 0 & $ \sqrt{2}$ & $\sqrt{2}$ & 0.125\\
\hline
Post-shock value (isothermal) & 1.0727 & 2.3305 & 1.3953 & $ \sqrt{2}$ & 3.8809 & 0.2681 \\
\hline
Post-shock value (non-isothermal) & 0.9880 & 2.5303 & 1.1415 & $ \sqrt{2}$ & 3.4327 & 1.4075 \\
\hline
\end{tabular}
\caption{Initial conditions used for the oblique C-shock test, as described in Section~\ref{chocoblDA}.}
\label{tabchocoblDA}
\end{center}
\end{table*}

For this test, we set $\gamma_{AD}=75$, $\rho_i=1$. The sonic Mach number is ${\cal M}=10$ and the Alfv\'en Mach number is ${\cal M}_A=1.8$. Outflow boundary conditions are used in the simulation. After a short transient phase, the shock becomes stationary. 

The isothermal shock is modeled through $P_n = \rho_n c_s^2$ with $c_s = 0.5$ the sound speed, and without solving the energy Equation~(\ref{eqenergDA}). Results are shown in Figure~\ref{figchoc2DA} and compared to the semi-analytical solution described in \citet{McLow} (see Appendix~\ref{maclo} for more details).

\begin{figure*}
\begin{center}
\includegraphics[width=0.33\textwidth,angle=270]{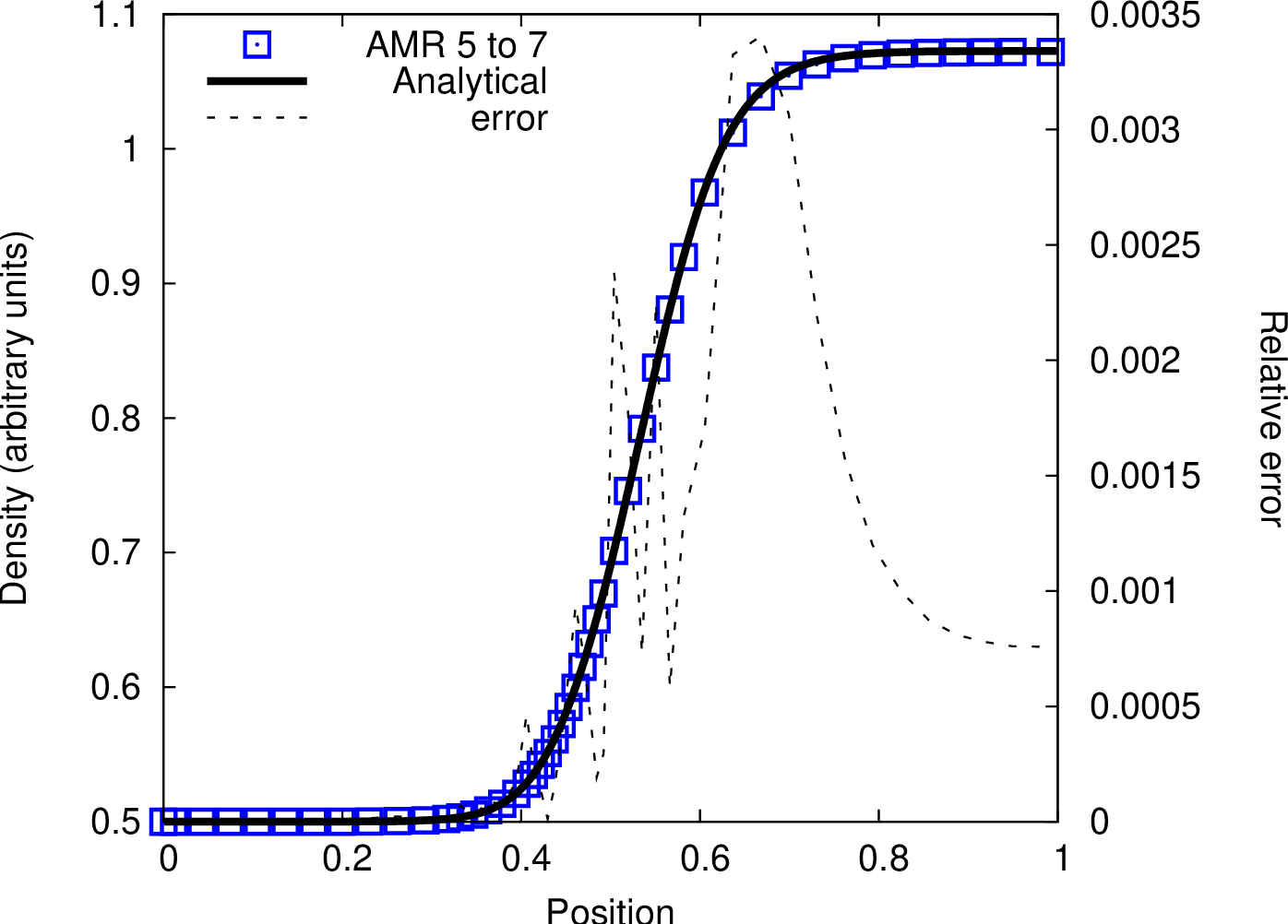}
\includegraphics[width=0.33\textwidth,angle=270]{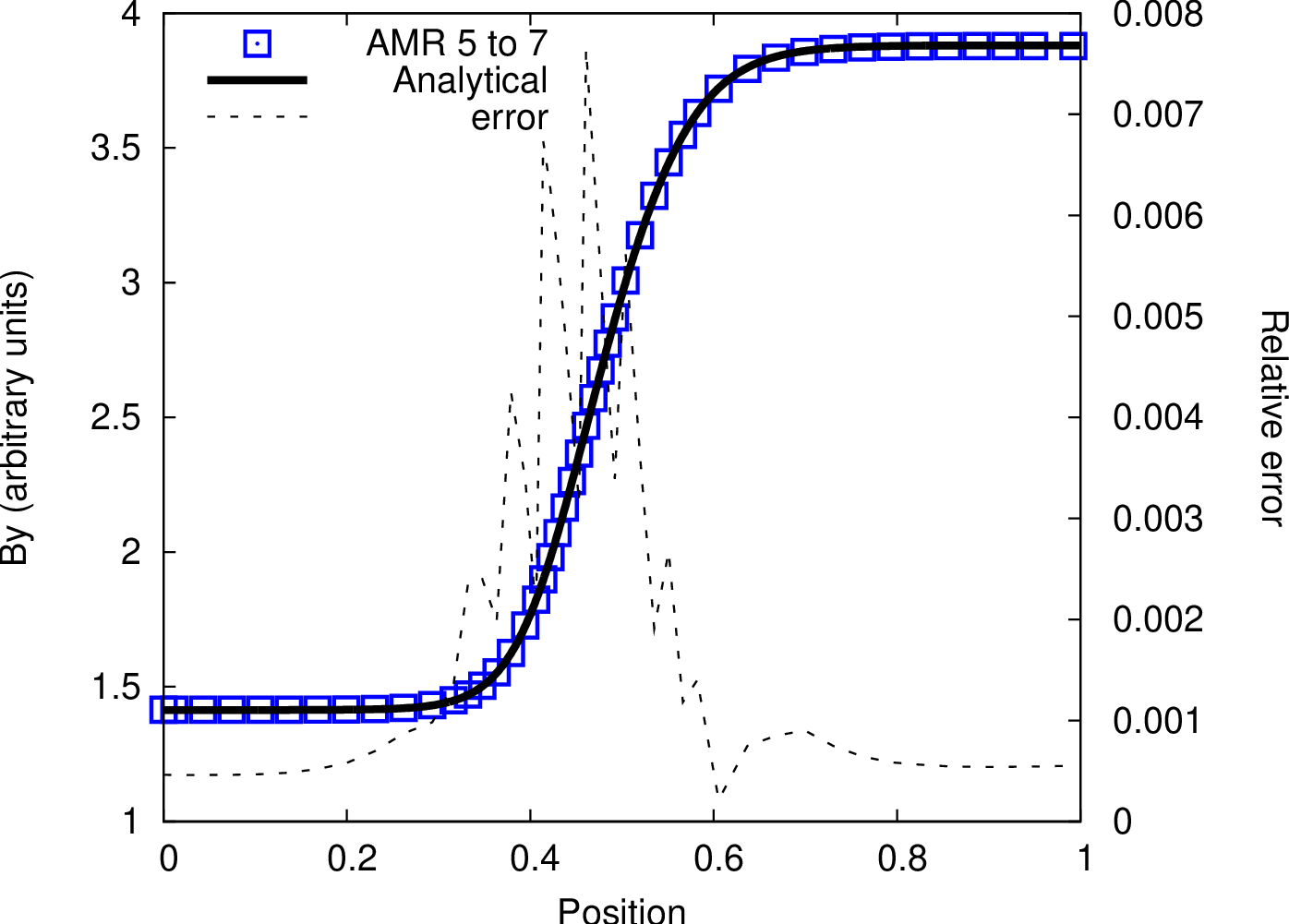}
\\
\includegraphics[width=0.33\textwidth,angle=270]{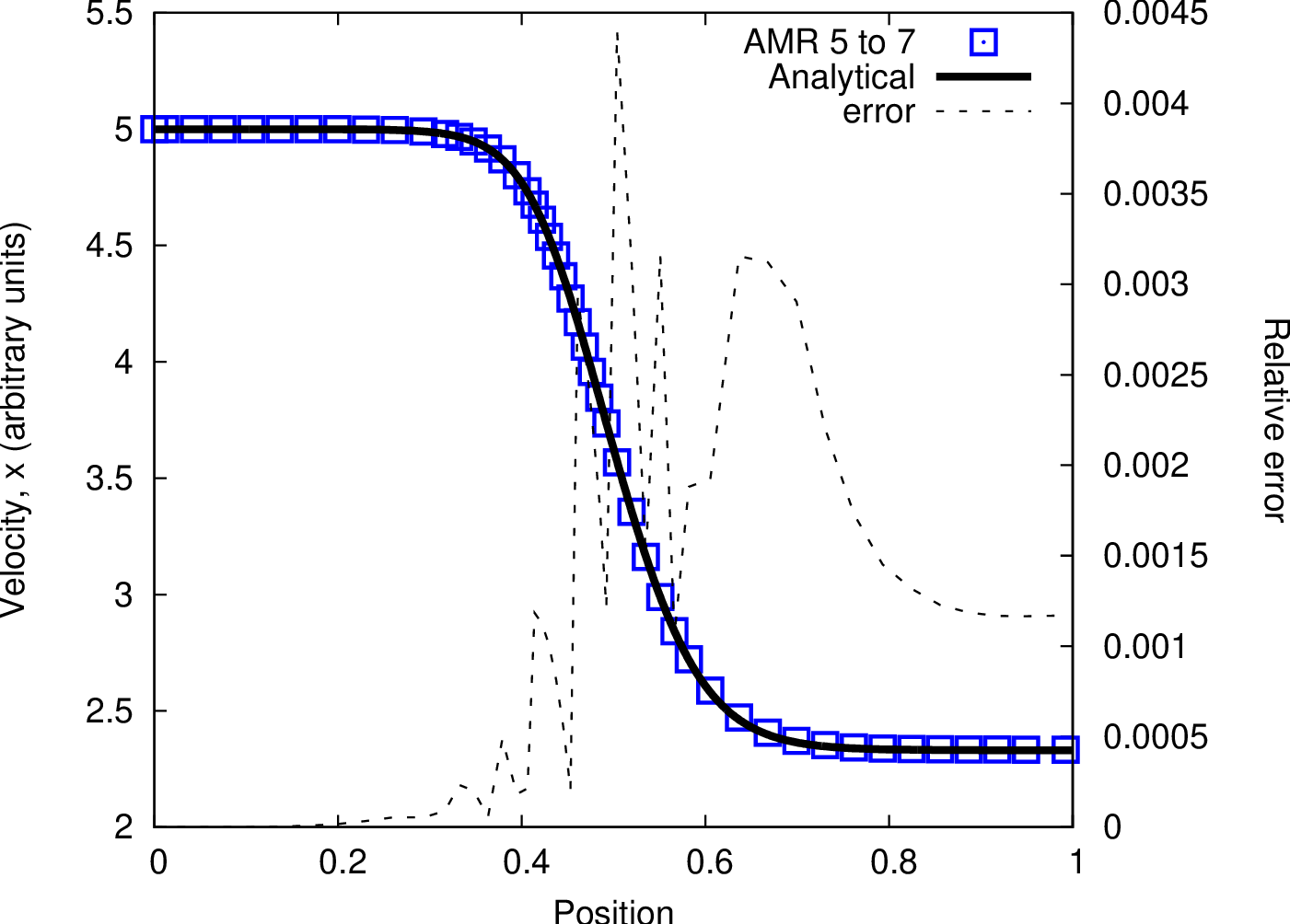}
\includegraphics[width=0.33\textwidth,angle=270]{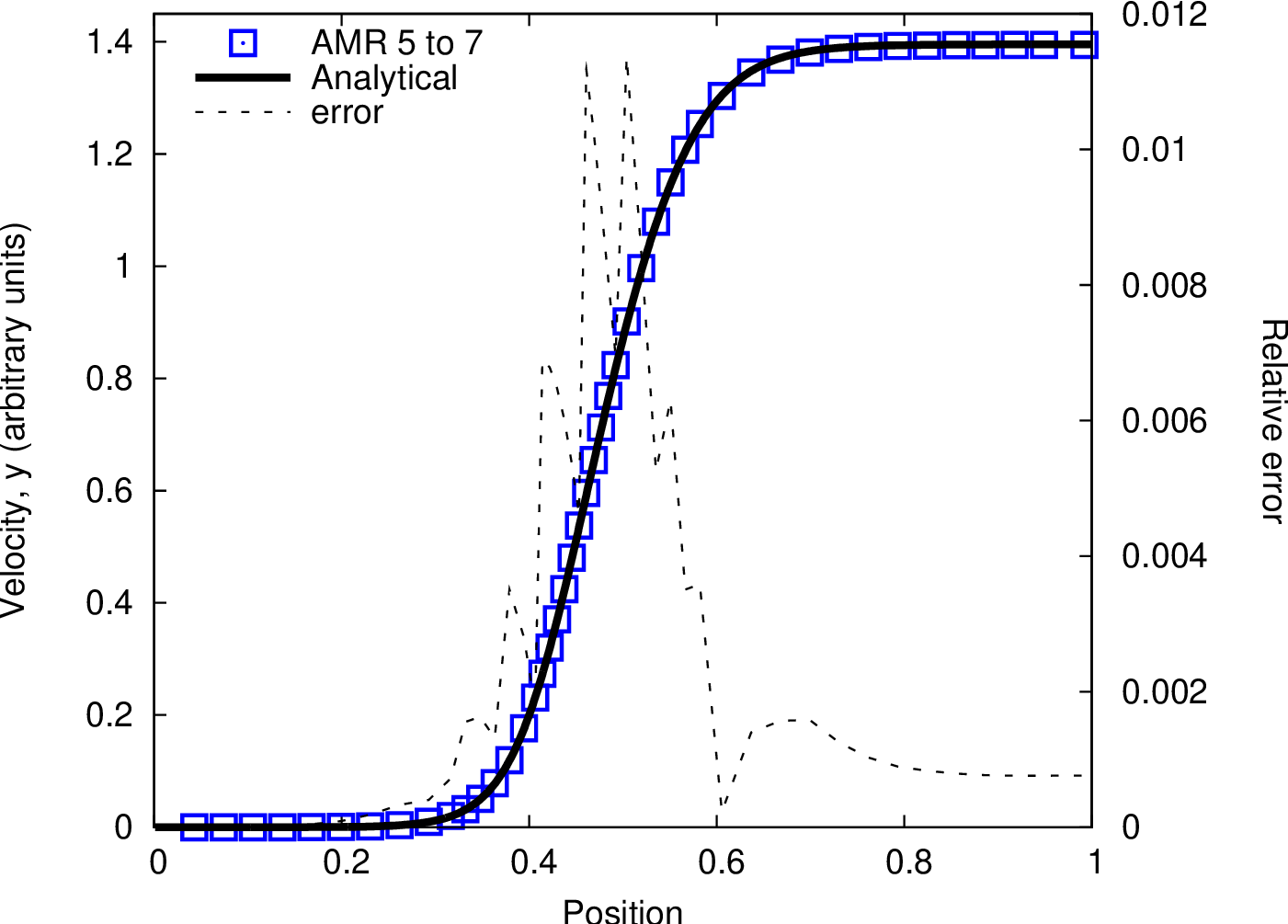}
\caption{Isothermal oblique shock with ambipolar diffusion. Lines and symbols are the same as in Figure~\ref{figbaren1}. The levels of refinement vary from 5 to 7.}
\label{figchoc2DA}
\end{center}
\end{figure*}

\begin{figure*}
%\begin{center}
\includegraphics[width=0.33\textwidth,angle=270]{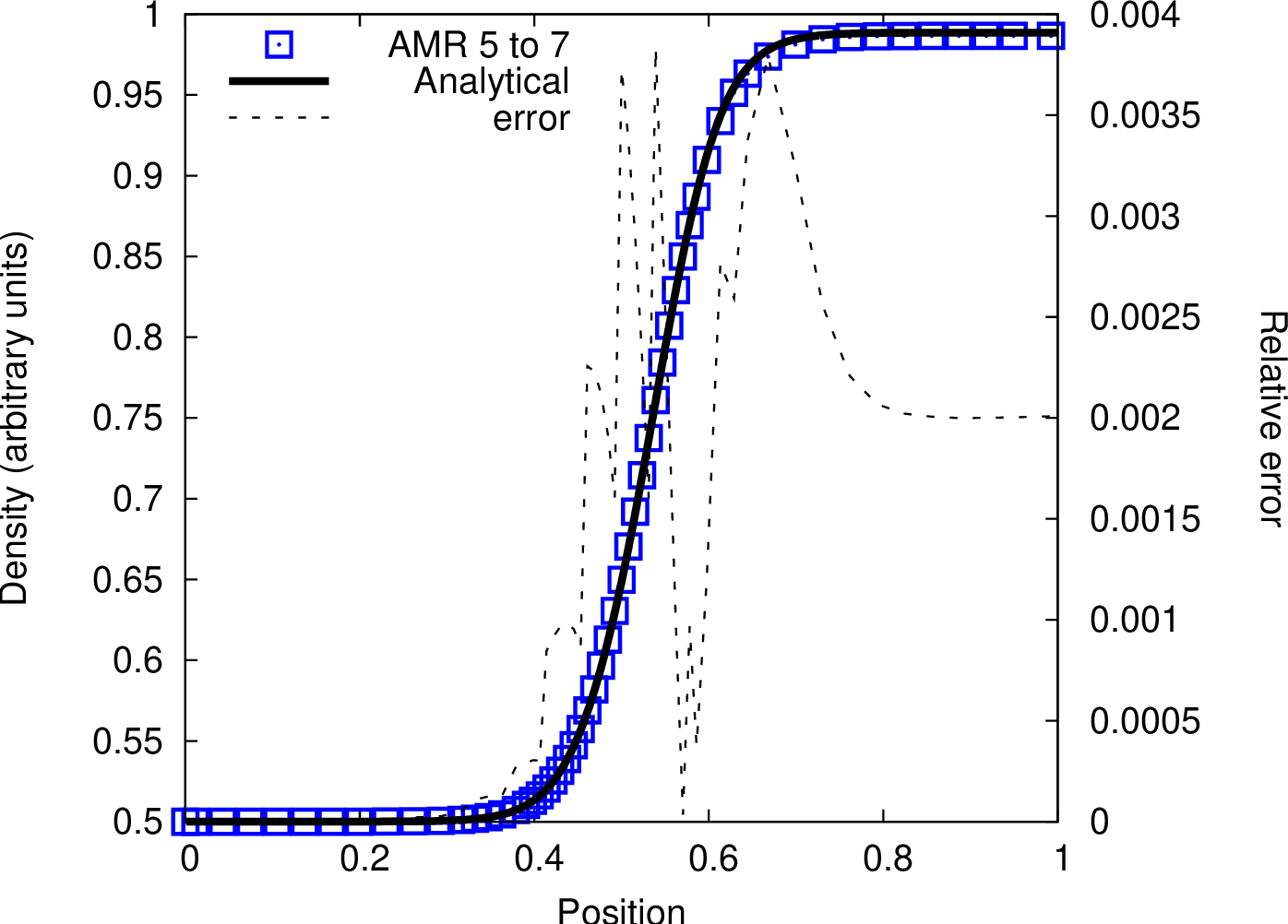}
\includegraphics[width=0.33\textwidth,angle=270]{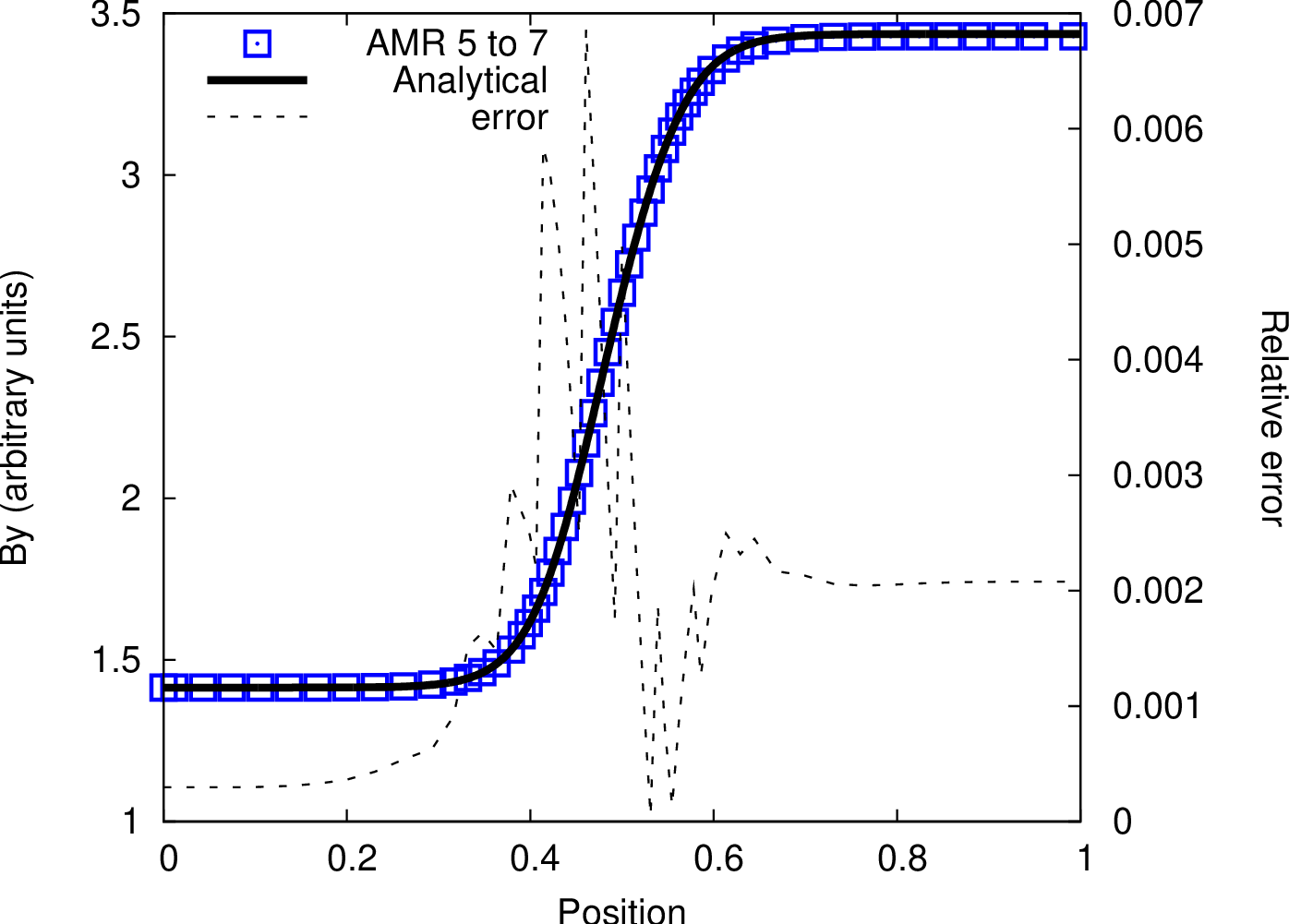}
\\
\includegraphics[width=0.33\textwidth,angle=270]{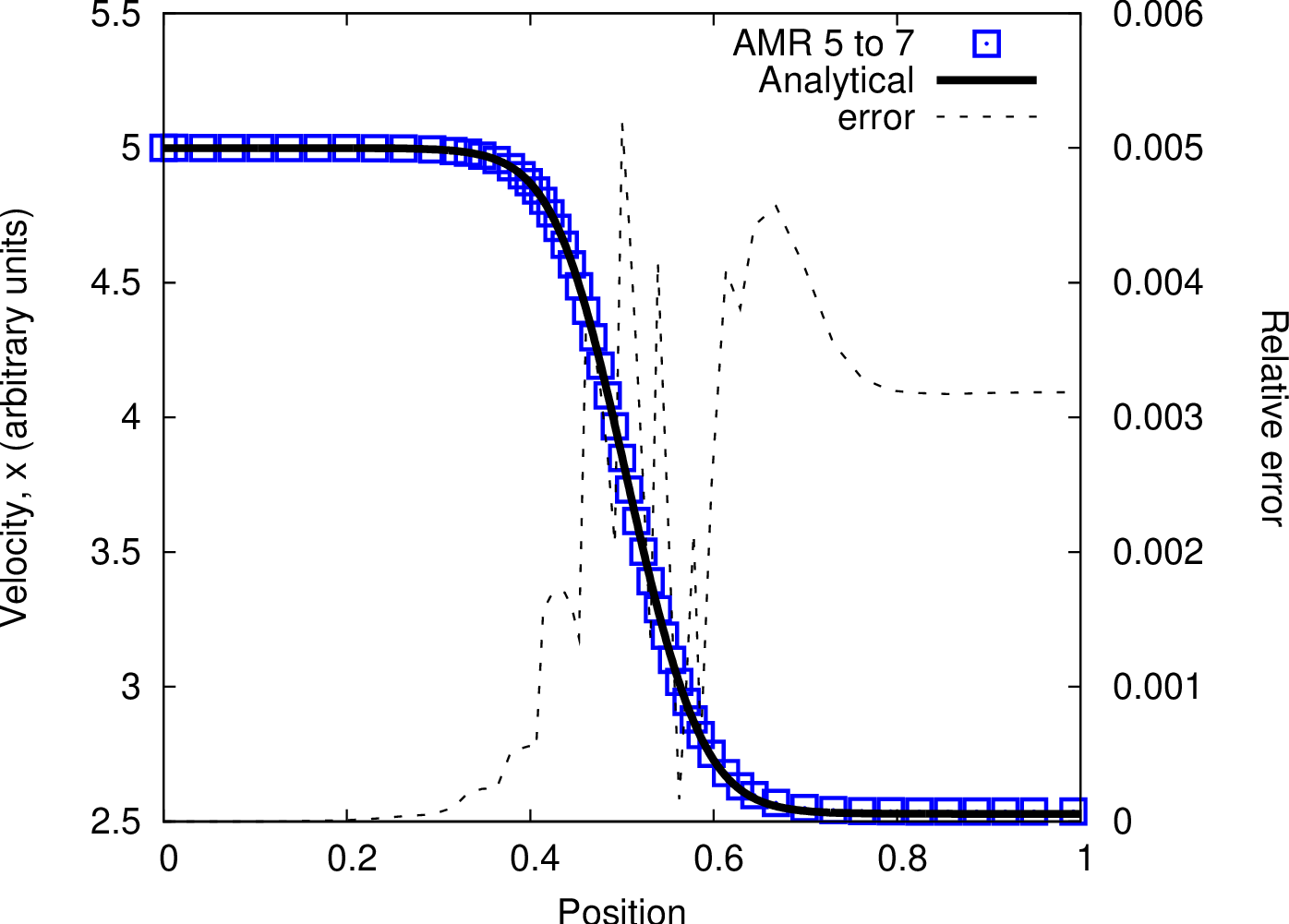}
\includegraphics[width=0.33\textwidth,angle=270]{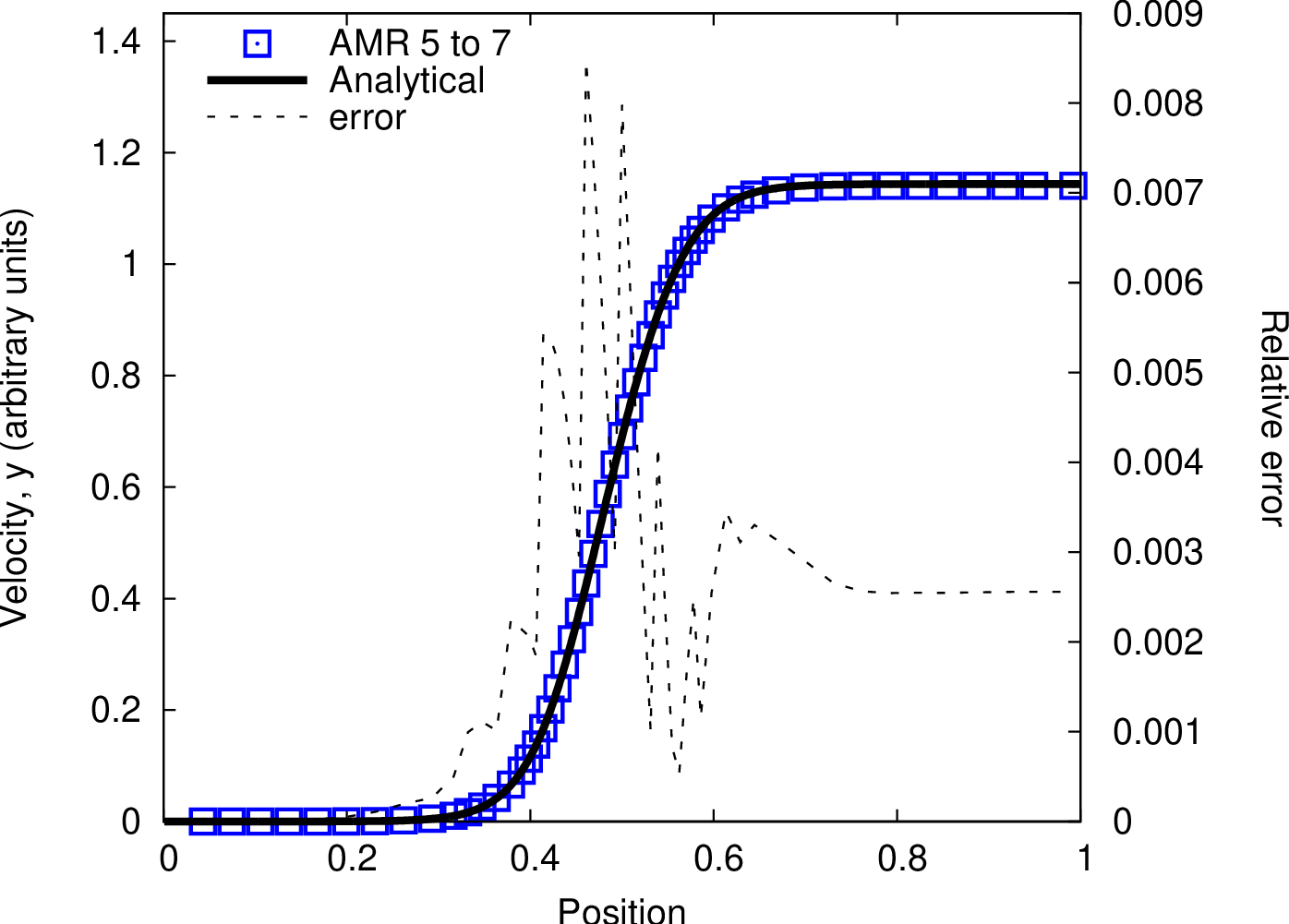}
\\
\includegraphics[width=0.33\textwidth,angle=270]{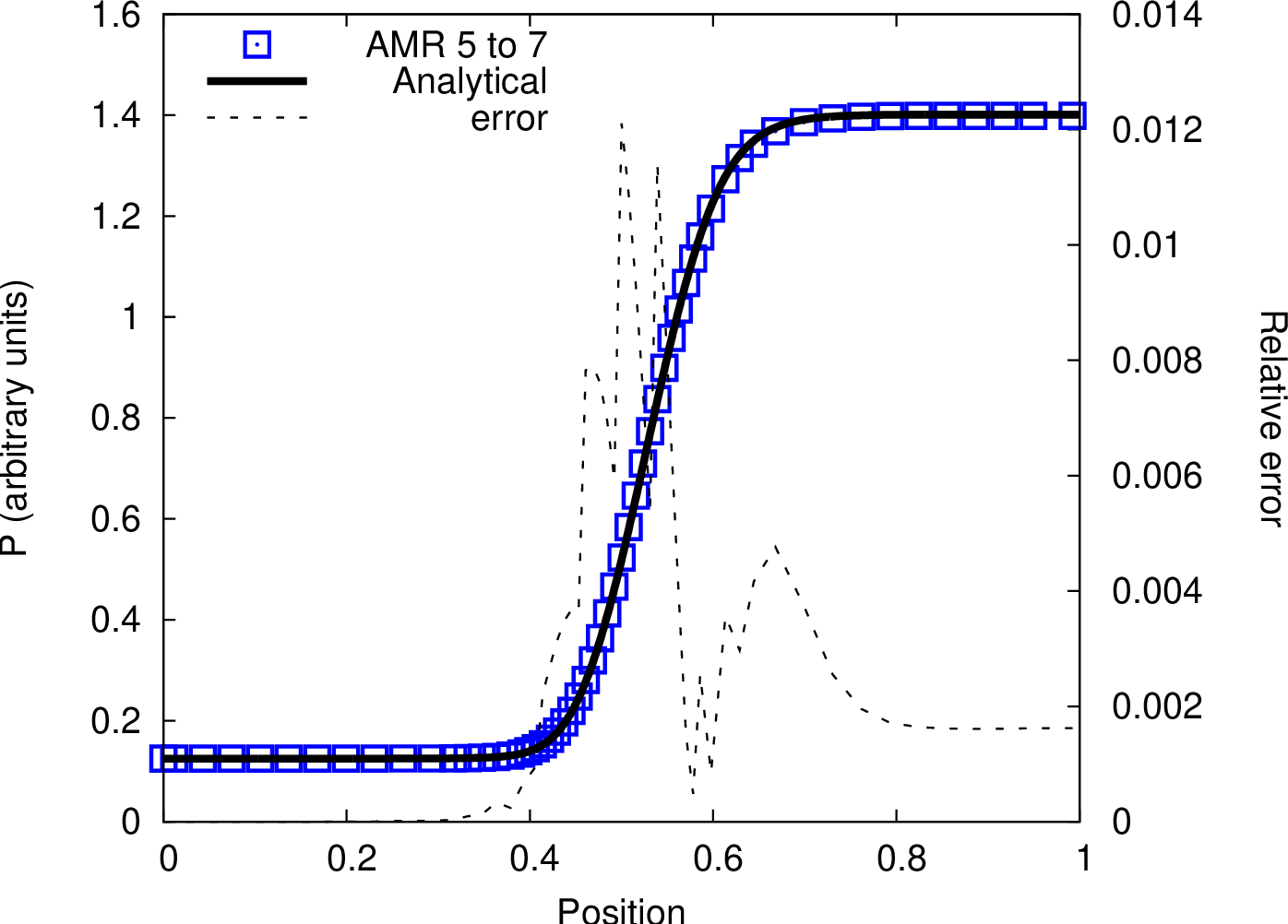}
\caption{Non-isothermal oblique shock with ambipolar diffusion. Lines and symbols are the same as in Figure~\ref{figbaren1}. The levels of refinement vary from 5 to 7.}
\label{figchoc2DANI}
%\end{center}
\end{figure*}

For the non-isothermal case the energy Equation~(\ref{eqenergDA}) is solved assuming a perfect gas with an adiabatic index $\gamma=\frac{5}{3}$ and without any additional cooling. The semi-analytical set of equations to be solved is derived from \citet{DuffinPudritz}, where we assume a constant ion density (see Appendix~\ref{duffpud} for more details). The steady-state is not very different from the isothermal case, except for the pressure.  The results for the non-isothermal case are shown Figure~\ref{figchoc2DANI}. \m{Our results are significantly different from} \citet{DuffinPudritz} in the pressure across the shock. This is explained by the additional heating term (and an artificial cooling term necessary for the equations to converge) in \m{their} set of equations. Therefore the equations tested are not exactly the same and thus neither are the semi-analytical solution \m{nor} the results.

In astrophysical simulations solving {\it non-isothermal} ambipolar diffusion only makes sense if cooling or heating of the gas is properly taken into account, i.e. if radiative transfer is solved. Otherwise, the set of MHD Equations~((\ref{eqcont}), (\ref{eqqtmvt}), (\ref{eqdbdt}) and (\ref{eqdivb})) is closed by an equation of state (a barotropic one in most cases).

We also checked that the results are similar for any orientation of the initial magnetic field and velocity field. Using AMR gives results almost as good as with a regular grid corresponding to the highest level of refinement (not displayed here for conciseness). 

The grid is refined if the gradient of magnetic field, pressure, density or velocity is greater than 0.1 (this insures for this test that the error on the AMR grid and on the regular grid are about the same). 

\subsubsection{The Alfv\'en wave test \label{theobalsara}}

Studying the decay of Alfv\'en waves in an ionized plasma provides a stringent test of the coupling between the flow and the magnetic field due to ambipolar diffusion. Following \citet{Choietal}, we have examined the behaviour of propagating and standing Alfv\'en waves in such a plasma. We closely follow the prescription and the notations defined by \citet{LesaffreBalbus} for the study of Alfv\'en waves in a plasma with Ohmic diffusion, and adapt them to the ambipolar diffusion case. Here we derive exact solutions for torsional Alfv\'en waves in a non-isothermal plasma with ambipolar diffusion.

The unperturbed state without Alfv\'en waves is defined as: 
\begin{eqnarray}
\rho_0 =1, \rho_{0n} &=&1, \rho_{0i}=1 \\ \nonumber
V_{0x}=0, V_{0y}&=&0, V_{0z}=0 \\ \nonumber
B_{0x}=0, B_{0y}&=&0, B_{0z}=1 \nonumber.
\end{eqnarray}

We seek for perturbed solutions of the form 
\begin{align}
{\bf u} &= {\bf \delta u} \exp{(st+ikz)} \\
{\rm and}~~~{\bf B} &= B_0{\bf \hat z} + {\bf b} = B_0{\bf \hat z} + {\bf \delta b}\exp{(st+ikz)},
\end{align}
where $\delta {{\bf b}} = \delta b_x\, {{\bf \hat x}} + \delta b_y\, {{\bf \hat y}}$ and $\delta {{\bf u}} = \delta u_x\, {{\bf \hat x}} + \delta u_y\, {{\bf \hat y}}$. $s$ is the wave angular frequency and $k$ the wave number. For a perturbation wavelength $\lambda$ along the $z$ direction, the wave vector $k$ is set to $k=2 \pi/\lambda$. For such solutions,
the mass density remains constant along the wave trajectory ($\rho \equiv \rho_0$). 

Following \citet{LesaffreBalbus}, we restrict ourselves to MHD flows satisfying $ {\nabla} (P+\frac{1}{2} B^2) \equiv {{\bf0}}$, so that the momentum equation reads: 
\begin{equation}
\partial_t {\bf u} = \frac{B_0 }{\rho_0}\partial_z {\bf b} 
\label{eqalfda1}
\end{equation}
and the induction equation simplifies to: 
\begin{equation}
\partial_t {\bf b} = B_0 \partial_z {\bf u} + \frac{B_0^2}{\gamma_{AD} \rho_{i0} \rho_0} \partial_z^2 {\bf b} 
\label{eqalfda2}.
\end{equation}
Combining Equation~(\ref{eqalfda1}) and Equation~(\ref{eqalfda2}) gives a quadratic dispersion relation: 
\begin{eqnarray}
s^2 + k^2 \eta_{AD} s + k^2 v_A^2 = 0 ,
\label{dispAD}
\end{eqnarray}
where the ambipolar diffusion coefficient is defined by $\eta_{AD}= v_A^2/\gamma_{AD}\rho_{i0}$ and the Alfv\'en velocity by $v_A=B_0/\sqrt{\rho_0}$. This equation is similar to the dispersion relation obtained by \citet{Balsara}, but we have derived it for the more general adiabatic, non-isothermal case, and also for any amplitude in $|\delta {{\bf b}}|$, provided that $ {\nabla} (P+\frac{1}{2} B^2) \equiv {{\bf0}}$. If we restrict ourselves to circularly polarized waves with e.g., $\delta b_y = i \delta b_x$ then ${\nabla} (\frac{1}{2} B^2) \equiv {{\bf 0}}$ will also ensure $ {\nabla} (P) \equiv {{\bf0}}$ as we now demonstrate.

It is clear from Equation~(\ref{dispAD}) that Alfv\'en waves propagate ($s_i \ne 0$, with $s_i$ the imaginary part of $s$) only for $v_A > k \eta_{AD}/2$. The solutions of Equation~(\ref{dispAD}) are given by: 
\begin{eqnarray}
s = -\frac{k^2 \eta_{AD} }{2}  \pm i \sqrt{ k^2v_A^2 - \left(\frac{k^2\eta_{AD}}{2}\right)^2} .
\end{eqnarray}
In the numerical tests that follow, we restrict ourselves to $\lambda=1$ and equal to the box size, so that $k=2\pi$. We will explore first a value $\gamma_{AD}=80$, yielding a diffusion coefficient $\eta_{AD}=1.25 \times 10^{-2}$, and resulting in a moderate damping with imaginary part $s_i=\pm 6.2783387$ and real part $s_r=-0.2467401$. We then consider the case $\gamma_{AD}=30$ ($\eta_{AD}=0.0333$), resulting in a stronger damping with $s_i=\pm 6.2486389$ and $s_r=-0.6579736$.

\paragraph{Estimating numerical diffusion}
\label{modifequation}

In order to estimate the quality of our numerical solution we need to compute the leading order error term in the ideal MHD scheme. This is done usually using the Modified Equation approach where a Taylor expansion of the numerical solution is performed. We restrict our analysis to the propagation of Alfv\'en waves since the Modified Equation is much simpler to handle in this case.  We use the characteristic variable $\alpha^{\pm} = {\bf u} \mp {\bf b}/\sqrt{\rho_0}$, so that the system describing the propagation of Alfv\'en waves becomes
\begin{equation}
\partial_t \alpha^{\pm} \pm v_A \partial_z \alpha^{\pm} = 0.
\end{equation}
We consider here only the right-propagating wave, dropping the superscript $+$. The conservative update writes
\begin{equation}
\frac{\alpha^{n+1}_i - \alpha^n_i}{\Delta t} + v_A \frac{\alpha^{n+\frac{1}{2}}_{i+\frac{1}{2}}-\alpha^{n+\frac{1}{2}}_{i-\frac{1}{2}}}{\Delta z}= 0.
\end{equation}
Since the Riemann solver accounts for Alfv\'en waves, the interface flux is given by the upwind value, solution of the predictor step.
\begin{equation}
\alpha^{n+\frac{1}{2}}_{i+\frac{1}{2}} = \alpha^n_i + \left( \partial_z \alpha \right)^n_i \frac{\Delta z}{2}.
\end{equation}
This entirely defines our second-order accurate numerical solution. We assumed here that the time-step is much smaller than the Courant time step, so that $v_A \Delta t/\Delta z \ll 1$. Taylor expanding the solution and its spatial derivative to the first non vanishing order in respect to $\alpha_i^n$ leads to the following Modified Equation with a second-order leading error term
\begin{equation}
\partial_t \alpha + v_A \partial_z \alpha \simeq \frac{v_A \Delta z^2}{12} \partial^3_z \alpha \simeq \eta_{num} \partial^2_z \alpha.
\end{equation}

The right-hand-side represents a third-order derivative of the solution, usually interpreted as a dispersive term. We nevertheless restrict ourselves to the test case studied in this paper, namely a sinusoidal wave of period equal to the box size L, and approximate the leading-order term as a diffusive term with numerical diffusion coefficient, namely:
\begin{equation}
\eta_{num} = \frac{2\pi v_A\Delta z^2}{12L}.
\end{equation}
From this analysis, we can estimate the amplitude of the diffusion due to the hyperbolic solver that needs to be added to the physical (whether ambipolar or Ohmic) diffusion to interpret the numerical solution. We also conclude that the leading order term coming from the ideal MHD solver scales as $\Delta x^2$. This sets the physical range of ambipolar and Ohmic diffusion one can expect to explore for a given mesh resolution. For a mesh of $16^3$ cells the numerical diffusivity is six times smaller than the ambipolar diffusion with $\gamma_{AD} = 80$: $\eta_{num} = 0.002$ and $\eta_{AD} = 0.0125$. This is a good test case in order to assess the accuracy of the correction: the dominant term is still coming from the physics, but the numerical contribution is not negligible.

For Alfv\'en standing waves, the same study can be done. Considering two waves: $\alpha^+$ and $\alpha^-$, one propagating to the right and the other to the left. The system describing the standing Alfv\'en waves is
\begin{equation}
\partial_t \alpha^{+} + v_A \partial_z \alpha^{+} + \partial_t \alpha^{-} - v_A \partial_z \alpha^{-}= 0.
\label{equastat}
\end{equation}
The interface flux are given by the upwind value for $\alpha^+$
\begin{align}
\alpha^{n+\frac{1}{2}}_{i+\frac{1}{2}} = \alpha^n_i + \left( \partial_z \alpha \right)^n_i \frac{\Delta z}{2} \\
\alpha^{n+\frac{1}{2}}_{i-\frac{1}{2}} = \alpha^n_{i-1} + \left( \partial_z \alpha \right)^n_{i-1} \frac{\Delta z}{2} ,
\end{align}
and the downwind value for $\alpha^-$
\begin{align}
\alpha^{n+\frac{1}{2}}_{i+\frac{1}{2}} = \alpha^n_{i+1} - \left( \partial_z \alpha \right)^n_{i+1} \frac{\Delta z}{2} \\
\alpha^{n+\frac{1}{2}}_{i-\frac{1}{2}} = \alpha^n_{i} - \left( \partial_z \alpha \right)^n_{i} \frac{\Delta z}{2} ,
\end{align}
where we then express each term (values and spatial derivatives) in \m{terms} of $\alpha^n_i$, \m{using a third order Taylor expansion in} $\Delta z$.

We then obtain for the two propagating waves:
\begin{align}
\partial_t \alpha^+ + v_A \partial_z \alpha^+ \simeq + \frac{v_A \Delta z^2}{12} \partial^3_z \alpha^+ - \frac{v_A \Delta z^3}{48} \partial^4_z \alpha^+ \\
\partial_t \alpha^- - v_A \partial_z \alpha^- \simeq - \frac{v_A \Delta z^2}{12} \partial^3_z \alpha^- - \frac{v_A \Delta z^3}{48} \partial^4_z \alpha^-.
\end{align}

Combining those two equations in order to obtain Equation~(\ref{equastat}) leads to the solution:
\begin{align}
\partial_t \alpha^+ + v_A \partial_z \alpha^+ + \partial_t \alpha^- - v_A \partial_z \alpha^- \simeq  - \frac{v_A \Delta z^3}{24} \partial^4_z \alpha^-.
\end{align}

\label{rev8}Again, we interpret this fourth order term as a diffusive term with numerical diffusion coefficient:
\begin{equation}
\eta_{num} = \frac{2\pi v_A\Delta z^3}{24L^2} \label{eqn08}.
\end{equation}

\begin{center}
\begin{figure*}
\includegraphics[width=0.35\textwidth,angle=270]{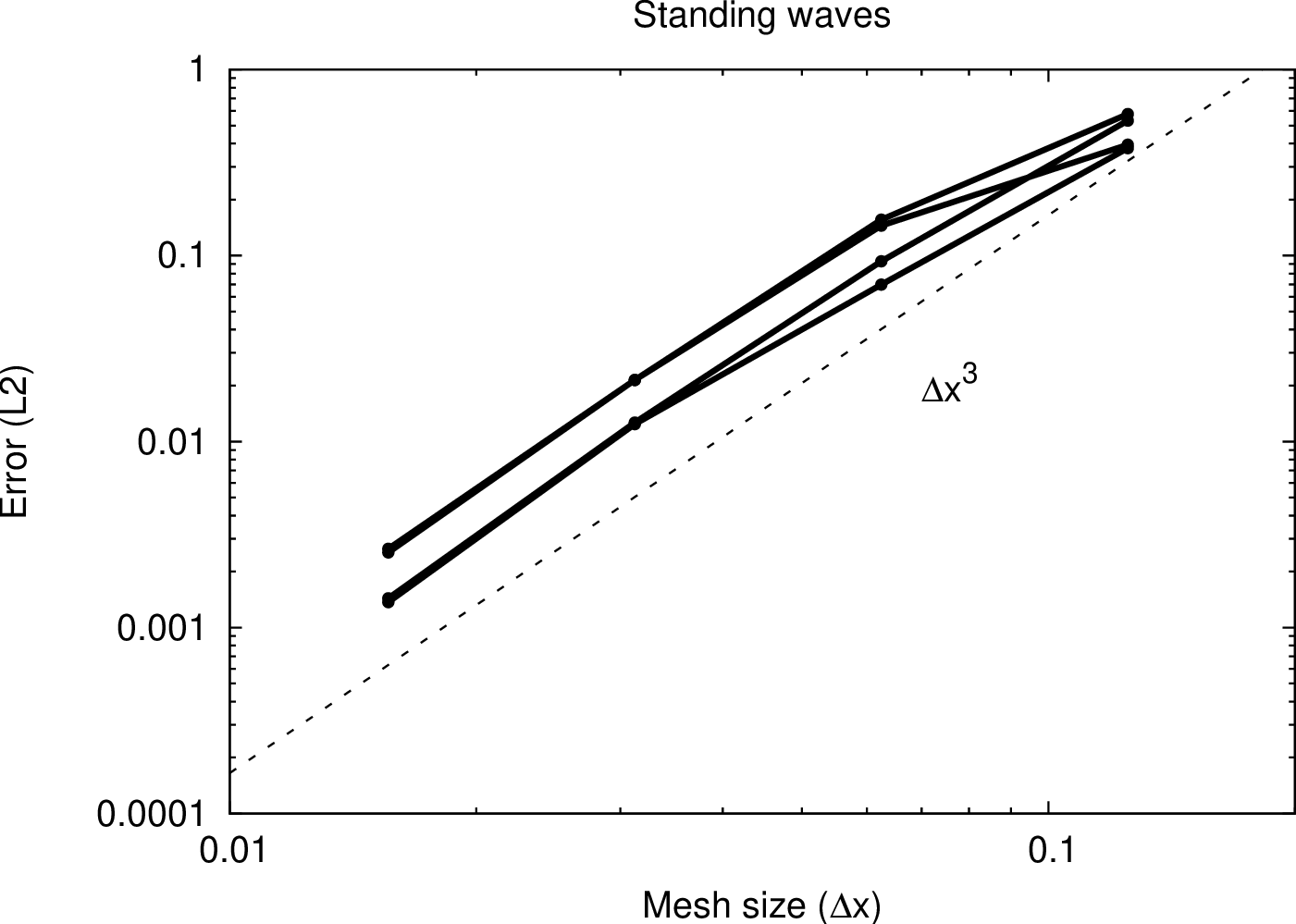}
\includegraphics[width=0.35\textwidth,angle=270]{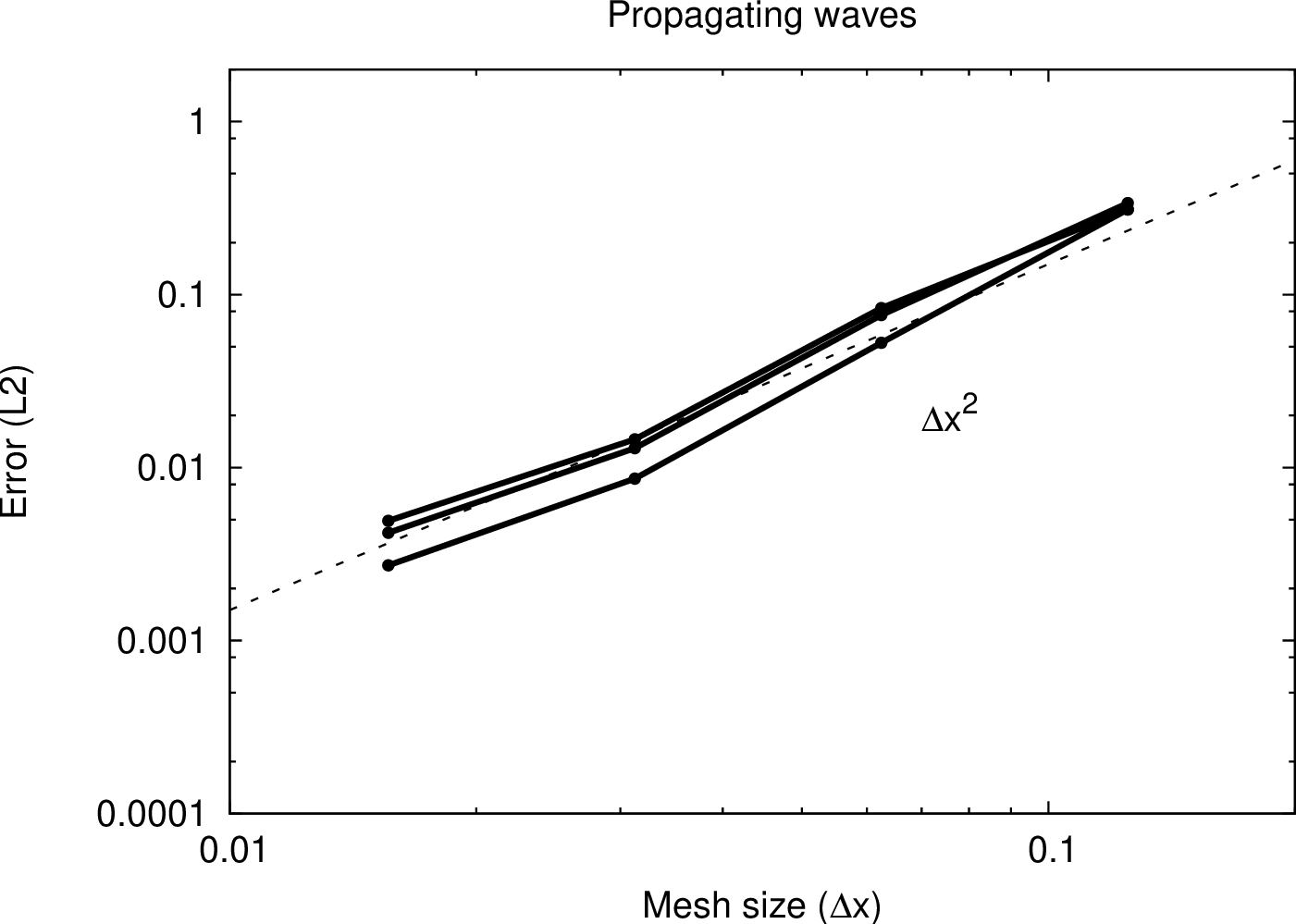}
\caption{Evolution of the error $\epsilon = \sqrt{\sum_{i=1}^{N} \, \frac{(By_{\textrm{numerical}}-By_{\textrm{analytical}})^2}{N}}$ with the mesh size $\Delta x$ for Alfv\'en standing waves (left plot) and Alfv\'en propagating waves (right plot) at different times. The dashed lines correspond to two slopes: $\epsilon \propto \Delta x^{3}$ for the standing waves and $\epsilon \propto \Delta x^{2}$ for the propagating waves.}
\label{figordDA}
\end{figure*}
\end{center}

These two expressions for numerical diffusion ($\propto \Delta z^2$ for propagating waves, and $\propto \Delta z^3$ for standing waves) are representative of the real diffusion, as confirmed by the study of the evolution of the error (Figure~\ref{figordDA}).

To take into account this numerical diffusivity, we solve again the equations of induction (Equation~\ref{eqalfda2}) and momentum (Equation~\ref{eqalfda1}) for \m{a dispersion equation} with an additional (numerical) diffusion:
\begin{align}
\partial_t {\bf u} &= \frac{B_0 }{\rho_0}\partial_z {\bf b}+\eta_{num} \partial_z^2 \mathbf{u}
\label{eqalfvenda1} \\
\partial_t {\bf b} &= B_0 \partial_z {\bf u} + \frac{B_0^2}{\gamma_{AD} \rho_{i0} \rho_0} \partial_z^2 {\bf b} + \eta_{num} \partial_z^2 {\bf b}
\label{eqalfvenda2}
\end{align}
yield
\begin{eqnarray}
s = -\frac{k^2 (\eta_{AD}+2 \eta_{num}) }{2}  \pm i \sqrt{ k^2v_A^2 - \left(\frac{k^2\eta_{AD}}{2}\right)^2}.
\label{dispersionavecnum}
\end{eqnarray}

As we have restricted the numerical effect to a diffusion, there is no contribution to the imaginary part of the pulsation, as can be seen in Equation~(\ref{dispersionavecnum}).

\paragraph{The propagating Alfv\'en waves test \label{alfvenpropa} }

We start the simulation with an initial perturbed state with $B_{1x}=  {\cal R}e( \delta b_x e^{ikx})$, $\delta b_x = 1$,  $B_{1y}=  {\cal R}e( i \delta b_x e^{ikx})$, and $v_{1nx}={\cal R}e(\frac{i k B_0 }{\rho s} B_{1x})$ and $v_{1ny}={\cal R}e(\frac{i k B_0 }{\rho s} B_{1y})$, where ${\cal R}e$ denotes the real part of a complex number. For the propagating wave test, we have chosen our initial conditions so that $s_i\ge 0$.

The internal energy equation (see \citet{Shu_1992}) can be written as: 
\begin{eqnarray}
\frac{\partial \rho \epsilon }{\partial t} + {\nabla} . (\rho \epsilon \mathbf{v}) = -P {\nabla} . \mathbf{v} + \frac{(({\nabla} \times  \mathbf{B}) \times \mathbf{B}) ^2}{\gamma_{AD} \rho_i \rho}.
\end{eqnarray}
In the case of perfect gases, we have $P = (\gamma -1) \rho \epsilon$. Since Alfv\'en waves are transverse waves, ${\nabla} . \mathbf{v} = 0$ and ${\nabla} . (\rho e \mathbf{v}) = 0$. The energy equation thus reduces to 
\begin{eqnarray}
\frac{\partial P }{\partial t} = \frac{\gamma -1}{\gamma_{AD} \rho_i \rho}(({\nabla} \times  \mathbf{B}) \times \mathbf{B}) ^2 \label{eqenergAlfDA}
\end{eqnarray}.
This last equation, combined with our choice $\delta b_y = i \delta b_x$, gives ${\nabla} (P) \equiv {{\bf0}}$.

Using Equation~(\ref{eqenergAlfDA}), the time evolution of the pressure writes 
\begin{align}
 P  = P_{init} + (\gamma -1)\frac{k^2 \, \eta_{AD}}{2s_r} (e^{2 s_r t}-1).
\end{align}

\begin{figure*}
\begin{center}
\includegraphics[width=0.33\textwidth,angle=270]{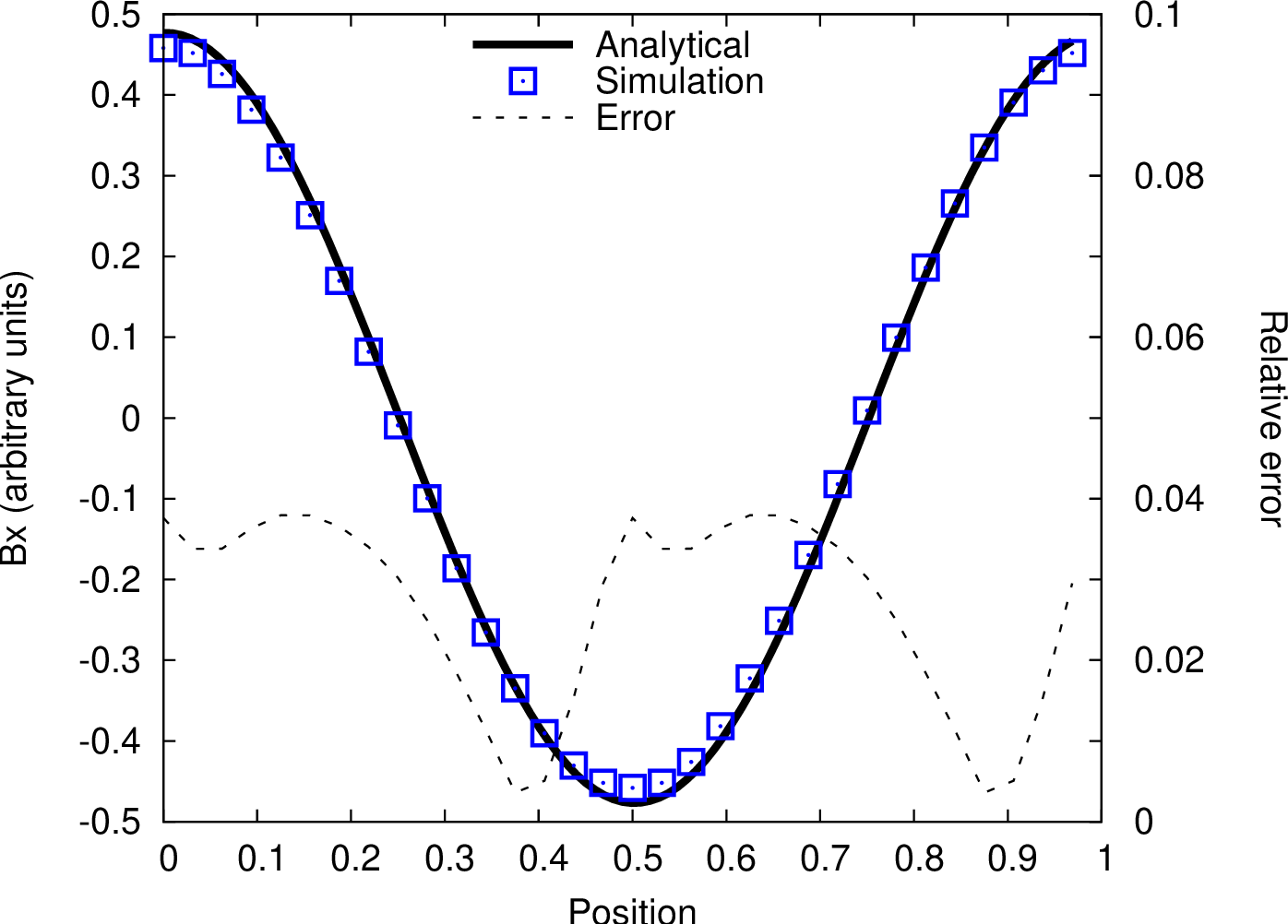}
\includegraphics[width=0.33\textwidth,angle=270]{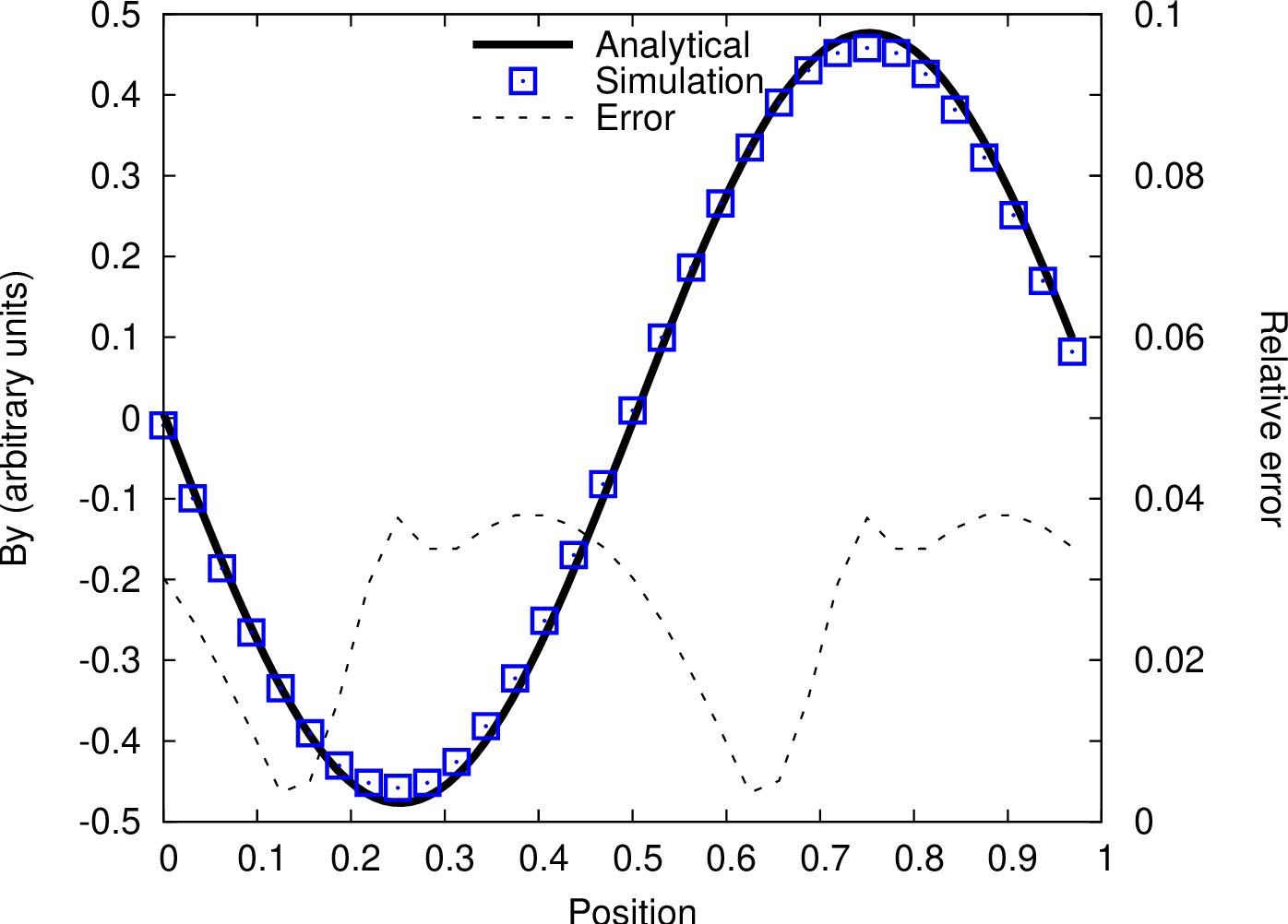}
\\
\includegraphics[width=0.33\textwidth,angle=270]{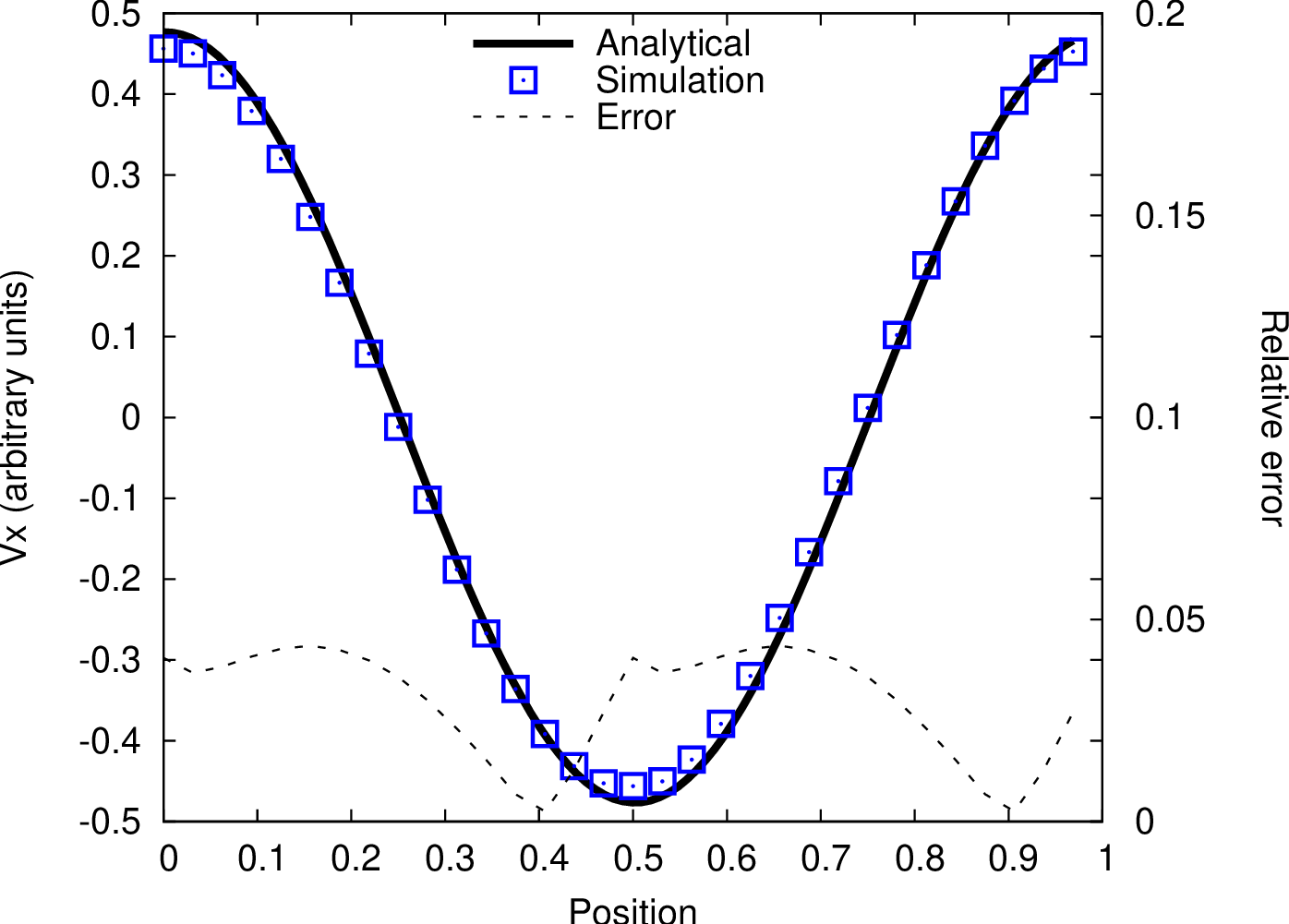}
\includegraphics[width=0.33\textwidth,angle=270]{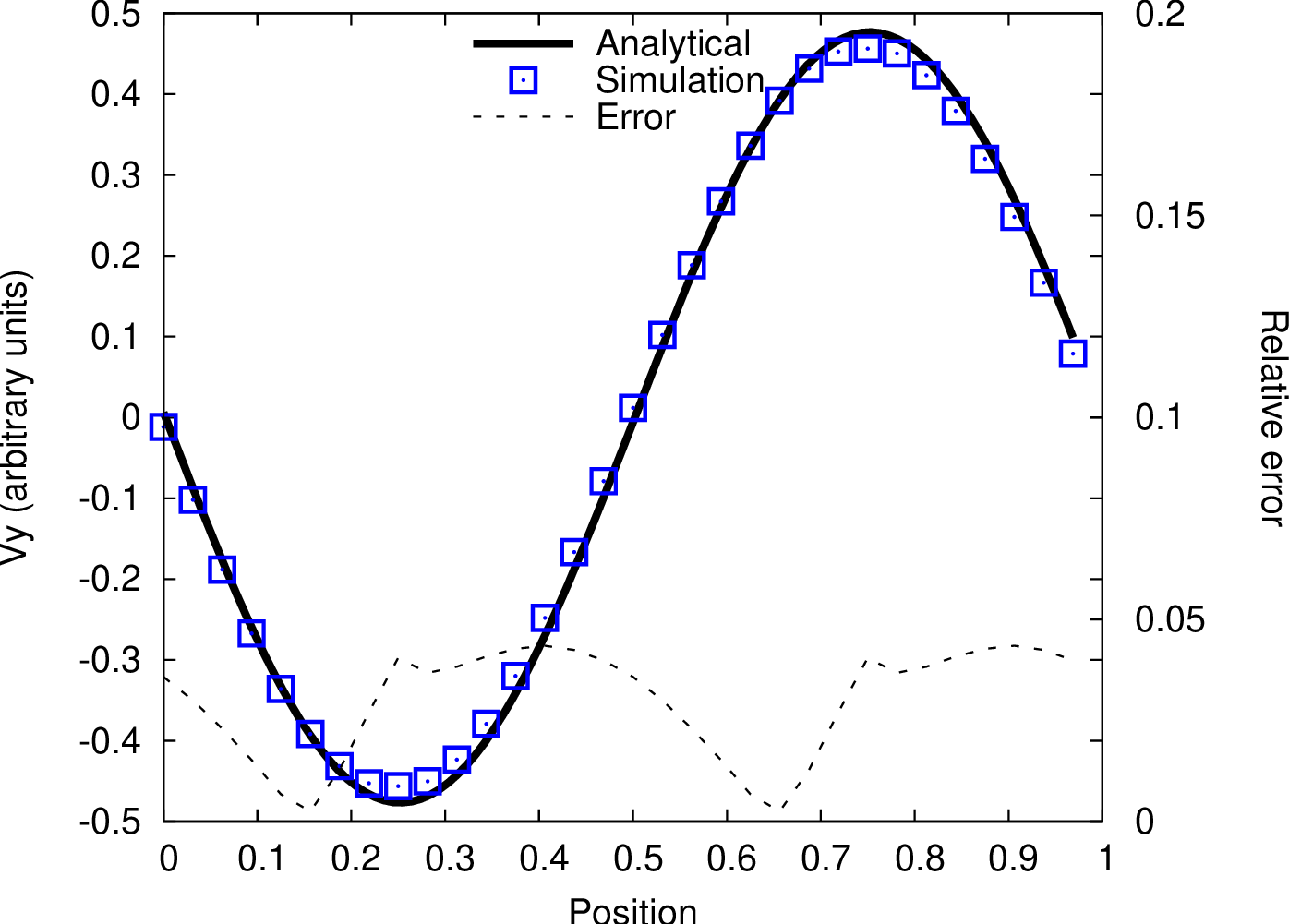}
\\
\includegraphics[width=0.33\textwidth,angle=270]{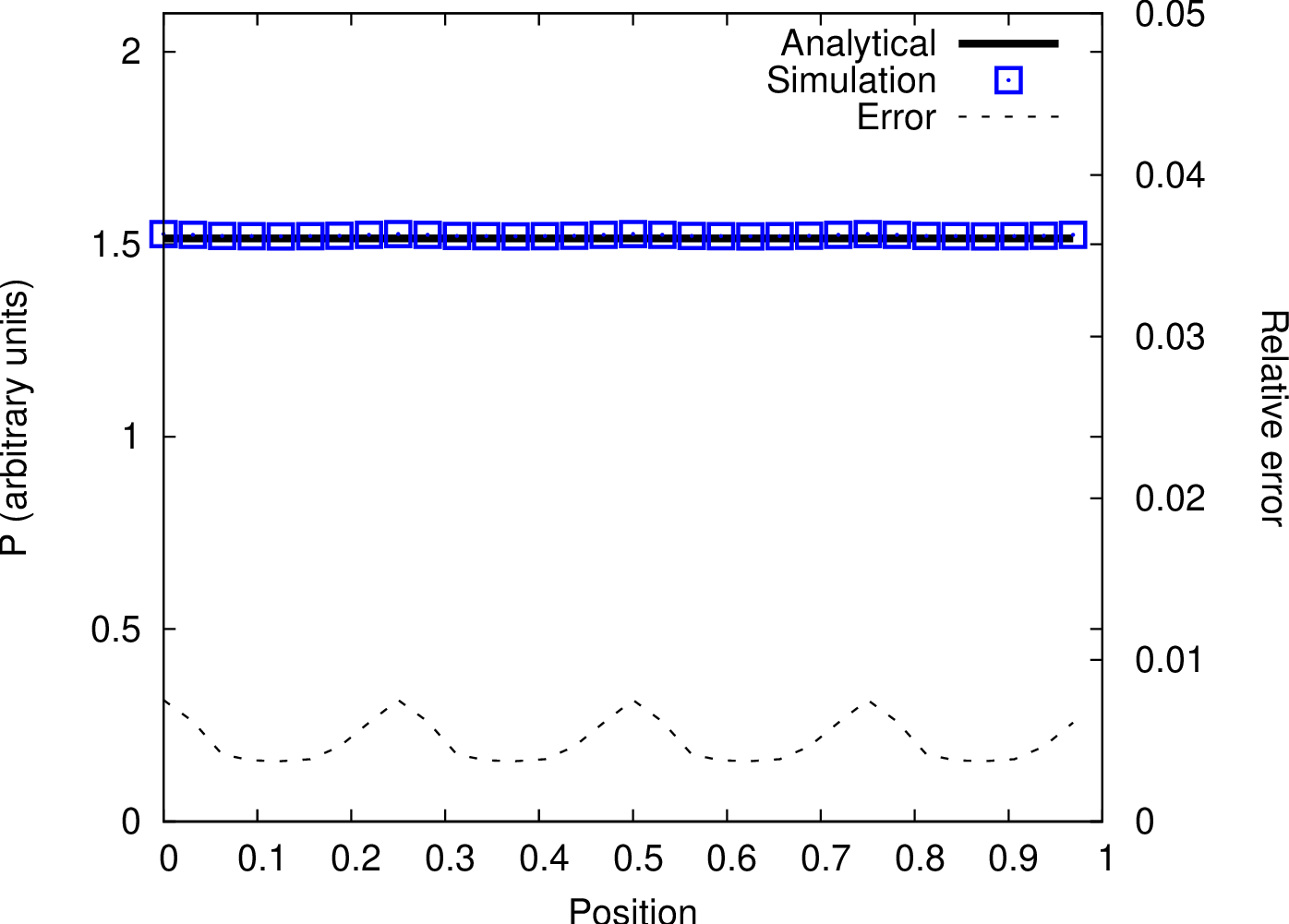}
\includegraphics[width=0.33\textwidth,angle=270]{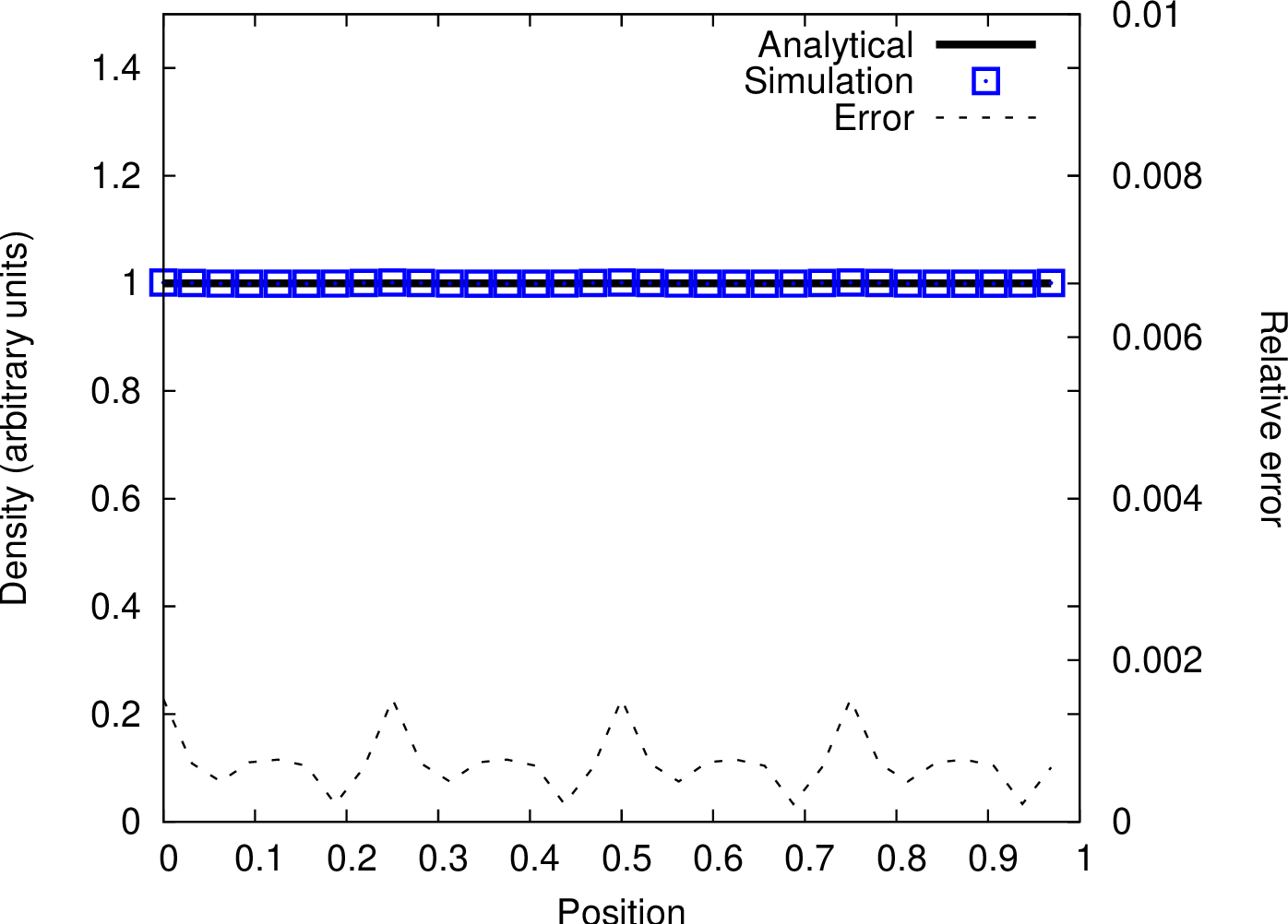}
\caption{The propagating Alfv\'en waves test with ambipolar diffusion ($\gamma_{AD}=80$) after about five periods. The simulation is represented by squares, while the solid-line is the analytical solution. The dotted line is the relative error. We use for this test a fully refined Cartesian grid with 32 cells.}
\label{figalfvenDA}
\end{center}
\end{figure*}

Figure~\ref{figalfvenDA} shows profiles of $B_{1x}$, $v_{1nx}$, $B_{1y}$, $v_{1ny}$, $\rho$ and $P$ along the $z$ direction after three wave periods ({\it i.e}, $t= 3 \times \frac{ 2\,\pi}{s_i}$ ), for $\gamma_{AD}=80$, with a fully refined grid using 32 cells. The solid line represents the analytical solution. The agreement between the numerical and the analytical solution is excellent (see the amplitude of the error on the figure), even after the wave amplitude has decreased by a factor of about 2.

\begin{figure*}
\begin{center}
\includegraphics[width=0.33\textwidth,angle=270]{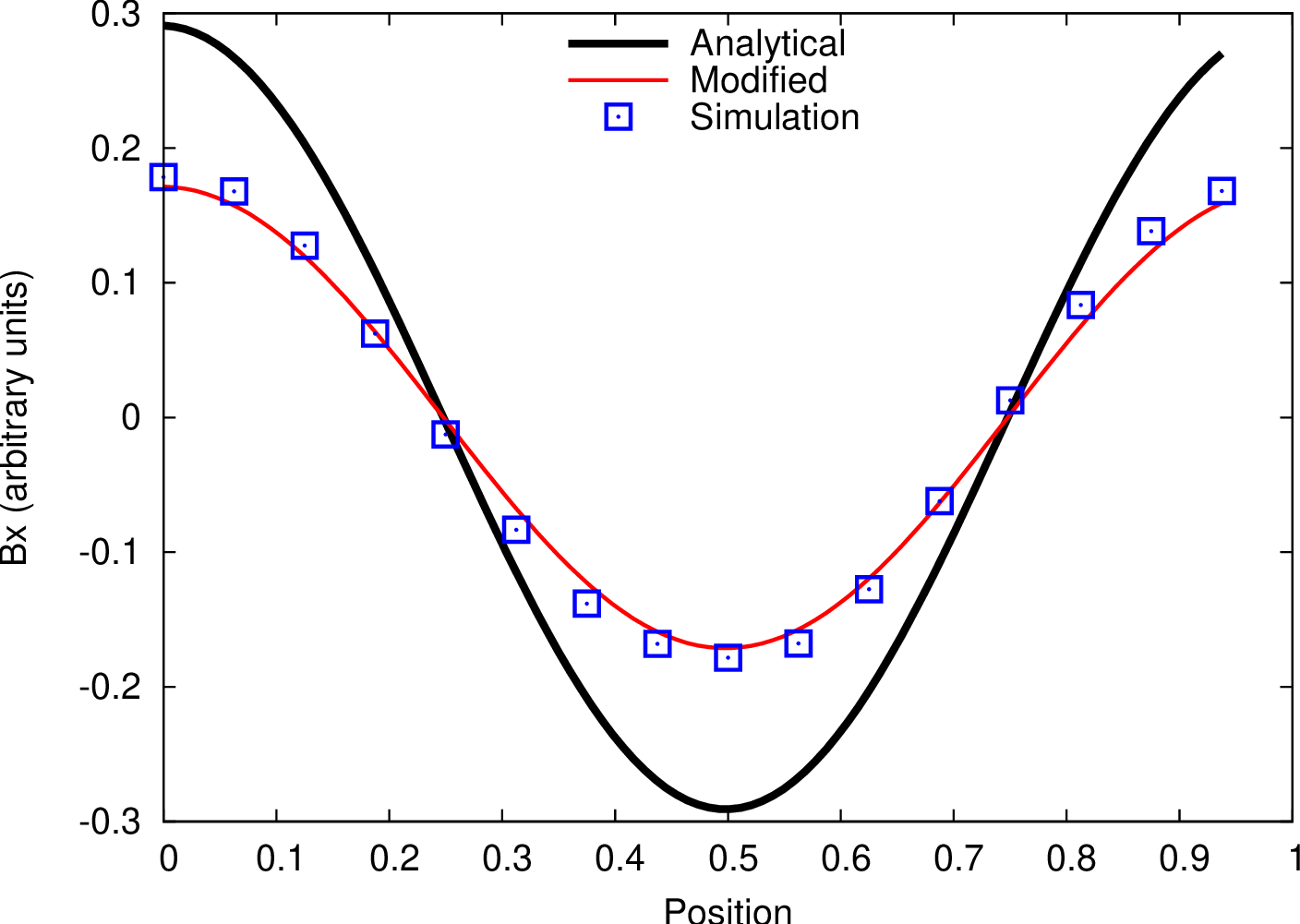}
\includegraphics[width=0.33\textwidth,angle=270]{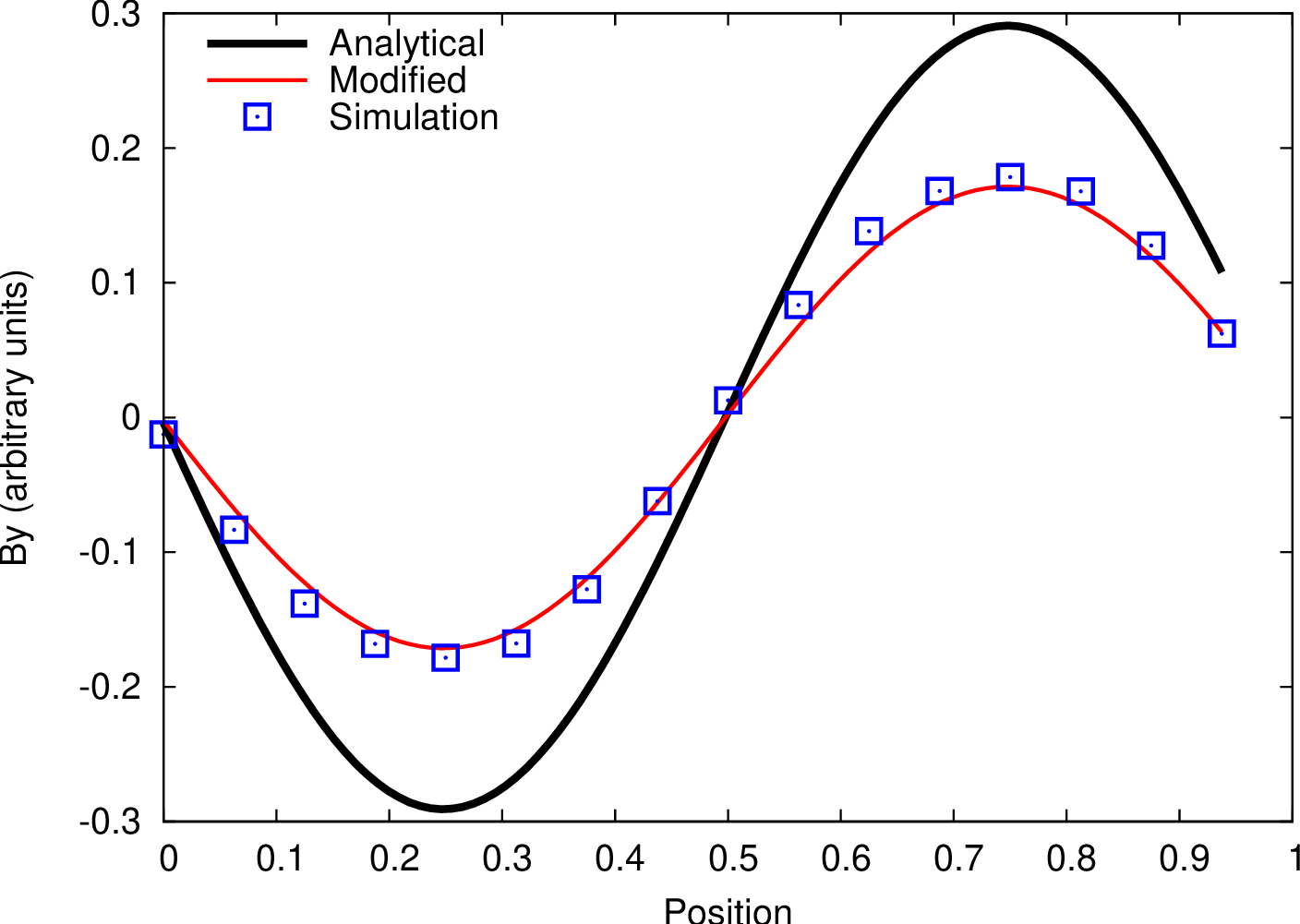}
\\
\includegraphics[width=0.33\textwidth,angle=270]{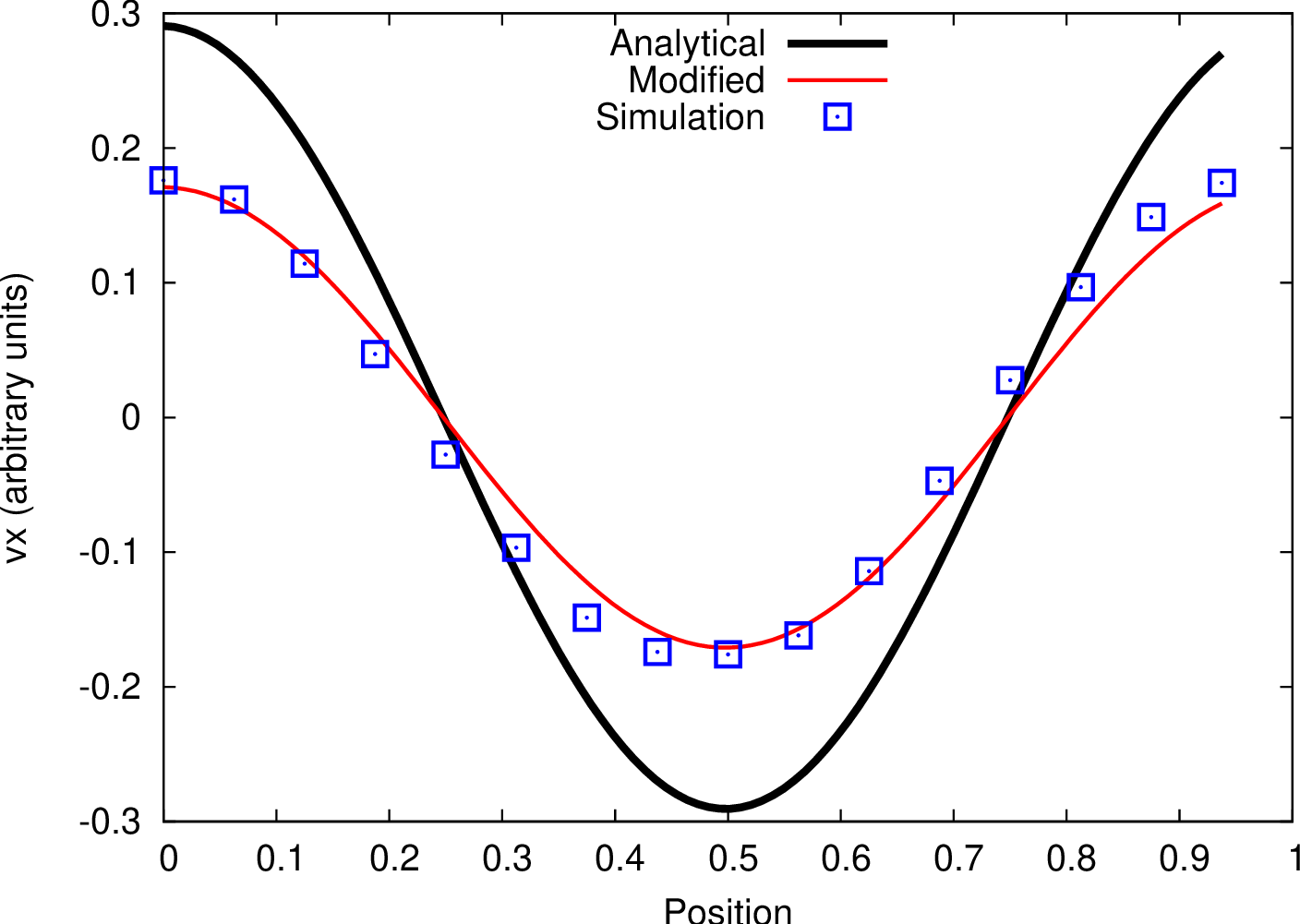}
\includegraphics[width=0.33\textwidth,angle=270]{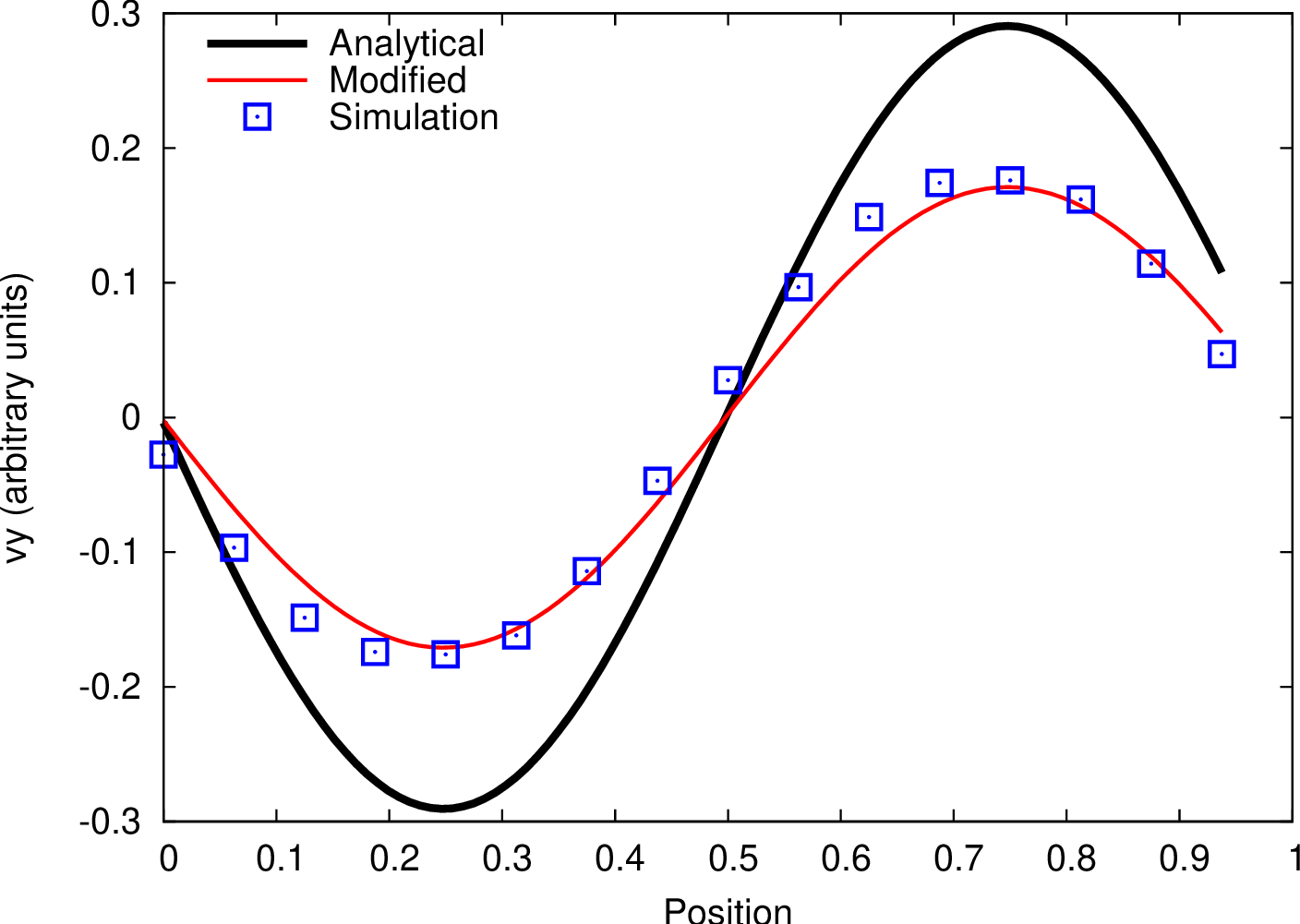}
%\\
%\includegraphics[width=0.33\textwidth,angle=270]{alfvenp_16}
%\includegraphics[width=0.33\textwidth,angle=270]{alfvend_16}
\caption{The propagating Alfv\'en waves test with ambipolar diffusion ($\gamma_{AD}=80$) after about five periods. The simulation is represented by squares, while the solid-lines are the two exact solutions (taking into account or not the effect of numerical diffusion according to Equation~(\ref{dispersionavecnum})). We use for this test a fully refined Cartesian grid with 16 cells.}
\label{figalfvenDA_16}
\end{center}
\end{figure*}

In order to check for the numerical diffusion as explained in Equation~(\ref{dispersionavecnum}) we need to perform the same simulation using less cells for the numerical diffusivity ($\eta_{num}$) to be not negligible compared to the physical diffusivity ($\eta_{AD}$). The profiles of $B_{1x}$, $v_{1nx}$, $B_{1y}$ and $v_{1ny}$ along the $z$ direction after five wave periods ({\it i.e}, $t= 5 \times \frac{ 2\,\pi}{s_i}$ ) for $\gamma_{AD}=80$ with a grid of 16 cells is represented Figure~\ref{figalfvenDA_16}. The solid lines represent the analytical solutions either without taking into account the numerical diffusivity (the {\it not corrected curves}), or correcting the damping factor according to Equation~(\ref{dispersionavecnum}) (the {\it corrected} curves). The agreement between the numerical and the analytical solution taking into account numerical diffusivity is excellent (see the amplitude of the error on the figure). %The numerical method used to solve for the pressure (or the energy) is not the same, so the numerical diffusivity will also differ, and the correction is not that good, as can be seen in Figure~\ref{figalfvenDA_16}. In order to get better results, one would have to do the same kind of study for this equation, which is beyond the scope of this paper.

\medskip
\paragraph{The standing Alfv\'en waves test \label{alfvenstatio}}
\label{standing}

\begin{figure*}
\includegraphics[width=0.33\textwidth,angle=270]{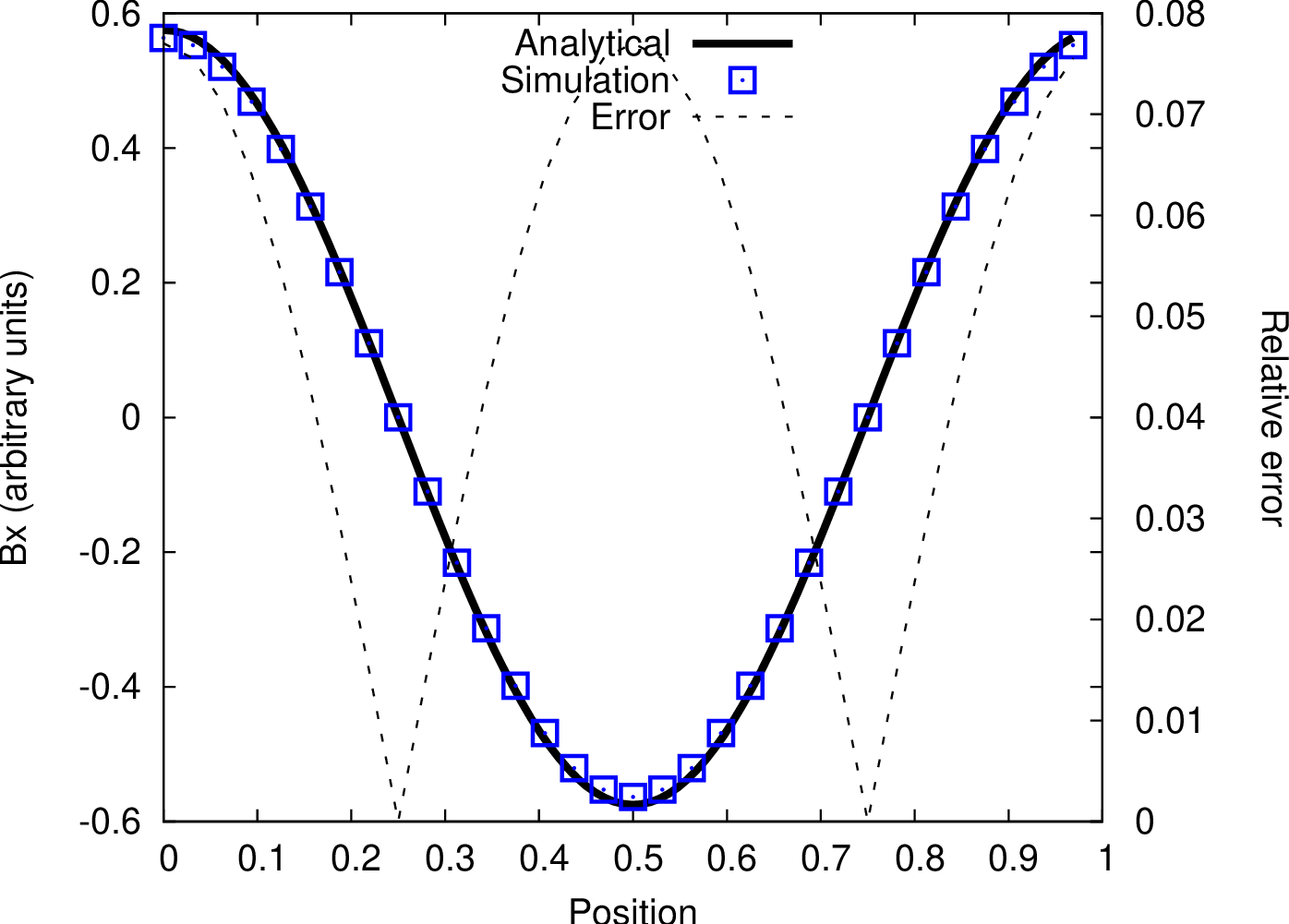}
\includegraphics[width=0.33\textwidth,angle=270]{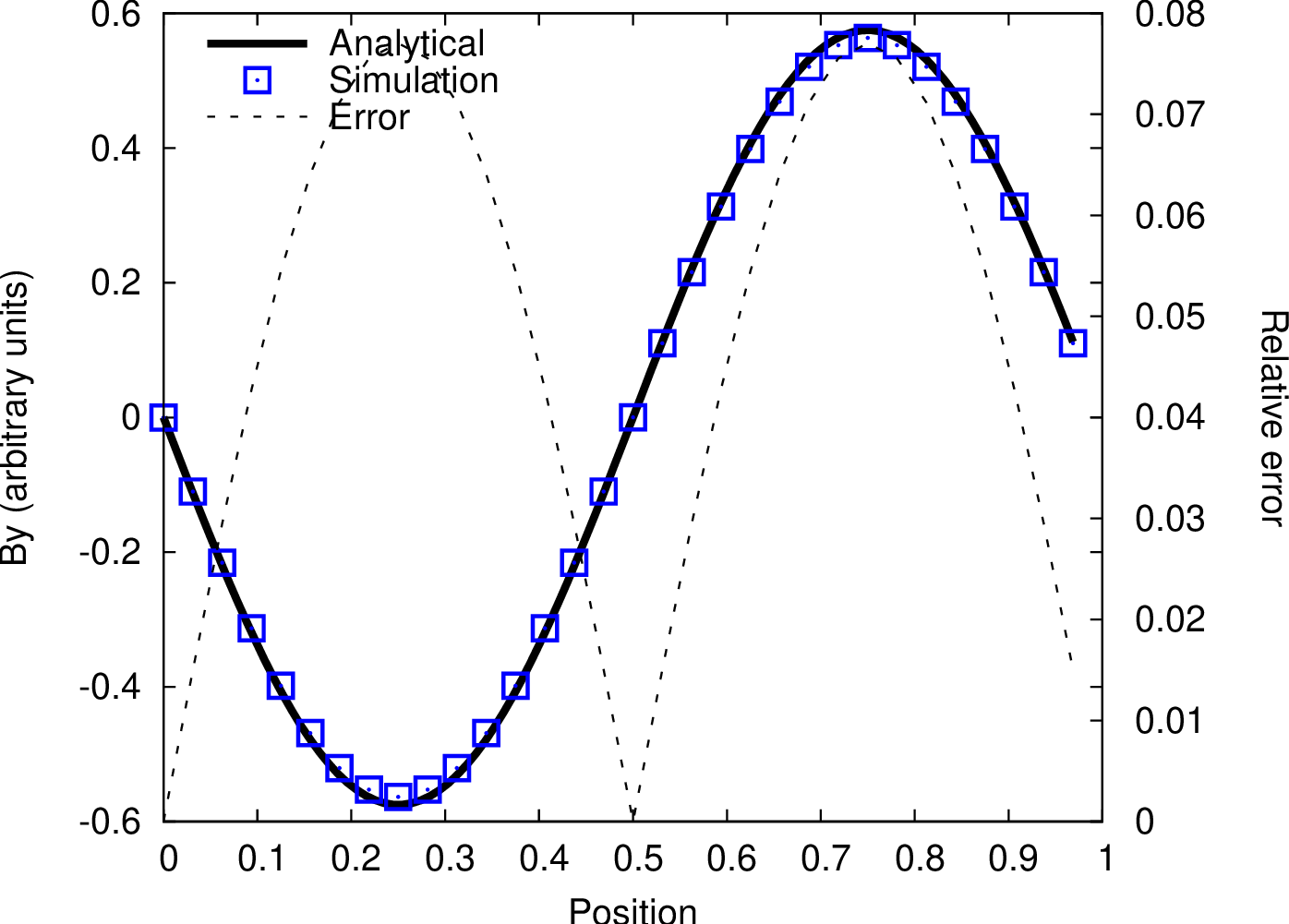}
\\
\includegraphics[width=0.33\textwidth,angle=270]{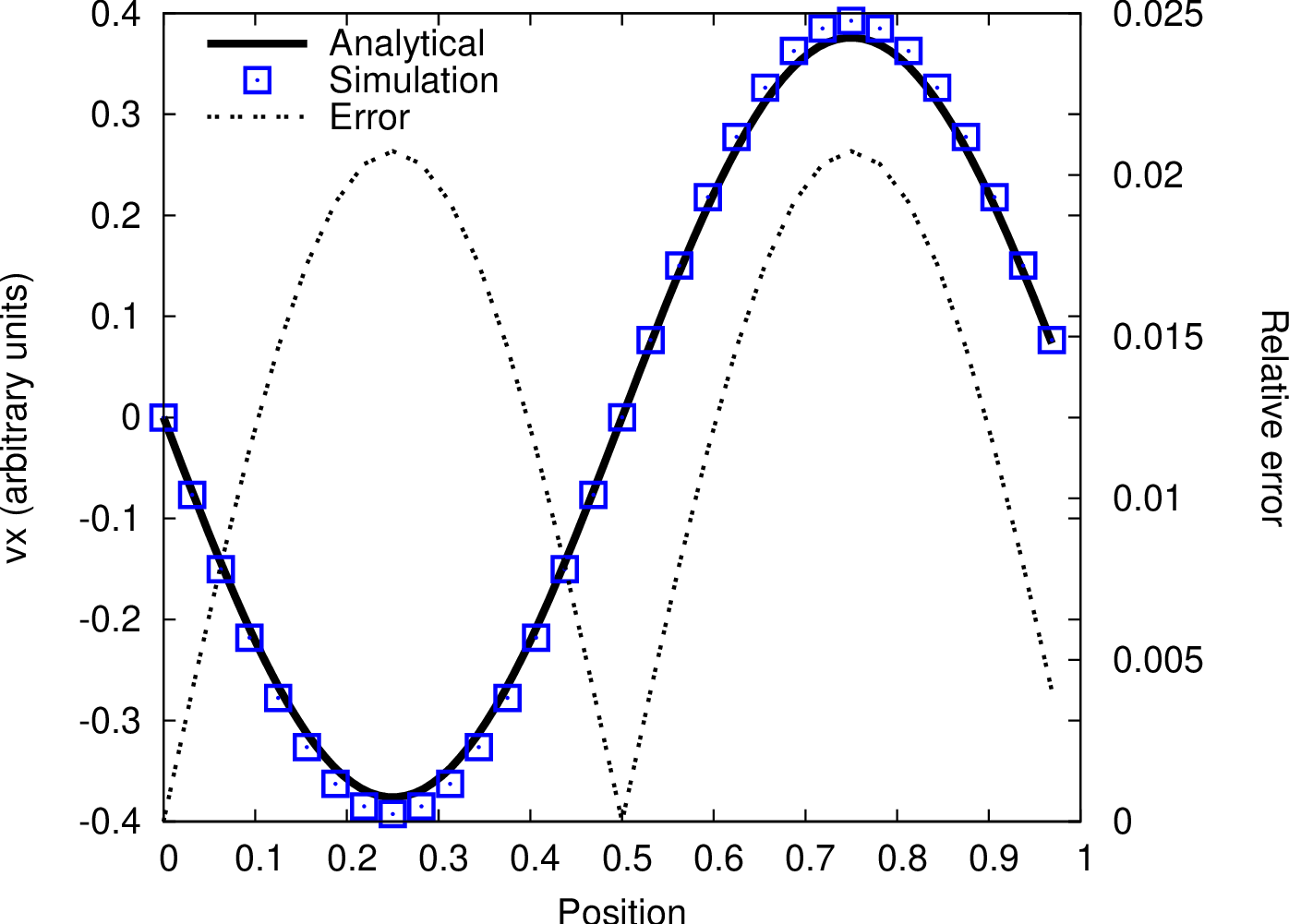}
\includegraphics[width=0.33\textwidth,angle=270]{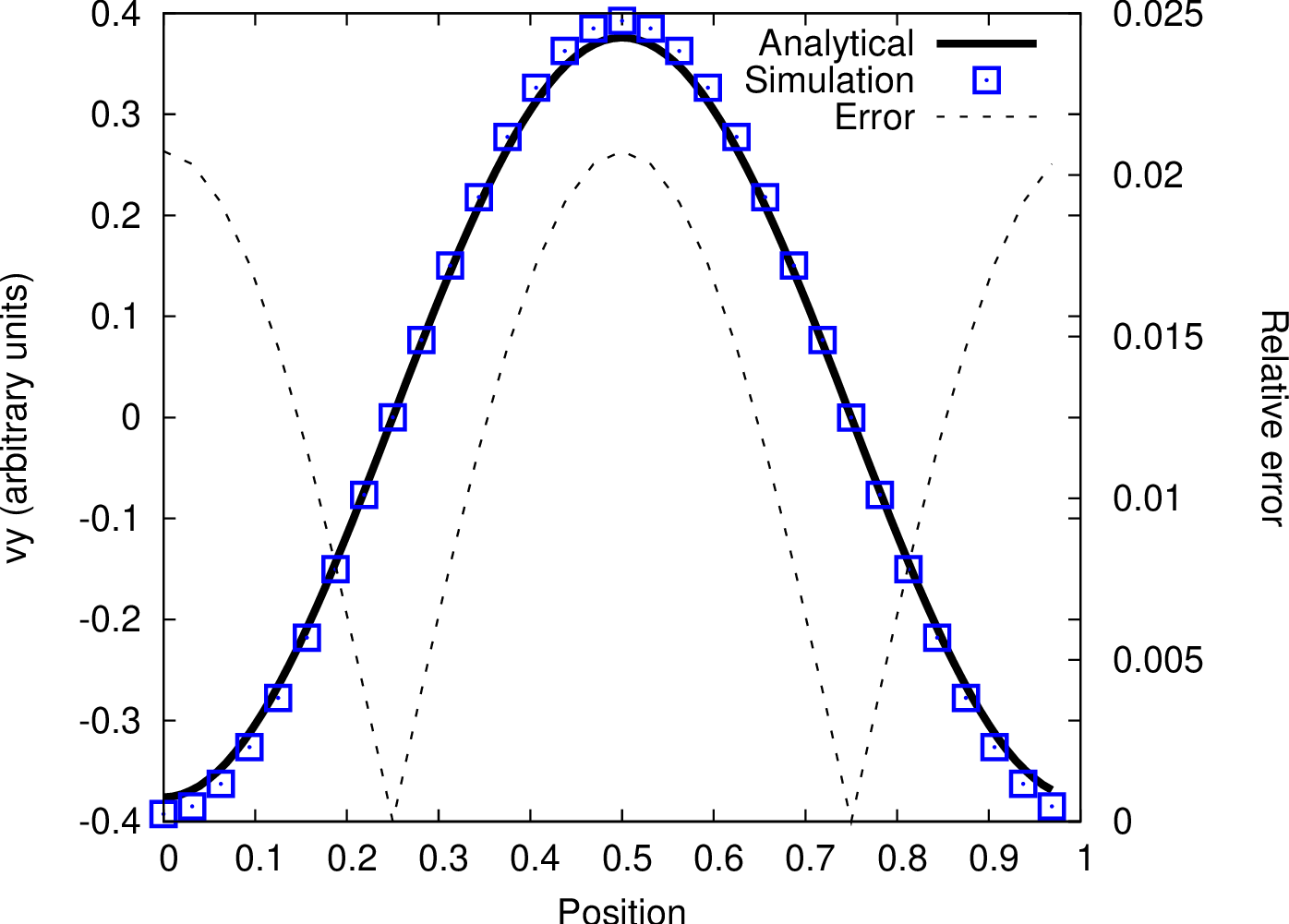}
\\
\includegraphics[width=0.33\textwidth,angle=270]{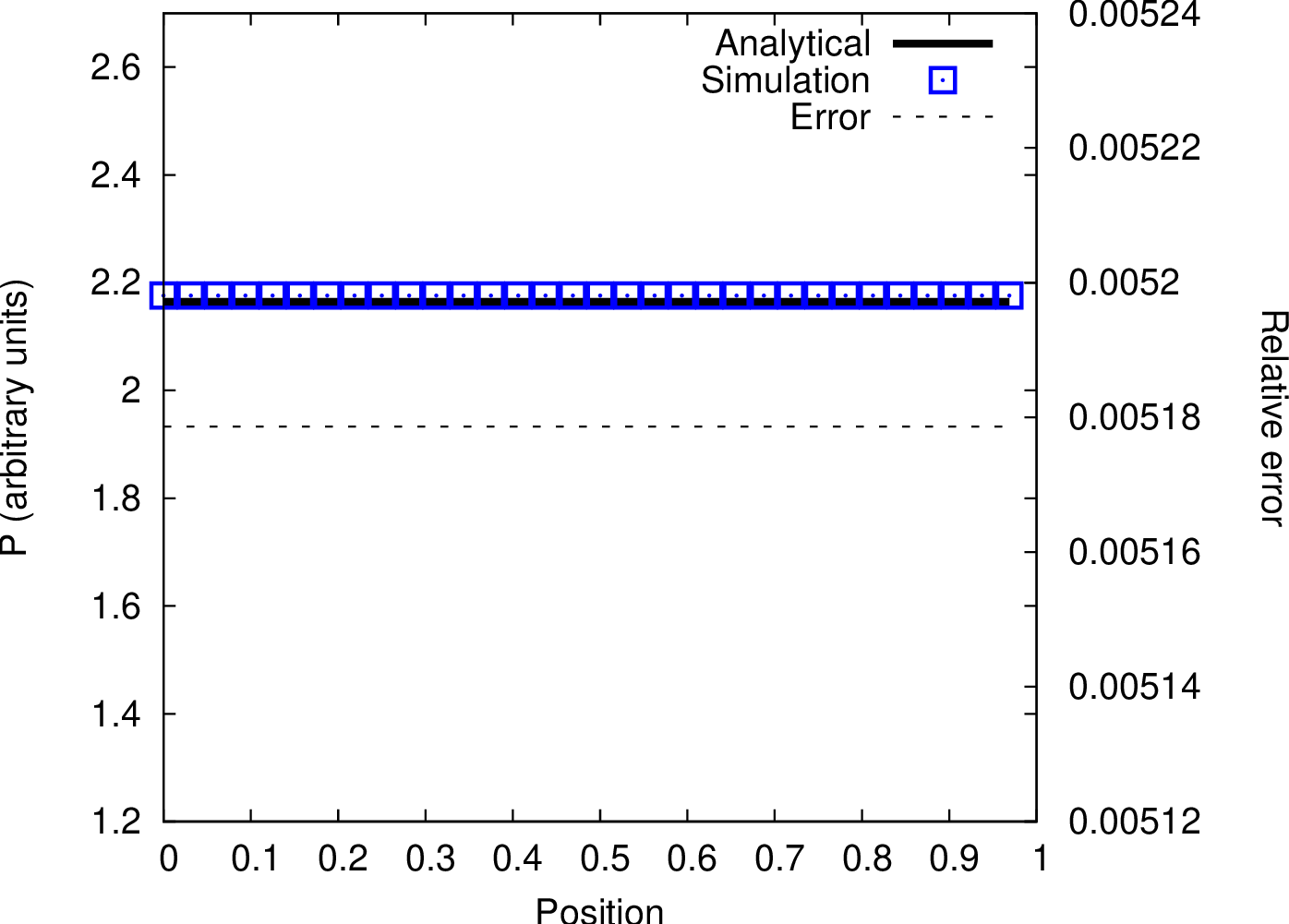}
\includegraphics[width=0.33\textwidth,angle=270]{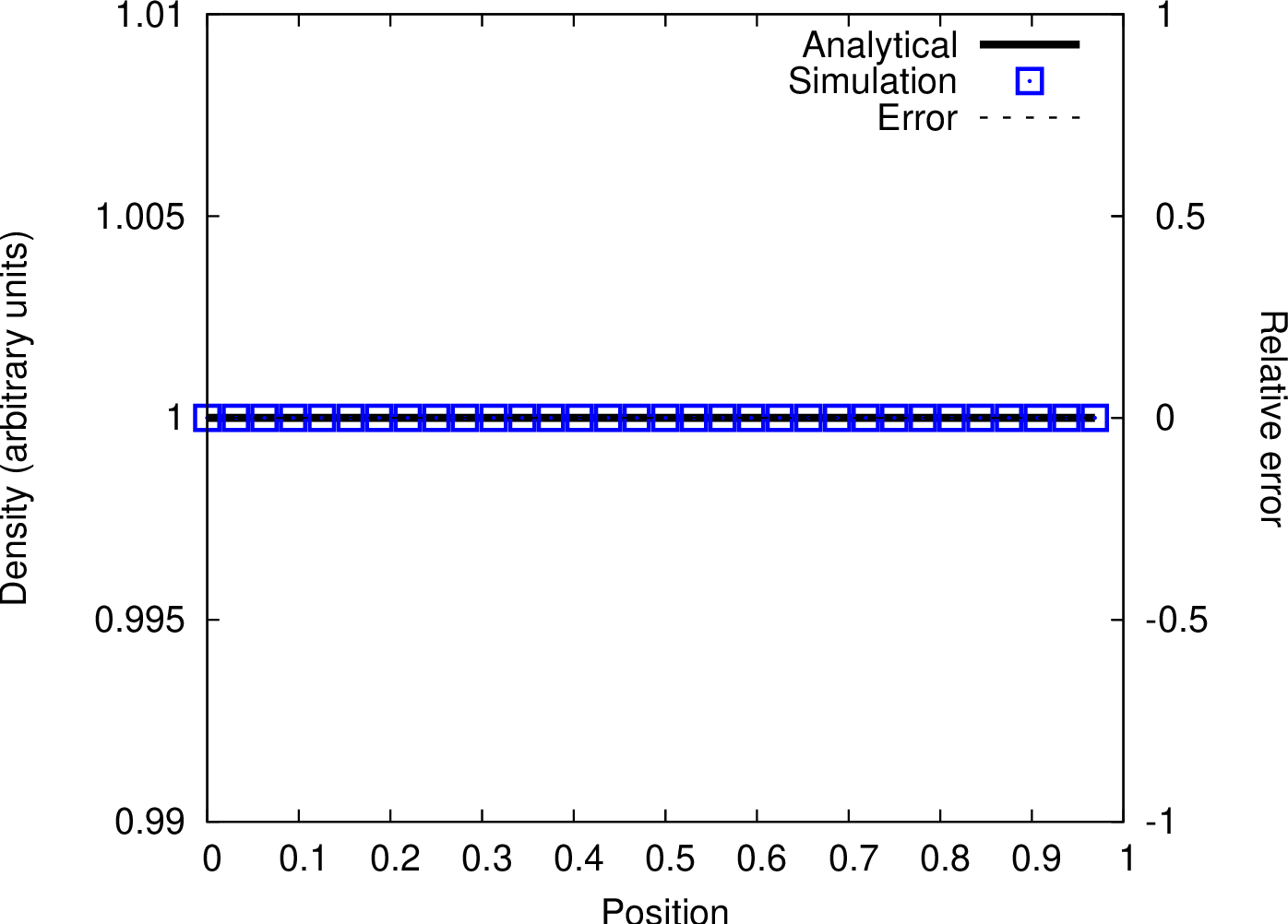}
\caption{The standing Alfv\'en wave test with ambipolar diffusion ($\gamma_{AD}=80$) after about four periods. The simulation is represented by squares, while the solid-lines are the exact solutions. The dotted lines represent the relative error. We use for this test a fully refined Cartesian grid with 32 cells.}
\label{figstandingDA}
\end{figure*}

We now start the simulation from an initial perturbed state obtained by adding two propagating waves in opposite directions with the same damping $s_r$, and let the system evolve. Figure~\ref{figstandingDA} displays a snapshot of the evolution of $B_{1x}$, $v_{1nx}$, $B_{1y}$, $v_{1ny}$, $\rho$ and $P$ along $x$, after about 4 periods (in order for $v_{1nx}$ and $v_{1ny}$ to be greater than zero) for $\gamma_{AD}=80$. The excellent agreement between the numerical and the analytical solution is confirmed.

As previously, we determine the time evolution of the pressure thanks to Equation~(\ref{eqenergAlfDA}) 
\begin{align}
 P  &= P_{init} + (\gamma -1) k^2 \eta_{AD}  \Big[   \frac{e^{2 s_r t} -1}{s_r}  + e^{2s_r t} \left( \frac{s_r \cos(2 s_i t) + s_i \sin(2s_i t) }{ |s|^2} \right) - \frac{s_r}{|s|^2} \Big] .
\end{align}

\begin{center}
\begin{figure}
\includegraphics[width=0.45\textwidth,angle=270]{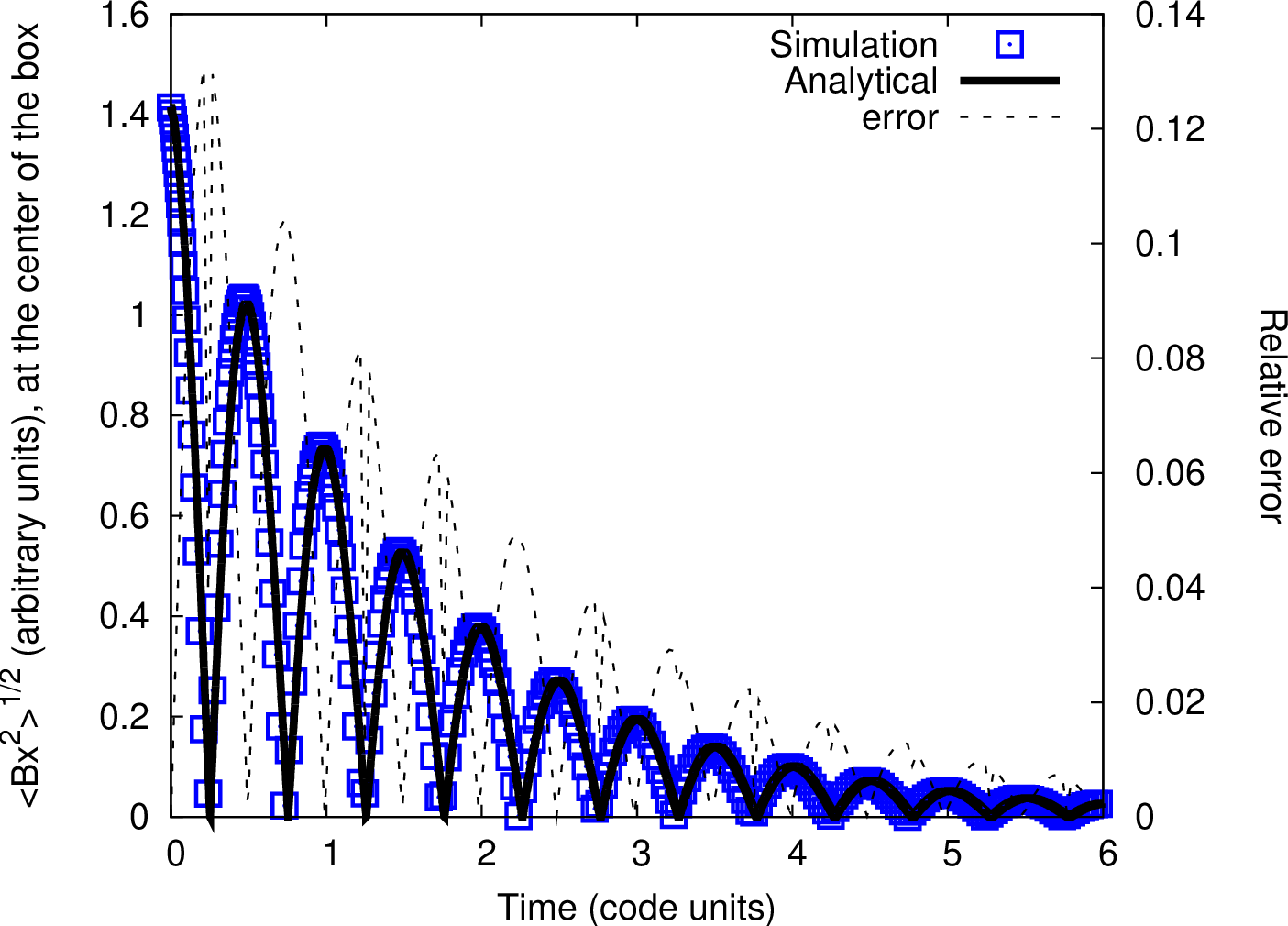}
\caption{Time evolution of $\sqrt{<Bx^2>}$, the root-mean-square of the magnetic field in the $x$ direction at the center of the box, for Alfv\'en standing waves with ambipolar diffusion ($\gamma_{AD} = 30$). The squares are the result of the simulation and the solid line is the analytical solution.}
\label{timebx}
\end{figure}
\end{center}

Following \citet{Choietal} it is interesting to study the time variation of the magnetic field in the $z$ direction, $B_x$, as represented Figure~\ref{timebx}. The analytical solution is represented by the solid line while the dotted line represents the error and the squares the simulation.

\subsubsection{Convergence order}

We tested the evolution of the precision of the implementation of ambipolar diffusion by examining the evolution of the error with the level of refinement, {\it i.e} with the mesh size $\Delta x$, for Alfv\'en standing waves and the Barenblatt test. The error $\epsilon$ is defined here as the maximum difference between the analytical values and the numerical solution, corrected by the damping factor for Alfv\'en waves. The error against the cell size follows a power-law, at least in the range studied here (up to 10 periods of the wave). For the standing waves we find
\begin{eqnarray}
\epsilon \varpropto \Delta x^{3}.
\end{eqnarray}
For the Alfv\'en propagating waves
\begin{eqnarray}
\epsilon \varpropto \Delta x^{2}.
\end{eqnarray}
For the Barenblatt test
\begin{eqnarray}
\epsilon \varpropto \Delta x^{2}.
\end{eqnarray}

A log-log plot of the error as a function of cell size $\Delta x$ for different times is shown on Figure~\ref{figordDA} for Alfv\'en standing waves and propagating waves, and on Figure~\ref{figordbaren} for the Barenblatt test. \label{rev1}\rev{Note that the evolution of the error follows the power laws found through the modified equation study, in Section~\ref{modifequation}.}

\subsubsection{Estimate of the numerical drift coefficient of ambipolar diffusion}

As seen in \S~\ref{alfvenstatio}, the dissipation of Alfv\'en waves is slightly larger than expected according to the analytical values. 
The spurious dissipation due to the numerical scheme can be  estimated as: 

\begin{eqnarray}
\frac{1}{\gamma_{mes}} = \frac{1}{\gamma_{AD}} + \frac{1}{\gamma_{num}},
\end{eqnarray}
where $\gamma_{mes}$ is the value  measured in the numerical simulation, with ${\gamma_{mes}}^{-1} = - \frac{2 s_r \rho_i}{k^2 v_A ^2}$, and ${\gamma_{num}}$ is the drift contribution due to numerical dissipation. Another way to proceed is to set $\gamma_{AD} = \infty $, to
examine how the Alfv\'en waves dissipate, and then to  estimate $\gamma_{num}$ as ${\gamma_{num}}^{-1} = - \frac{2 s_r \rho_i}{k^2 v_A ^2}$. Both methods give about the same value for ${\gamma_{num}}$. For a level of AMR refinement of $2^4$, we get ${\gamma_{num}}^{-1} = 3\times 10^{-3}$; for $2^5$,  ${\gamma_{num}}^{-1} = 5\times10^{-4}$ and for $2^6$,  ${\gamma_{num}}^{-1} = 6\times10^{-5}$, to be compared with ${\gamma_{AD}}^{-1}=0.0125$ or 0.033 for the present simulations. As expected, the better the resolution, the smaller the numerical diffusion.

\begin{center}
\begin{figure}
\includegraphics[width=0.43\textwidth,angle=270]{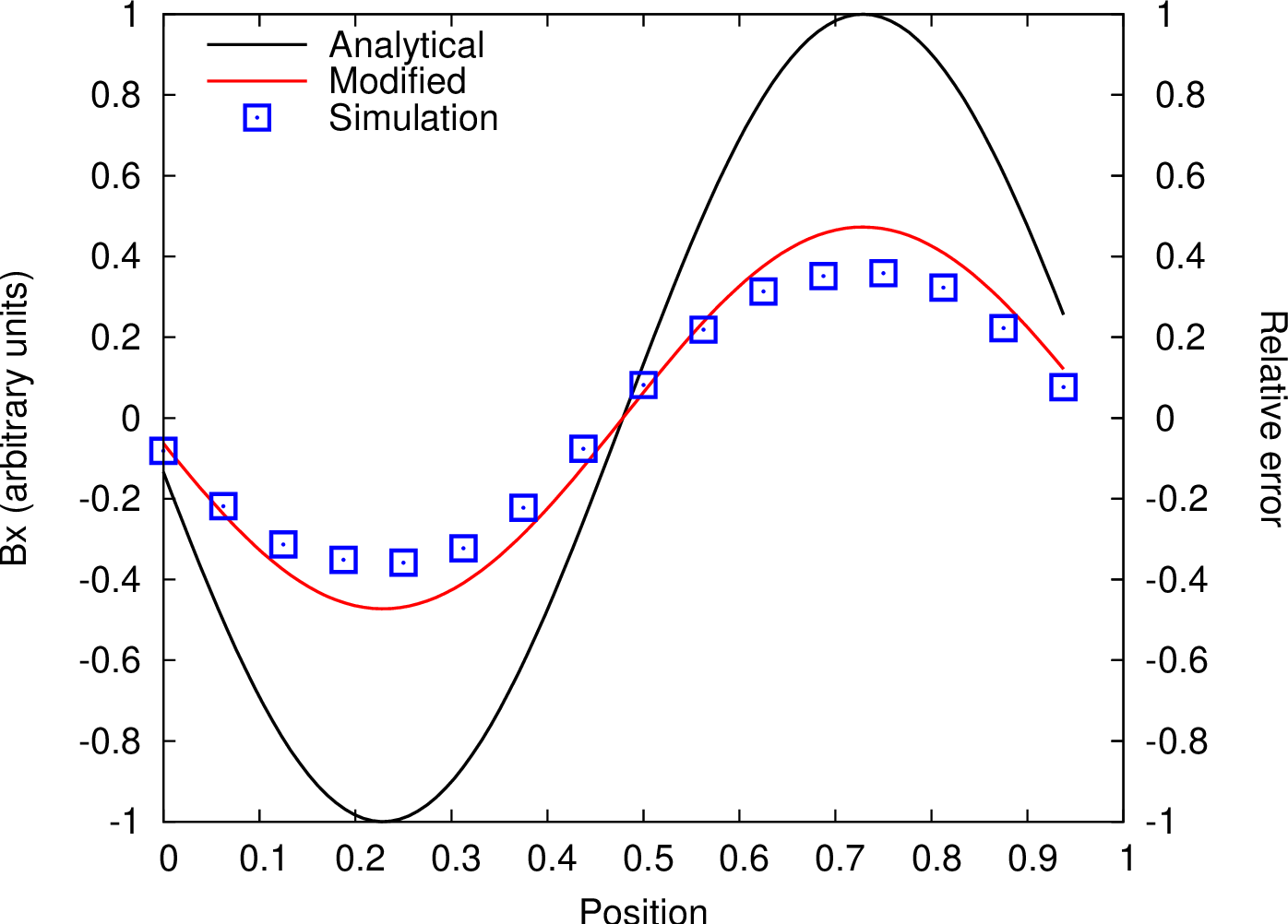}
\caption{Plot of the magnetic field without ambipolar diffusion: $\gamma_{AD} = \infty $. The black solid line shows the analytical solution of the unperturbed Alfven wave, while the red solid line shows the analytical solution with numerical diffusion taken into account (corrected as explained in Equation~(\ref{dispersionavecnum}) for a level of refinement of 4).}
\label{fitcoarse}
\end{figure}
\end{center}

Figure~\ref{fitcoarse} is a plot of the dissipation of Alfv\'en waves with $\gamma_{AD} = \infty $, as explained previously. The red solid line corresponds to the analytical solution corrected with our estimate of the magnitude of the numerical diffusion, as explained in Equation~(\ref{dispersionavecnum}), while the black solid line corresponds to the uncorrected analytical solution (no diffusion).

\section{Ohmic diffusion}
\label{mag}

\subsection{Equations}

We now turn to the case of Ohmic diffusion in the MHD equations. Equations~(\ref{eqcont}), (\ref{eqqtmvt}),  (\ref{eqdivb}) and (\ref{lorentz}) remain the same. The energy equation is now: 
\begin{align}
\frac{\partial E_{tot}}{\partial t}  +{\nabla} . \Big( \mathbf{v} ( E_{tot} + P_{tot}) - \mathbf{B} (\mathbf{v}.\mathbf{B}) -\mathbf{E}_{\Omega} \times \mathbf{B}  \Big) = 0,
\end{align}
where $E_{tot}$ and $P_{tot}$ denote the total energy and pressure:
\begin{align}
E_{tot} = \rho \epsilon + \frac{1}{2}\rho v^2 + \frac{1}{2}B^2 \\
P_{tot} = (\gamma-1)\rho \epsilon +  \frac{1}{2}B^2.
\end{align}

The time evolution of $\mathbf{B}$ reads: 
\begin{align}
\frac{\partial \mathbf{B}}{\partial t} &= {\nabla} \times ( \mathbf{v} \times \mathbf{B} - \eta_{\Omega} {\nabla} \times   \mathbf{B} ).
\end{align}
The Lorentz force and the Ohmic diffusivity EMF read:
\begin{align}
\mathbf{F}_{Lorentz}&= ({\nabla} \times  \mathbf{B}) \times \mathbf{B} \\
\mathbf{E}_{\Omega} &= - \eta_{\Omega} {\nabla} \times   \mathbf{B} ,
\end{align}
where $\eta_{\Omega}$ denotes the Ohmic diffusivity.

\subsection{Computation of Ohmic diffusivity}

Various authors (\citet{Machida_etal06}, \citet{Machida_etal07}, \citet{Machida_etal08}, \citet{Machida_etal09}) have studied the influence of Ohmic diffusion, in particular in the context of molecular cloud's collapse. Their work assumes that the heating from Ohmic resistivity is negligible, and that the approximation $ {\nabla} \times (- \eta_{\Omega} {\nabla} \times   \mathbf{B} ) \simeq \eta_{\Omega} \Delta \mathbf{B} $ is valid. We choose a more general framework and do not assume either of these two assumptions. We implement in {\ttfamily RAMSES} non-isothermal Ohmic diffusivity, with the exact EMF $ \mathbf{E_{\Omega}} = - \eta_{\Omega} {\nabla} \times   \mathbf{B} $

To compute the term of Ohmic diffusivity we proceed exactly as in \S~\ref{ambterm}.

\subsubsection{The Ohmic diffusion EMF}

The EMF in the $z$ direction $\mathbf{E_{\Omega}} \cdot \mathbf{e_z} = - \eta_{\Omega}( {\nabla} \times   \mathbf{B})_z $ is to be computed at $x_{i-\frac{1}{2}},y_{j-\frac{1}{2}},z_k$. Since the EMF writes:
\begin{align}
E^{\Omega}_{z;i-\frac{1}{2},j-\frac{1}{2},k} = -\eta_{\Omega} &\Big( \frac{B_{y;i,j-\frac{1}{2},k}-B_{y;i-1,j-\frac{1}{2},k}}{\Delta x} - \frac{B_{x;i-\frac{1}{2},j,k}-B_{x;i-\frac{1}{2},j-1,k}}{\Delta y} \Big) \label{eqn00003},
\end{align}
it is naturally defined at the right position using the natural definition of the Ohmic field components (see Figure~\ref{schemas12}). $\eta_{\Omega}$ is computed at  $x_{i-\frac{1}{2}},y_{j-\frac{1}{2}},z_k$ using the procedure described in \S~\ref{ambterm} to compute $\gamma_{AD}$, $\rho$ and $\rho_i$.

\subsubsection{The Ohmic diffusion energy flux}

This flux writes $ \mathcal{F}_{\Omega} = \eta_{\Omega} (\mathbf{J} \times  \mathbf{B})$. As explained in \S~\ref{fluxenergAD} the flux has to be evaluated on each face of the cell, that is at locations  $(x_{i \pm \frac{1}{2}},y_j,z_k)$, $(x_i,y_{j \pm \frac{1}{2}},z_{k})$ and $(x_i,y_j,z_{k \pm \frac{1}{2}})$. The computation of $\mathbf{J}$ and $\mathbf{B}$ at these locations is already explained in \S~\ref{fluxenergAD}.

\subsubsection{Computation of the time step in presence of Ohmic diffusion}

The characteristic Ohmic diffusivity time step, $t_{\Omega}$,  is computed according to 
\begin{equation}
t_{\Omega} =  0.1 \times \frac{ \Delta x ^ 2}{\eta_{\Omega}},
\end{equation}
where, as for the ambipolar diffusion case, the coefficient 0.1 yields a small enough time step to ensure good code convergence.
The computational time step is the minimum between $t_{\Omega}$ and the time step obtained for the ideal MHD case.

\subsection{Tests for the Ohmic diffusion}

\subsubsection{Test of Ohmic diffusivity alone}

We first examine the accuracy of the treatment of Ohmic diffusivity alone. We take exactly the same conditions as in \S~\ref{ambalone} for ambipolar diffusion. We further assume that $\eta_{\Omega}$ is constant. In that case the induction equation reduces to a diffusion equation with a constant diffusion coefficient, using the divergence-free condition ${\nabla} \cdot \mathbf{B}=0 $:
\begin{eqnarray}
\frac{\partial \mathbf{B}}{\partial t} = \eta_{\Omega} \Delta \mathbf{B} \label{eqn09}.
\end{eqnarray}

\begin{figure*}
\includegraphics[width=0.33\textwidth,angle=270]{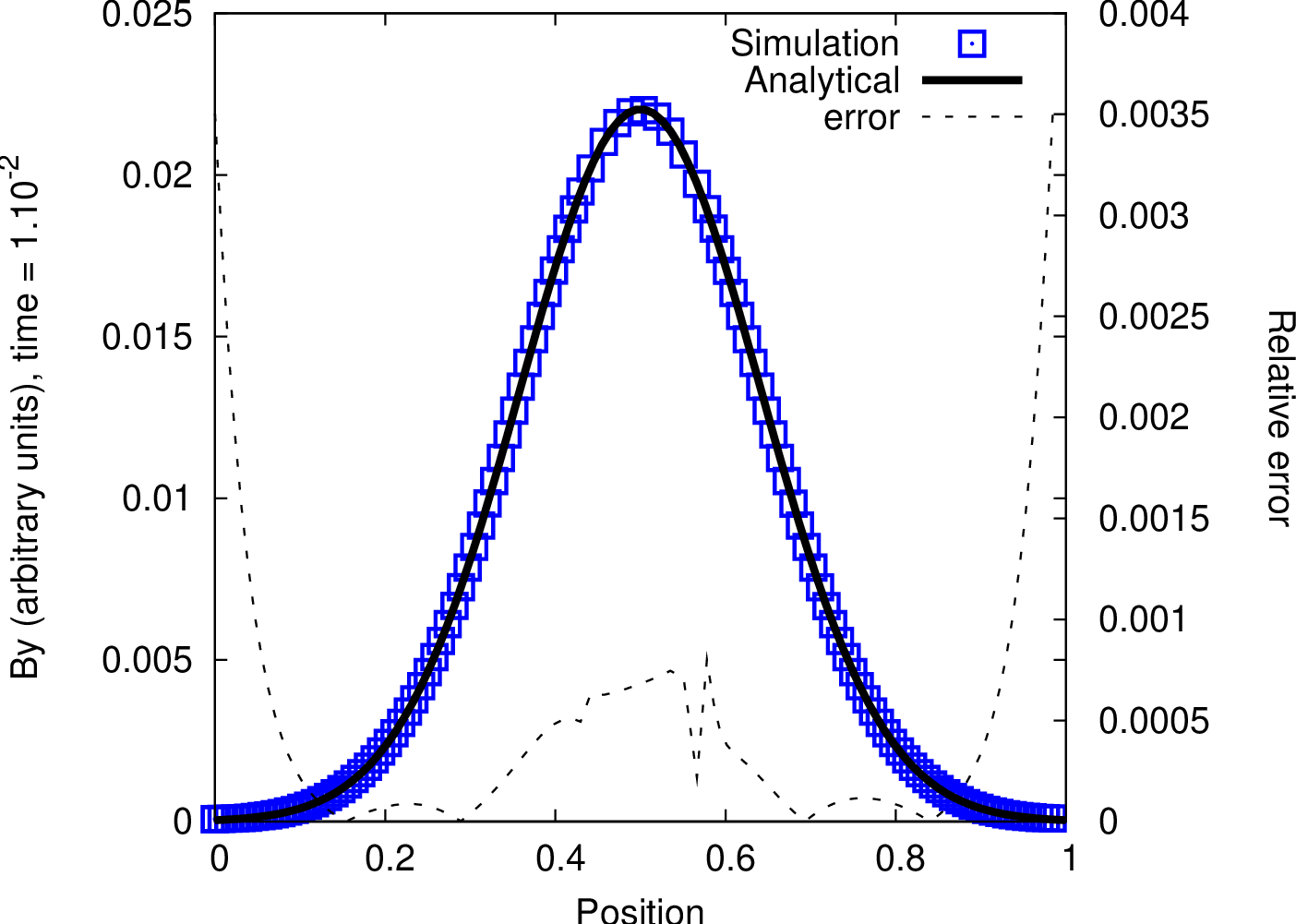}
\includegraphics[width=0.33\textwidth,angle=270]{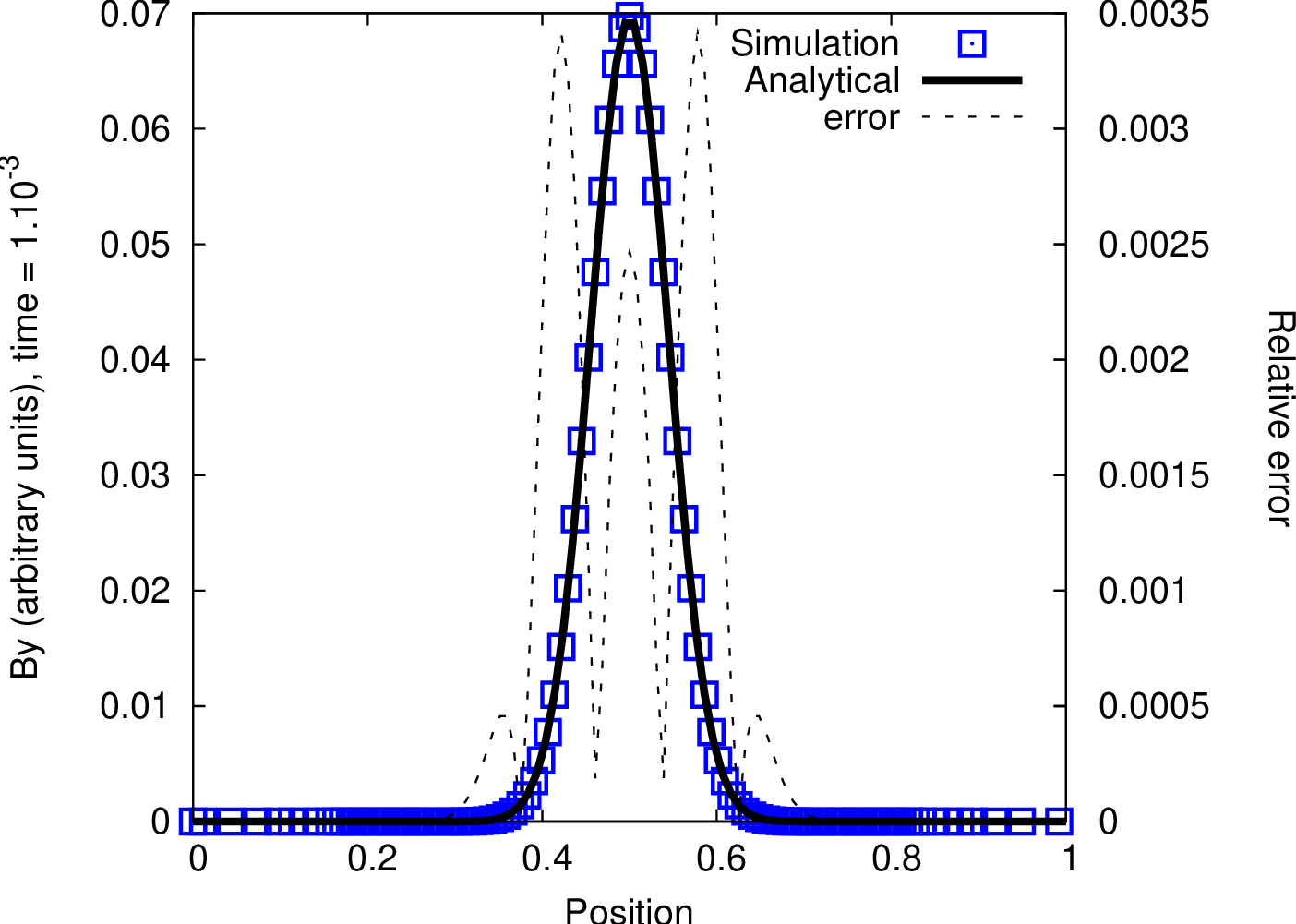}
\\
\includegraphics[width=0.33\textwidth,angle=270]{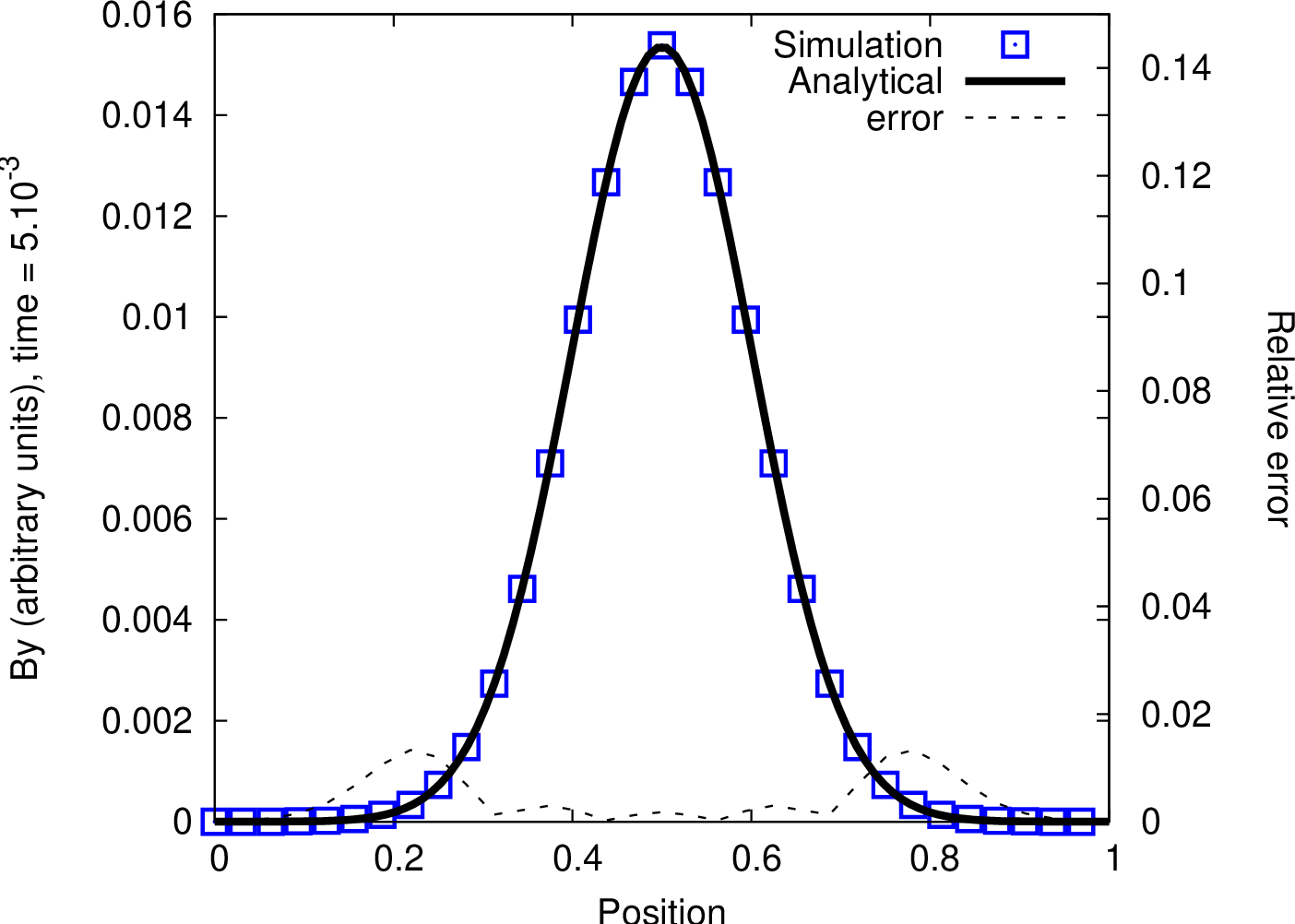}
\includegraphics[width=0.38\textwidth,angle=270]{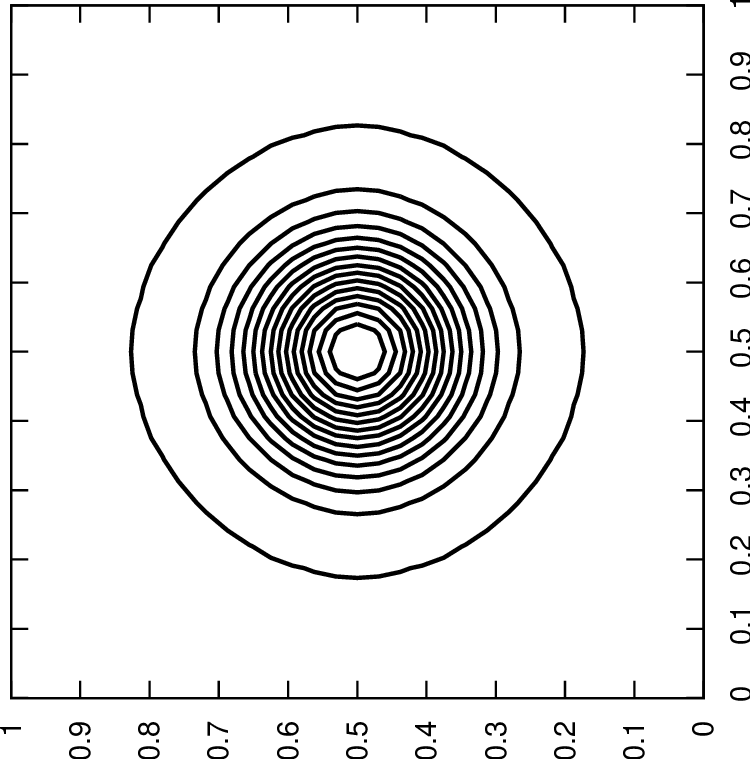}
\caption{Test for Ohmic diffusion only, assuming a Laplacian. The upper panels are snapshots for the 1D test, with an AMR grid with levels from 5 to 7, at times $t=1.10^{-3}$ on the top left and $t=1.10^{-2}$ one the top right panel. The solid lines are the analytical solution, while the dashed lines are the relative error. The lower panels represent the 2D test, on a fully refined grid up to level 5, with a contour snapshot on the right and a transverse cut on the left, at $t=5.10^{-3}$: the symmetry is well preserved.}
\label{diffbaren}
\end{figure*}

The solution to this equation for an initial state given by a Dirac pulse is the well known heat diffusion equation which yields a gaussian distribution with a width spreading as $\sigma \propto {\sqrt t}$. This can easily be studied either for a one dimensional pulse (e.g. $B_y(x)$, $B_x=0$, $B_z=0$) or a two dimensional pulse (e.g. $B_y(x,z)$, $B_x=0$, $B_z=0$). The results (setting $\eta_{\Omega}=1$) are displayed on Figure~\ref{diffbaren}. The agreement between the numerical and the analytical results is excellent, always better than about 0.5\%. We checked that the results obtained on an AMR grid are as good as the ones obtained on a regular grid corresponding to the highest level of refinement, and that exactly the same results are obtained for any orientation of the magnetic field.

\begin{figure}
\begin{center}
\includegraphics[width=0.35\textwidth,angle=270]{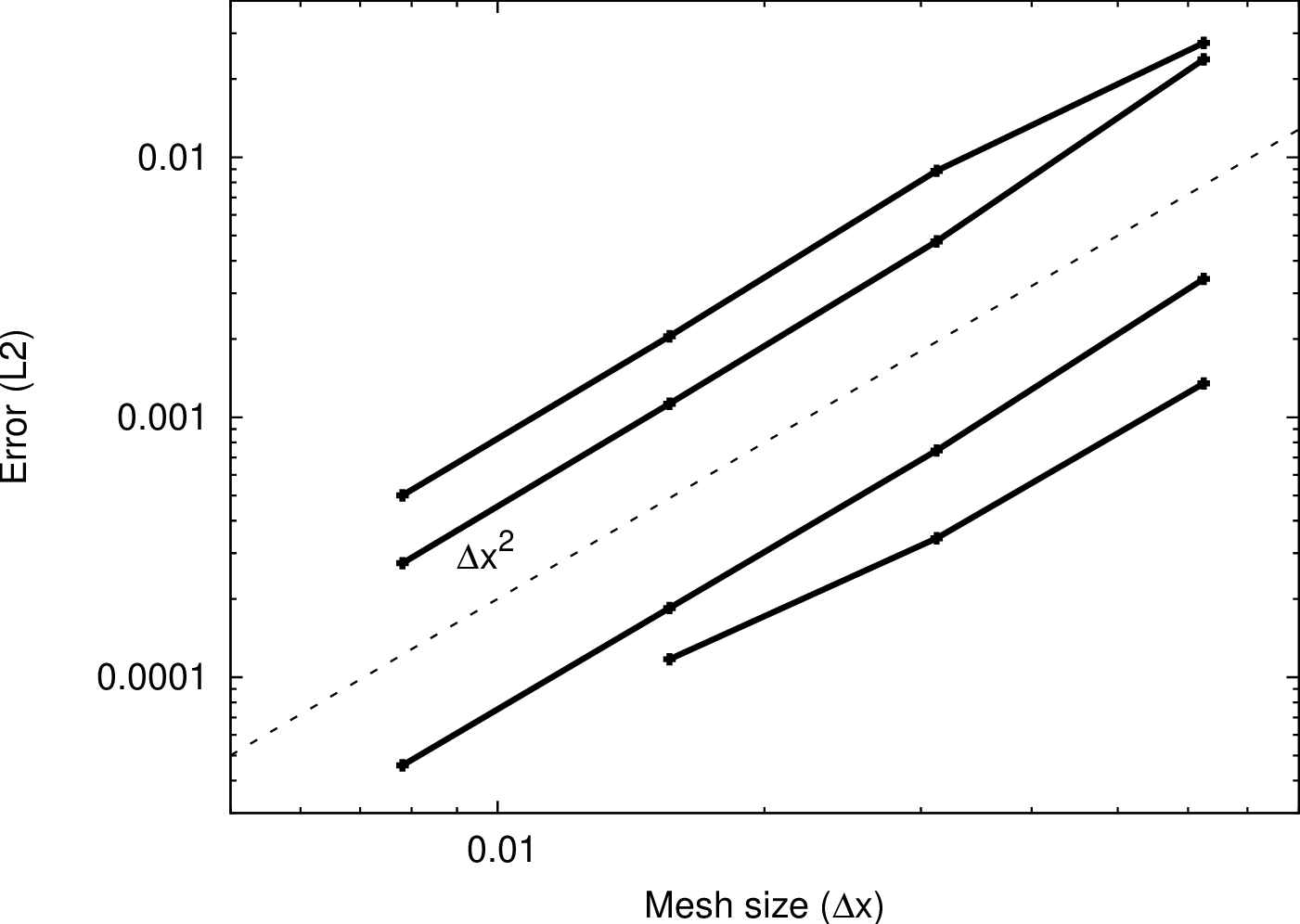}
\caption{Evolution of the error $\epsilon$ for the barenblatt test ($\epsilon = \sqrt{\sum_{i=1}^{N} \, \frac{(By_{\textrm{numerical}}-By_{\textrm{analytical}})^2}{N}}$). The dashed line correspond to: $\epsilon \propto \Delta x^{2}$.}
\label{errorohmicbaren}
\end{center}
\end{figure}

The evolution of the error as a function of the resolution is represented Figure~\ref{errorohmicbaren}. For this particular test (heating equation) the spatial scheme is of order 2: $\epsilon \propto \Delta x^{2}$. 

In this case, due to the smoothly varying magnetic field, using a refinement strategy based on $\nabla B$ is not very efficient: the AMR runs are using typically the same number of cells as a fully refined grid (at least for our tests between level 5 and 9).

\subsubsection{C-shock \label{chocoblDM}}

Proceeding as in \S~\ref{Cshock} we have tested the accuracy of our treatment of Ohmic diffusivity for the case of an oblique C-shock. For a stationary shock in the $x$ direction (all quantities are supposed to only depend on $x$) the equations of mass, momentum, energy, magnetic field conservation and the condition ${\bf \nabla } . {\bf B}=0$ read:
\begin{align}
\partial_x(\rho v_x)  &=0 \\
\partial_x(\rho v_x^2 + P_{gaz} + \frac{1}{2}B_y^2 ) &= 0 \\
\partial_x(\rho v_x v_y - B_x B_y) &=0 \\
\partial_x \left( (E_{tot}+P_{tot}) v_x- (\mathbf{v} \cdot \mathbf{B})B_x - \eta_{\Omega} B_y  \partial_x B_y  \right)   &= 0   \\
\partial_x(v_x B_y - v_y B_x - \eta_{\Omega} \partial_x B_y)  &= 0   \\
\partial_x(B_x) &=  0 .
\end{align}
This set of equations is solved numerically and provides the benchmark to which the simulation with the {\ttfamily RAMSES} code will be compared to assess the accuracy of the numerical treatment in the code.

\begin{figure*}
\includegraphics[width=0.33\textwidth,angle=270]{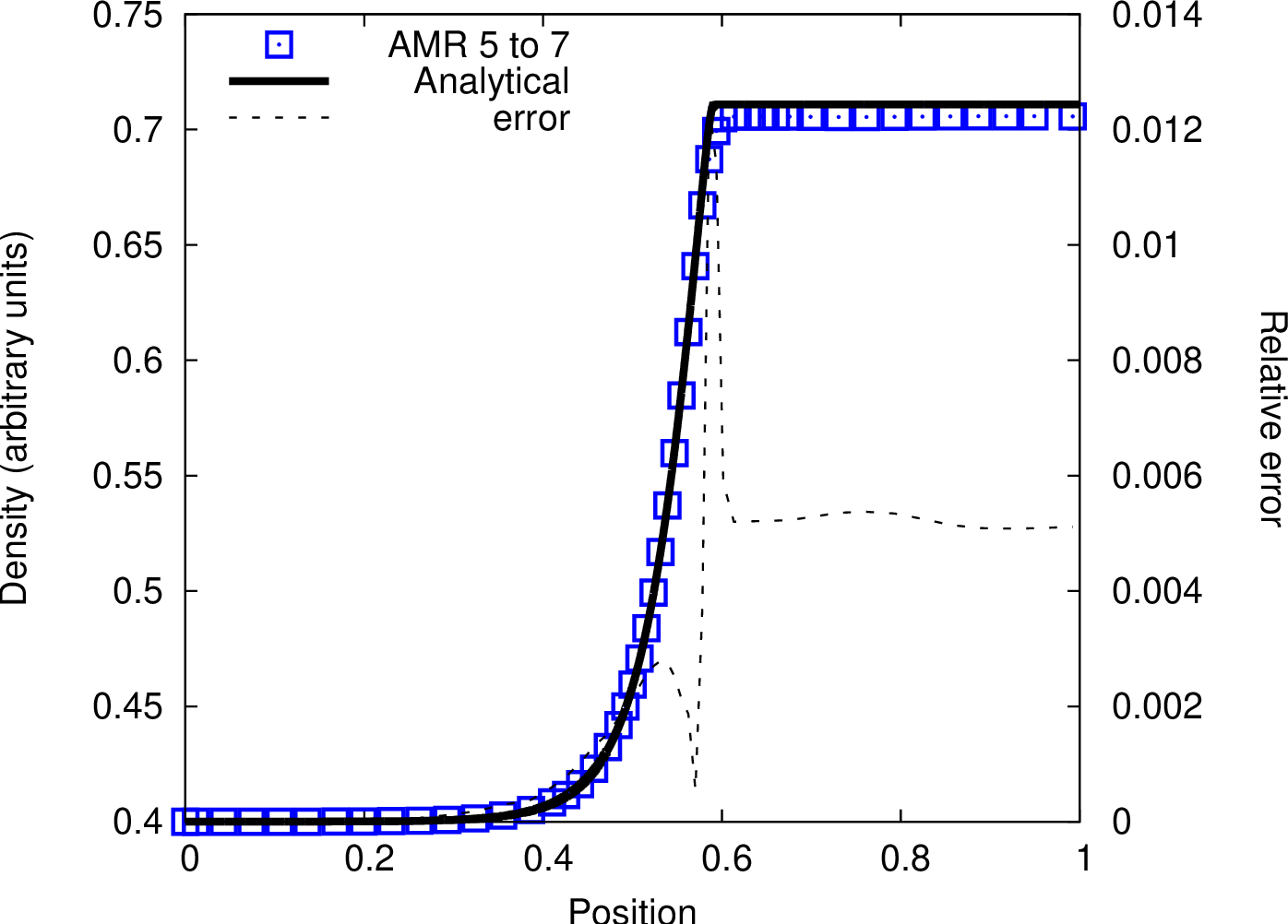}
\includegraphics[width=0.33\textwidth,angle=270]{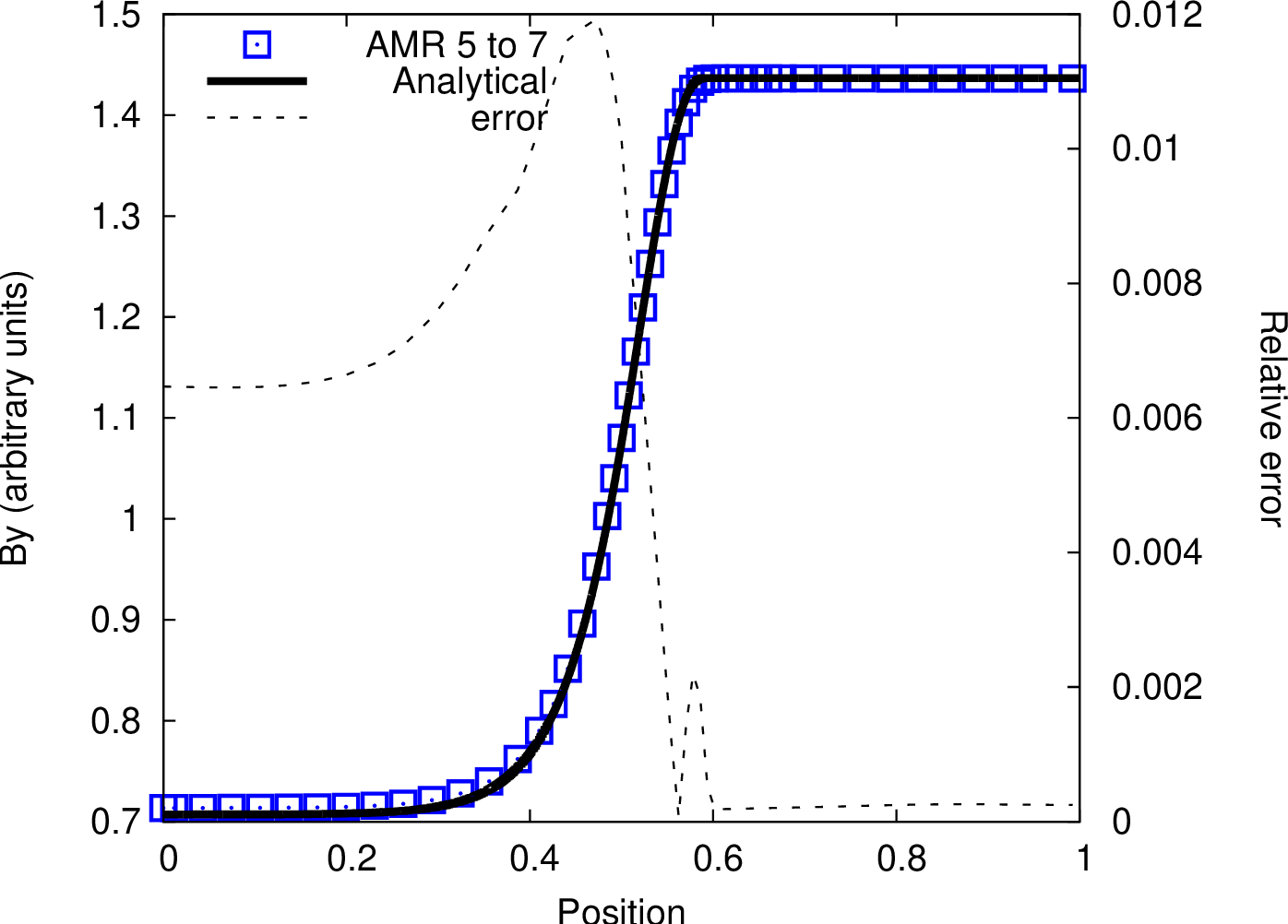}
\\
\includegraphics[width=0.33\textwidth,angle=270]{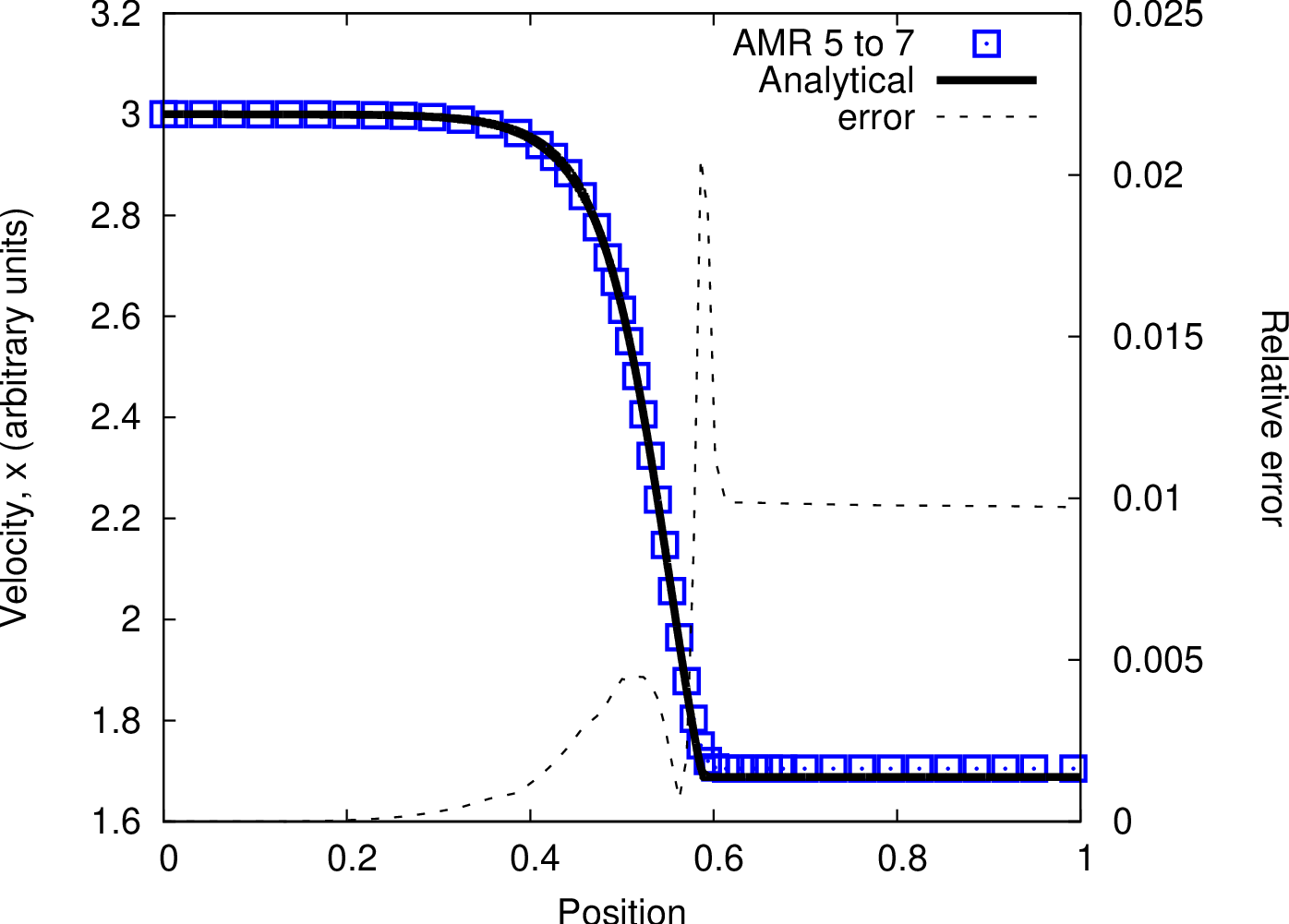}
\includegraphics[width=0.33\textwidth,angle=270]{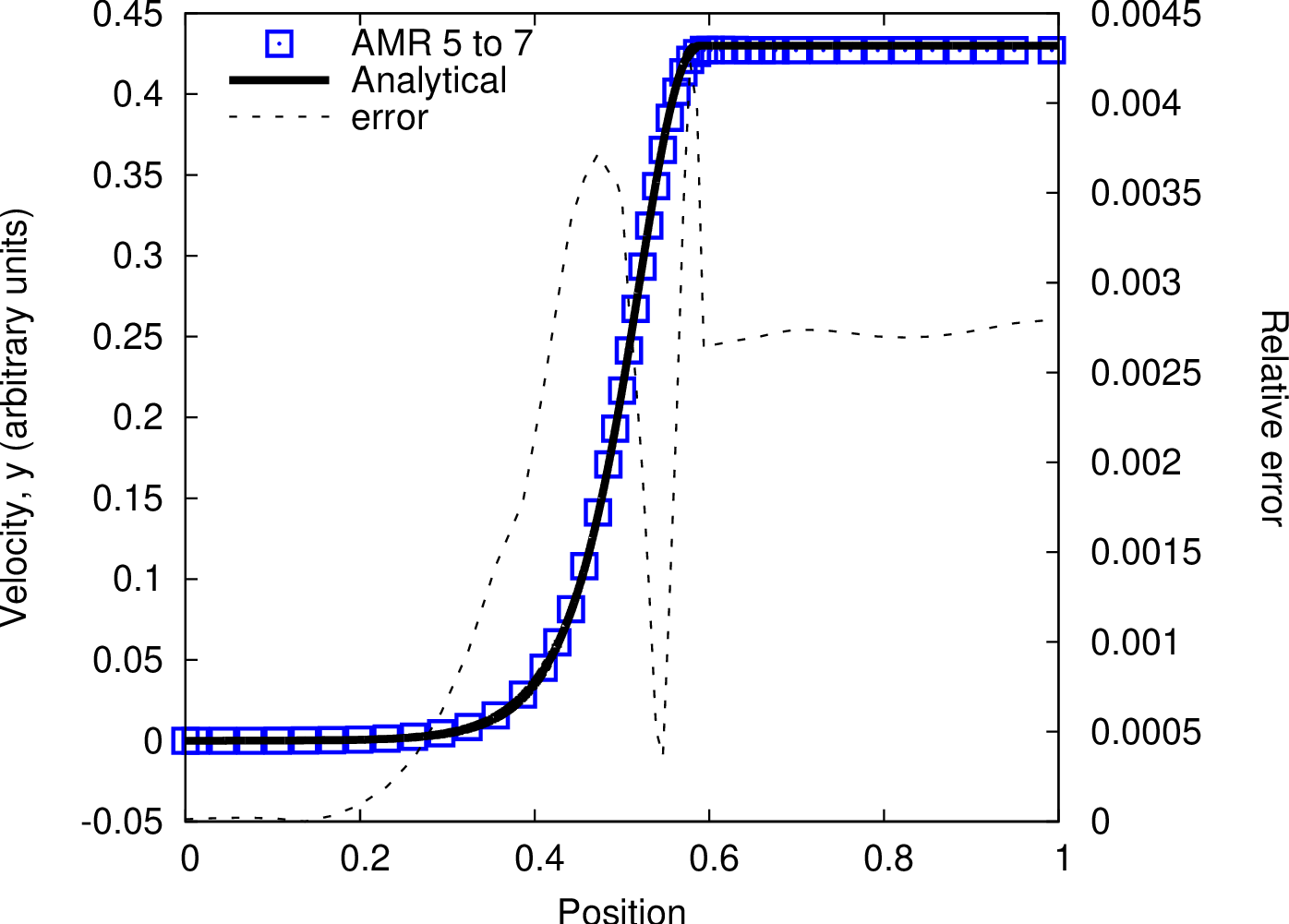}
\\
\includegraphics[width=0.33\textwidth,angle=270]{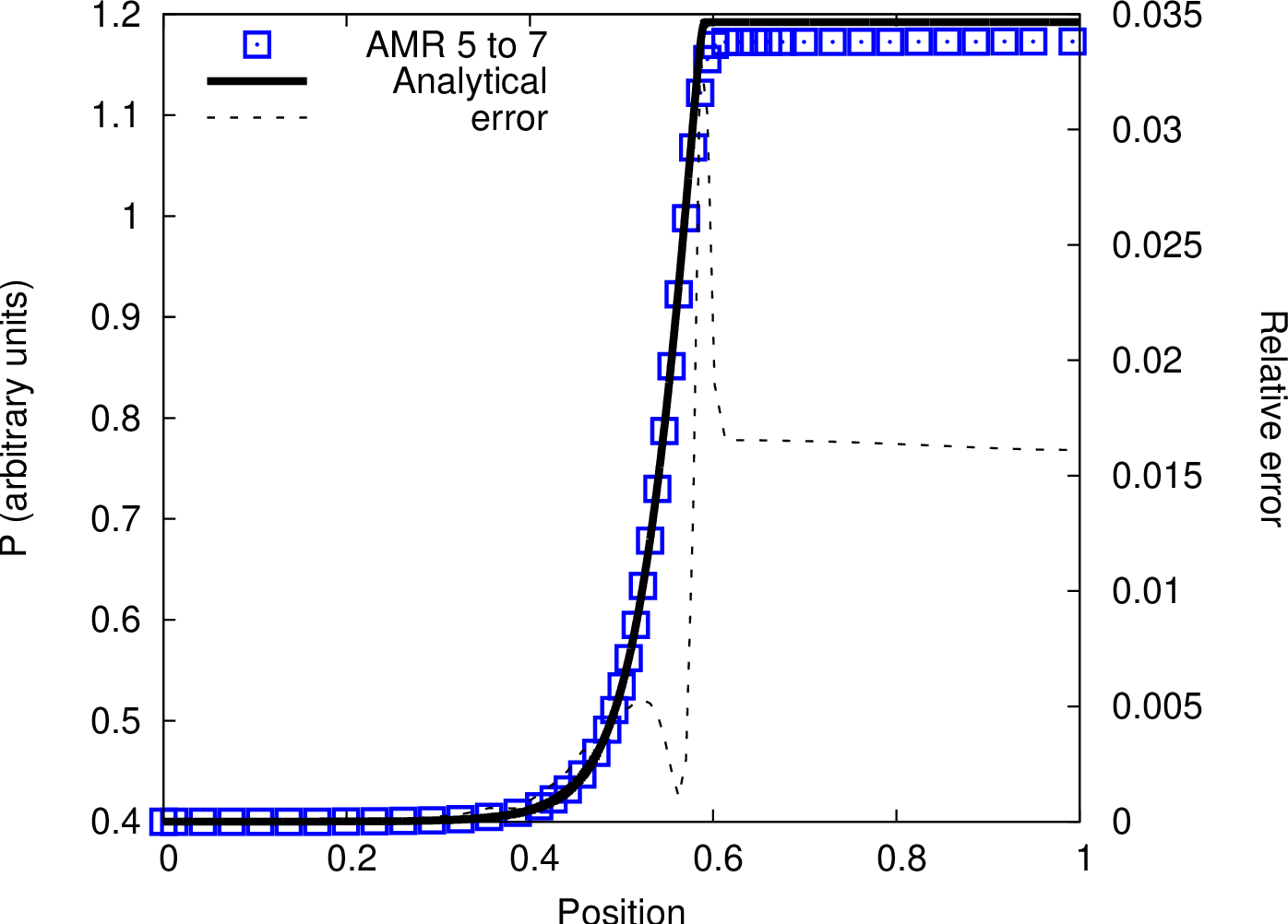}
\caption{Non-isothermal oblique shock with Ohmic diffusion ($\eta_{md}=0.1$). Same caption as in the previous figures. The level of refinement is from 5 to 7.}
\label{figchoc2dohm}
\end{figure*}

We start from a steep function as initial state for the different variables whose values are the ones taken at infinity ahead of and behind the shock, respectively, in the frame of the shock. These values are displayed in Table~\ref{tabchocoblDM}. For this test the Ohmic diffusivity coefficient is set to $\eta_{\Omega}=0.1$. The results are portrayed on Figure~\ref{figchoc2dohm}. As seen in the figure, after a transitory regime the shock becomes stationary, as expected. A very small drift velocity of the shock front persists, of the order of 0.25\% of the minimum value of $v_x$.
Identical results are obtained for any orientation of the magnetic field and of the initial velocity. 

\begin{table*}
\begin{center}
\begin{tabular}{|c|c|c|c|c|c|c|} 
\hline
Variable & $\rho$ & $v_x$ & $v_y$ & $B_x$ & $B_y$ & $P$ \\
\hline
Pre-shock value & 0.4 & 3 & 0 & $\frac{\sqrt{2}}{2}$ & $\frac{\sqrt{2}}{2}$ & 0.4 \\
\hline
Post-shock value & 0.71084 & 1.68814 & 0.4299 & $\frac{\sqrt{2}}{2}$ & 1.43667 & 1.19222 \\
\hline
\end{tabular}
\caption{Initial state used to generate an oblique C-shock, as described in \S~\ref{chocoblDM}.}
\label{tabchocoblDM}
\end{center}
\end{table*} 

Such an agreement between the numerical and the analytical solutions, within about 0.2\% (except a few points where it can reach 1\%) can be considered as very satisfactory and asseses the validity of our treatment when hydrodynamics and Ohmic diffusion are coupled.

The grid is refined if the gradient of magnetic field, pressure, density or velocity is greater than 0.1 (this insures for this test that the error on the AMR grid and on the regular grid are about the same).

\subsubsection{Alfv\'en waves \label{alfvenDM} }

Proceeding as for the ambipolar diffusion study we have examined the behaviour of propagating Alfv\'en waves as well as of standing waves in an non-isothermal ionized plasma in the case of Ohmic diffusion. Lesaffre and Balbus (2007) derived analytical solutions for the general case of MHD flows with shear, non-zero resistivity $ \eta_{\Omega}$, viscosity and cooling. In the absence of shear and rotation, these authors showed that torsional Afv\'en waves are a solution for such flows. 

Following closely the notations of  Lesaffre and Balbus (2007), the unperturbed state in both studies for a wave propagating in the $x$ direction is defined as: 
\begin{eqnarray}
\begin{split}
P_0 &=0.625, &\rho_{0n} &=1, &\rho_{0i}&=1, \\ 
V_{0x}&=0, &V_{0y}&=0, &V_{0z}&=0, \\ \nonumber
B_{0x}&=0, &B_{0y}&=0, &B_{0z}&=1. \nonumber
\end{split}
\end{eqnarray}

For the propagating wave study, the perturbed state is chosen such as $\delta b_x=1$ and $\delta b_y=i \,\delta b_x$ (we necessarily have $\delta b_z=0$). Furthermore, we have $\delta \rho =0$ (constant density), but the pressure varies with time, so that $\delta P \ne 0$ (see Lesaffre \& Balbus 2007). In the absence of shear, viscosity and rotation, the relation between the perturbed magnetic field, $\delta {{\bf b}} = (\delta b_x\, {{\bf x}} + \delta b_y\, {{\bf y}} ) e^{st+ikz}$, and the perturbed velocity, $\delta {{\bf u}}$, reads
\begin{equation}
s \,\delta {{\bf u}} = i\frac{B_0 k}{\rho}\delta {{\bf b}},
\end{equation}
with $s$ the wave angular frequency and $k$ the wave number.

The time evolution of the gaz pressure $P$ is governed by the equation:
\begin{equation}
\partial_t(\frac{P}{\gamma - 1} ) + \nabla \cdot (\frac{P}{\gamma - 1}\delta {{\bf u}}) = -P \nabla \cdot (\delta {{\bf u}}) + \eta_{\Omega} \,{{\bf J}}^2,
\end{equation}
with $\gamma$ the adiabatic coefficient of the gaz and ${{\bf J}}={{\bf \nabla}} \times {{\bf B}}$ the current.

Since $\delta {{\bf u}}$ only depends on $z$ and has components only in the $x$ and $y$ direction, ${\rm div} \,\delta {{\bf u}} = 0$. We finally get 

\begin{equation}
\partial_t P = (\gamma - 1)\eta_{\Omega} \, {{\bf J}}^2.
\end{equation}

The solutions of the dispersion relation read:  
\begin{equation}
s = -\frac{\hat{\eta}}{2} \pm \sqrt{ (\frac{ \hat{\eta} }{2}) ^2-k^2 v_A ^2 },
\end{equation}
with $\hat{\eta} = k^2 \eta_{\Omega}$.
\noindent A value $\eta_{\Omega}=5\times 10^{-3}$ yields a moderate damping, with $s_r=-9.8696\times10^{-2}$ and $s_i=\pm 1.9844$, whereas a value $\eta_{\Omega}=2\times 10^{-2}$ produces a stronger damping, with $s_r=-3.9478\times10^{-1}$ and $s_i=\pm 1.9473$.

\paragraph{Estimating numerical diffusion}

Proceeding exactly as in \S~\ref{modifequation} we can derive the leading order error term in the ideal MHD scheme for Alfv\'en standing, and propagating waves.

\paragraph{Propagating waves \label{alfvenpropaDM}}

We start the simulation from an initial perturbed state with $\delta {{\bf b}}= \delta b_x . {\cal R}e \left( e^{ikx}({{\bf x}} + i {{\bf y}}) \right) $ and $\delta {{\bf u}}=\frac{B_0 k}{\rho} {\cal R}e \left( \frac{i}{s} e^{ikx}({{\bf x}} + i {{\bf y}}) \right) $. The time evolution of the pressure is: 
\begin{equation}
P = P_{ini} + \frac{(\gamma - 1)\eta_{\Omega} \, k^2 \, \delta b_x^2}{2 s_r}(e^{2s_r t}-1).
\end{equation}

\begin{figure*}
\begin{center}
\includegraphics[width=0.33\textwidth,angle=270]{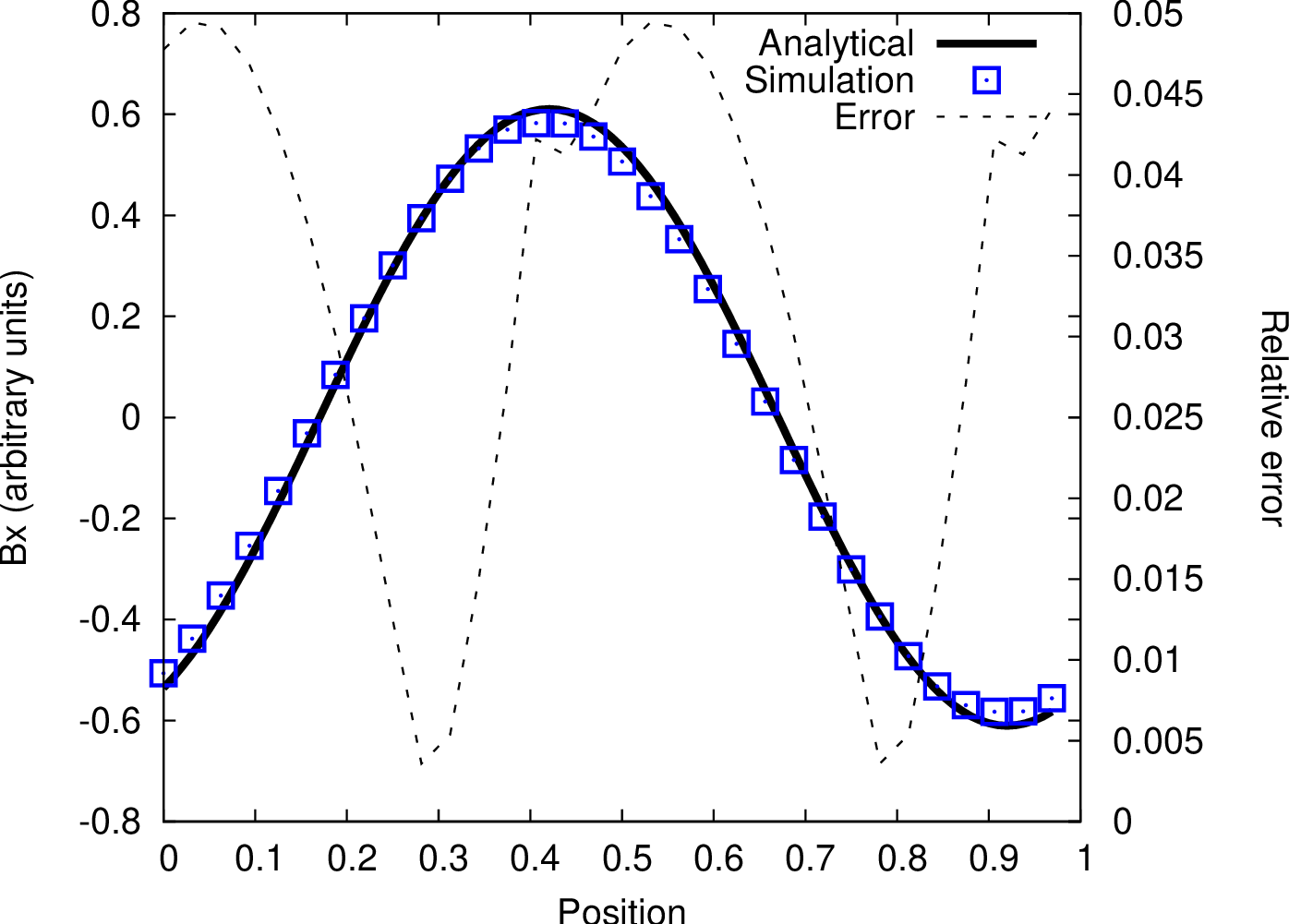}
\includegraphics[width=0.33\textwidth,angle=270]{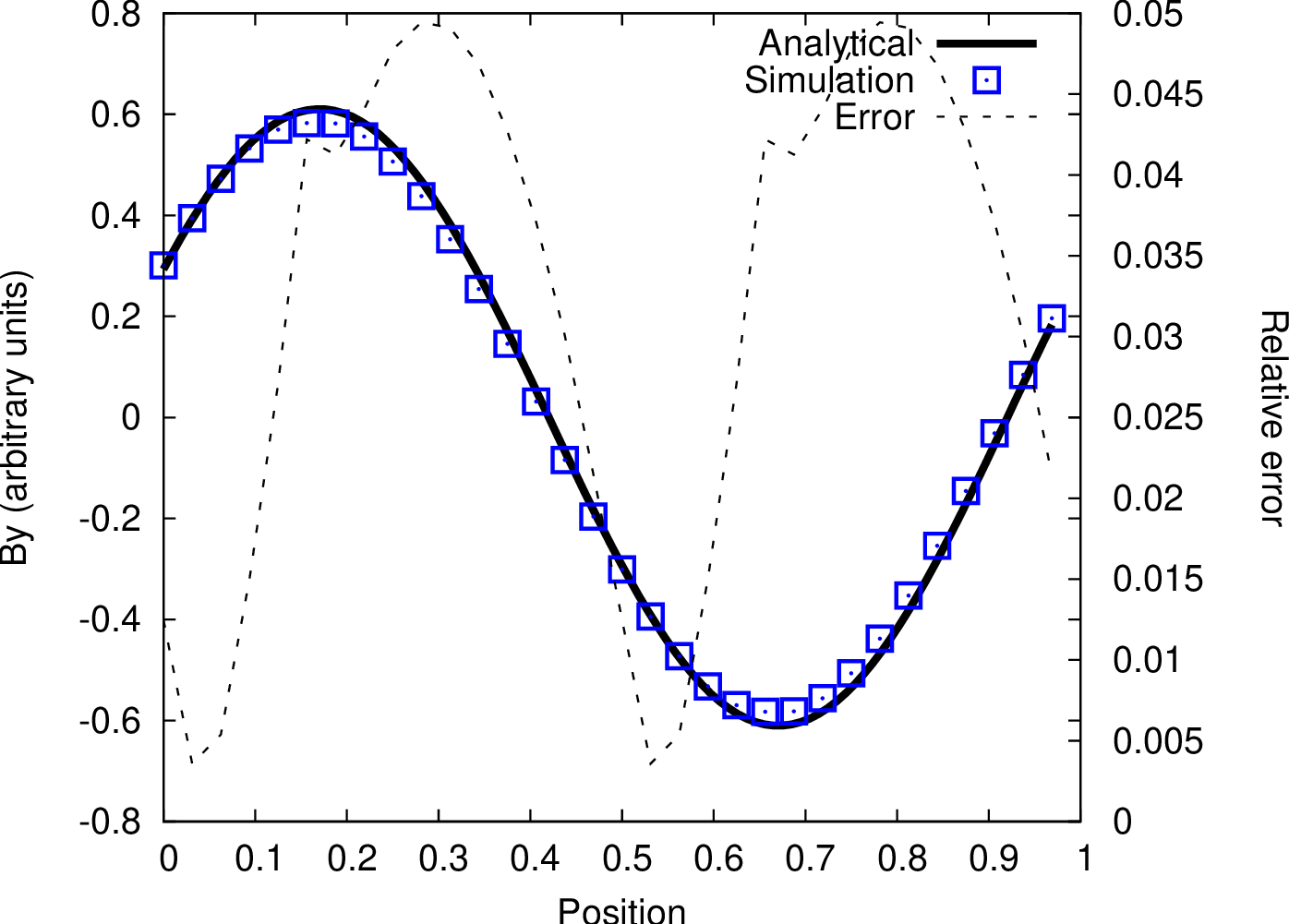}
\\
\includegraphics[width=0.33\textwidth,angle=270]{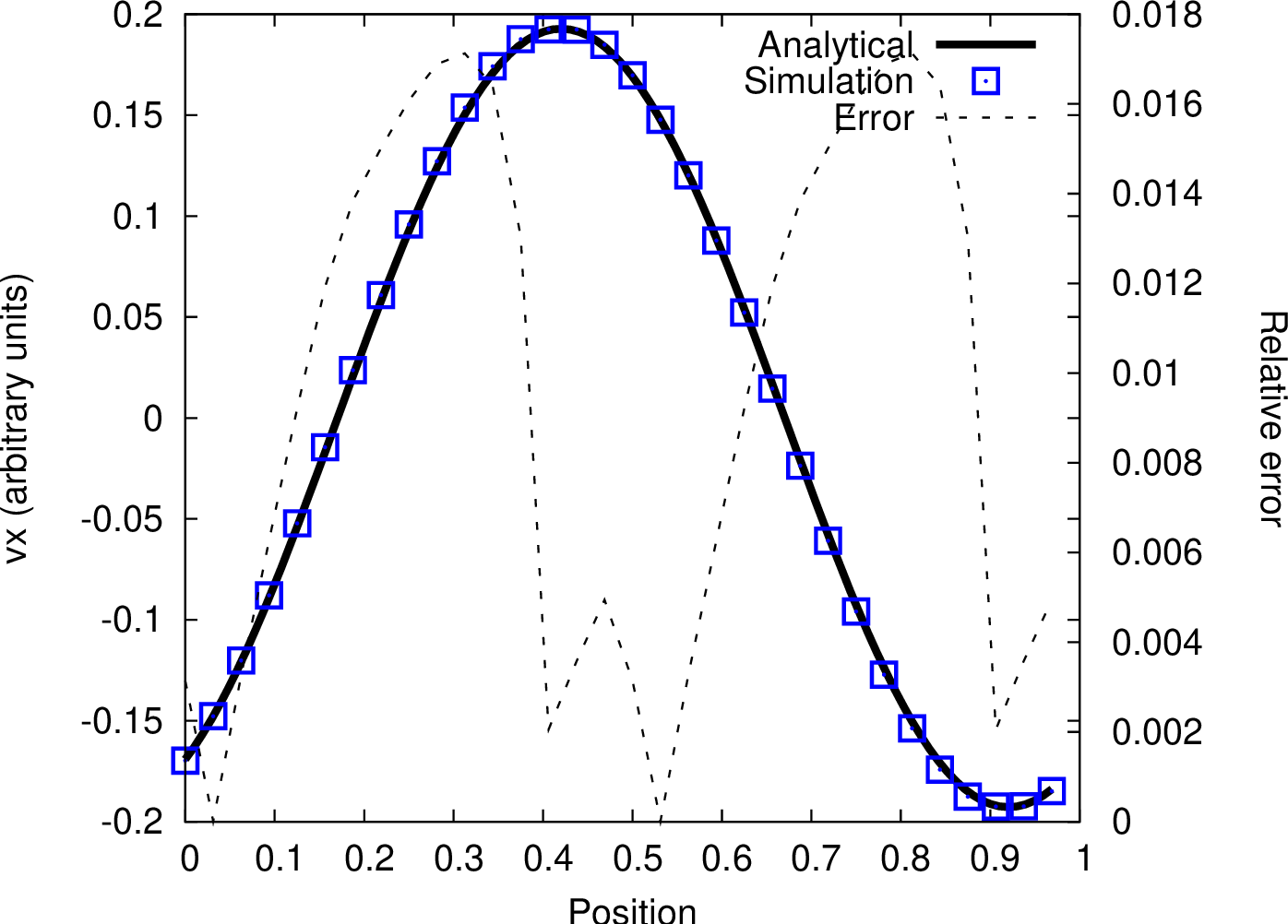}
\includegraphics[width=0.33\textwidth,angle=270]{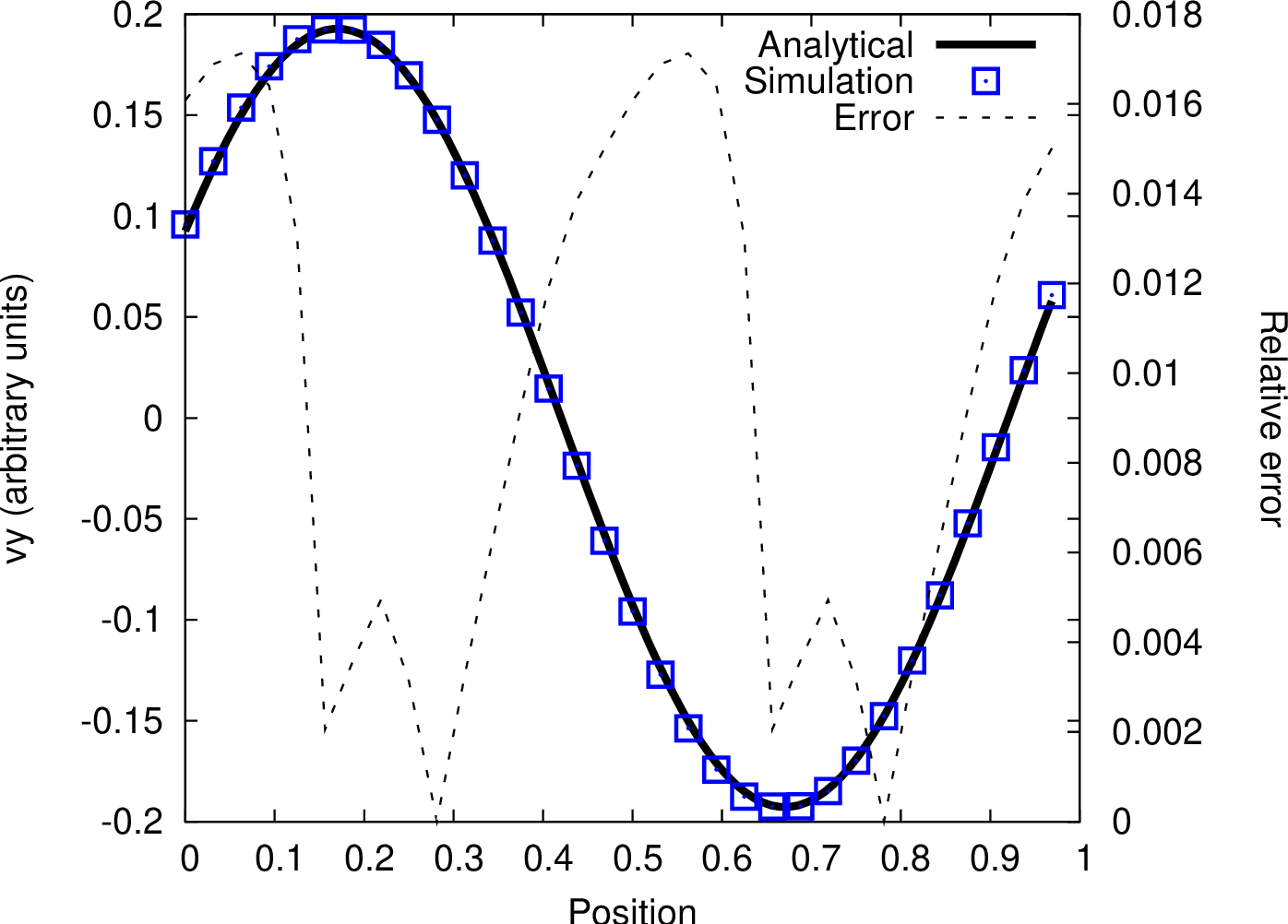}
\\
\includegraphics[width=0.33\textwidth,angle=270]{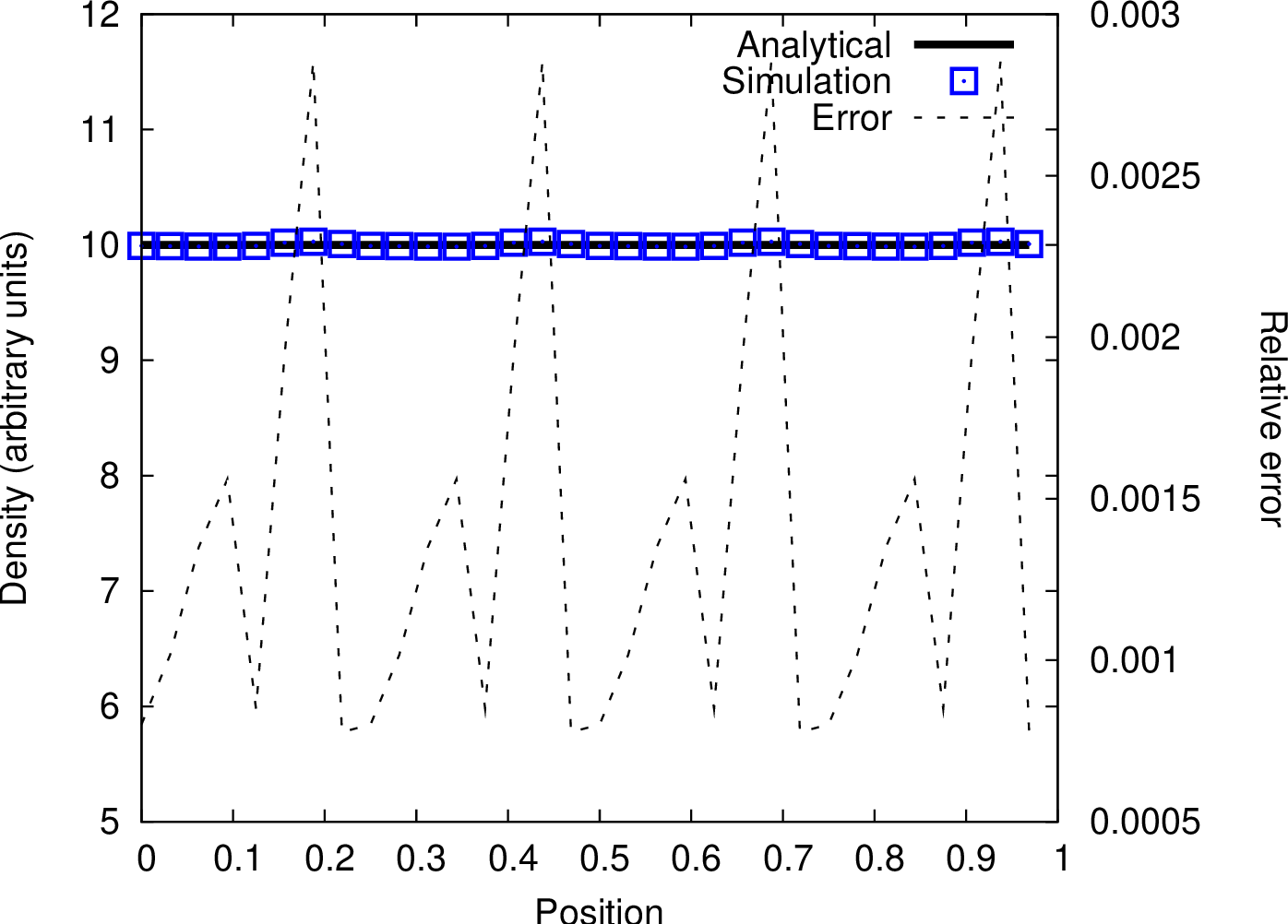}
\includegraphics[width=0.33\textwidth,angle=270]{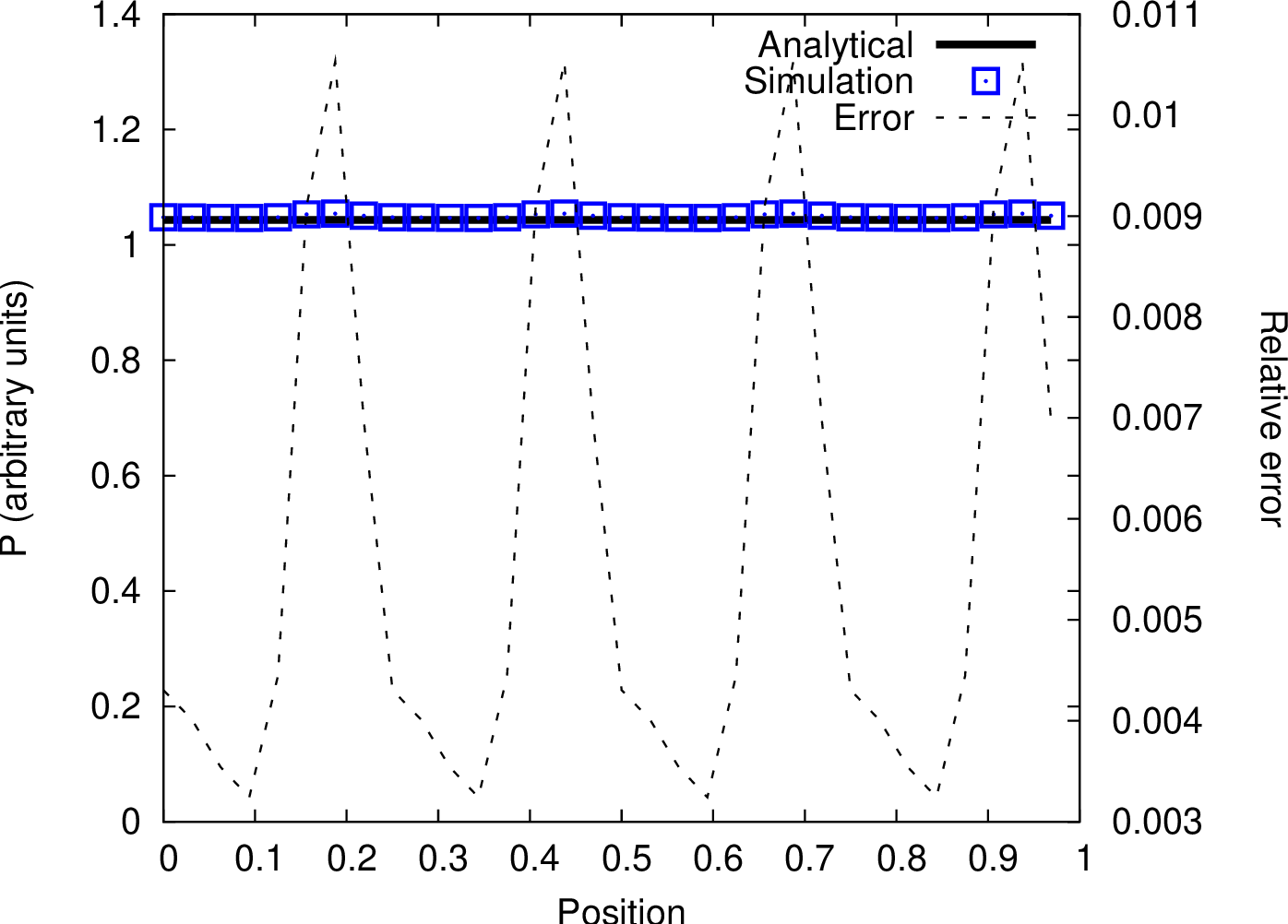}
\caption{Alfv\'en propagating waves after five periods. The level of AMR refinement is constant and equal to $2^5$. The Ohmic diffusivity is $\eta_{md} = 5. 10^{-3}$.}
\label{figalfvpropaDM}
\end{center}
\end{figure*}

For the propagating wave test, we have arbitrarily chosen $s_i > 0$ and let the system evolve from the initial state. Figure~\ref{figalfvpropaDM} portrays a snapshot of the evolution of $\delta {{\bf b}}_x$, $\delta {{\bf u}}_x$, $\delta {{\bf b}}_y$, $\delta {{\bf u}}_y$, $\rho$ and $P$ along the $z$-direction after five wave periods ({\it i.e}, $t= \frac{5 \times 2\pi}{s_i}$), for $\eta_{\Omega} = 5. 10^{-3}$. Once again, the agreement between the numerical and the analytical solution is very satisfactory, at most of the order of a few percents.

\paragraph{Standing waves \label{alfvenstatioDM}}

\begin{figure*}
\begin{center}
\includegraphics[width=0.33\textwidth,angle=270]{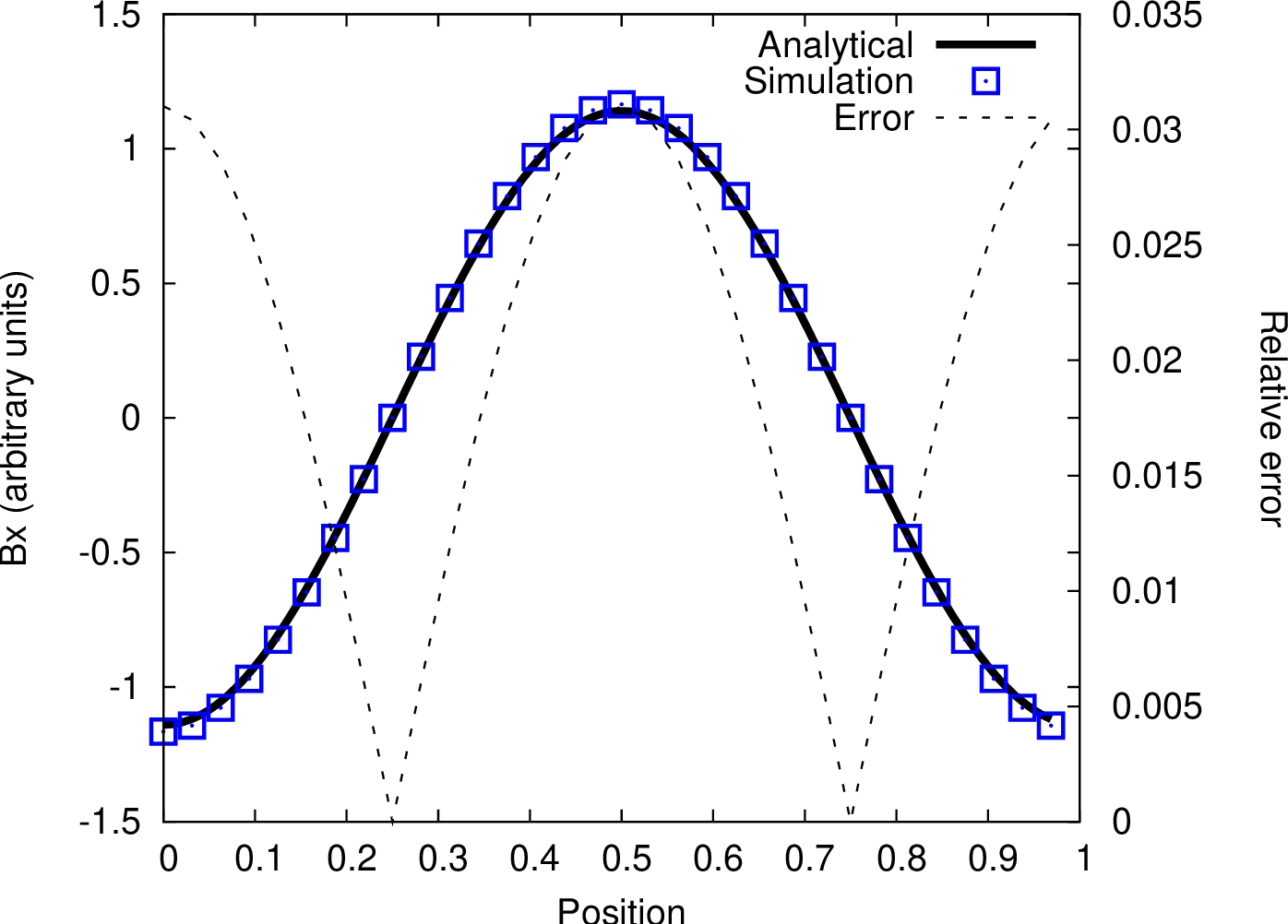}
\includegraphics[width=0.33\textwidth,angle=270]{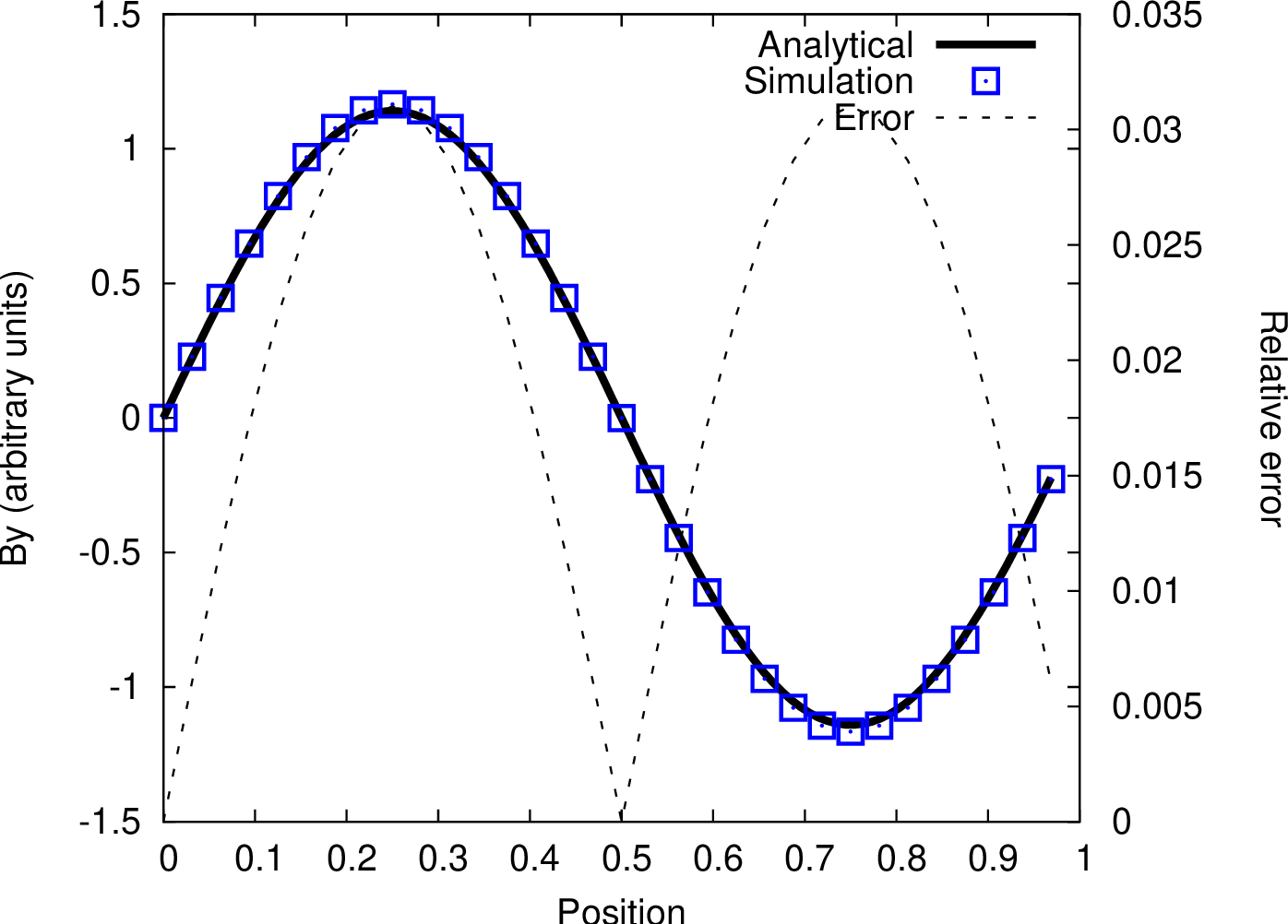}
\\
\includegraphics[width=0.33\textwidth,angle=270]{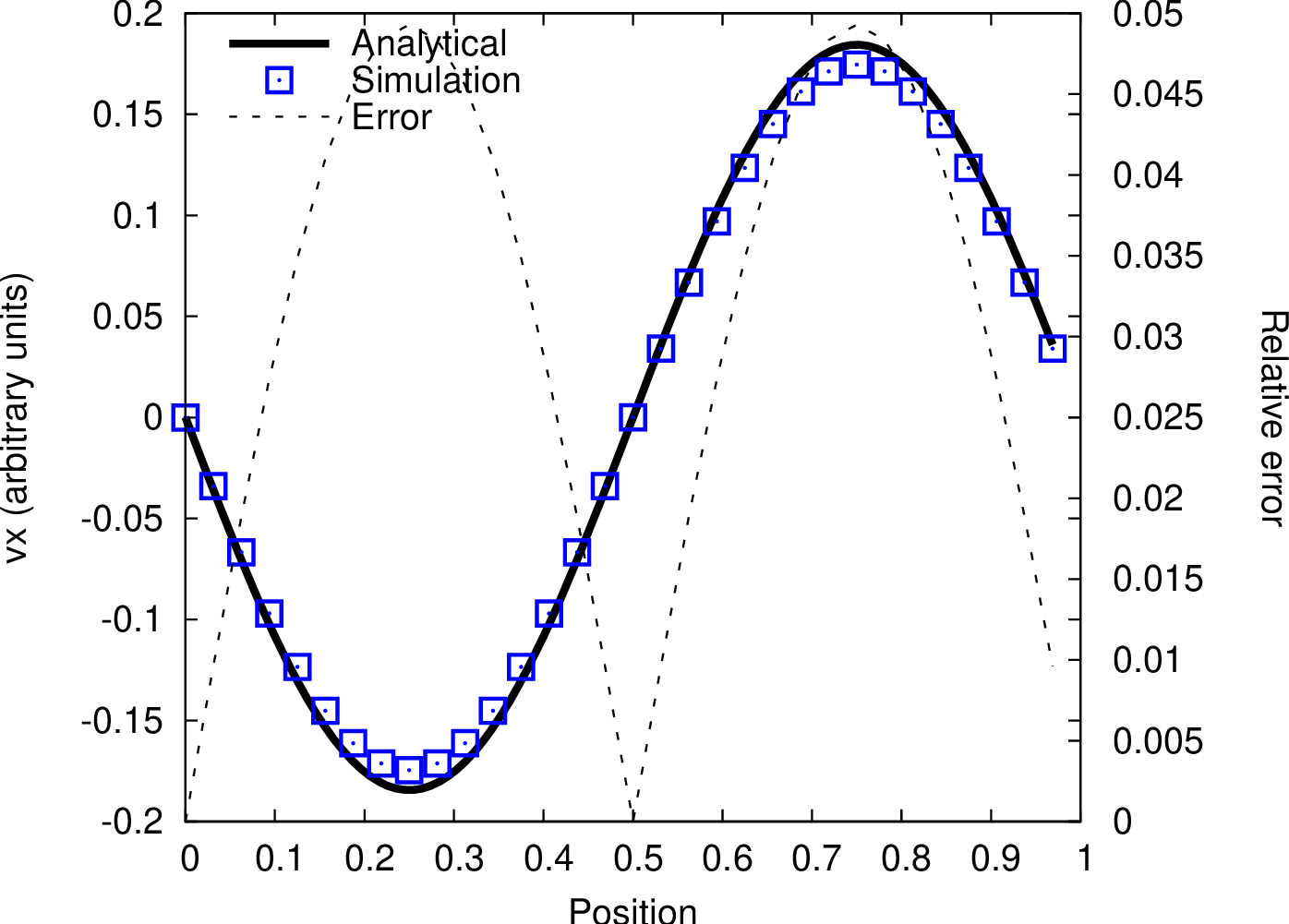}
\includegraphics[width=0.33\textwidth,angle=270]{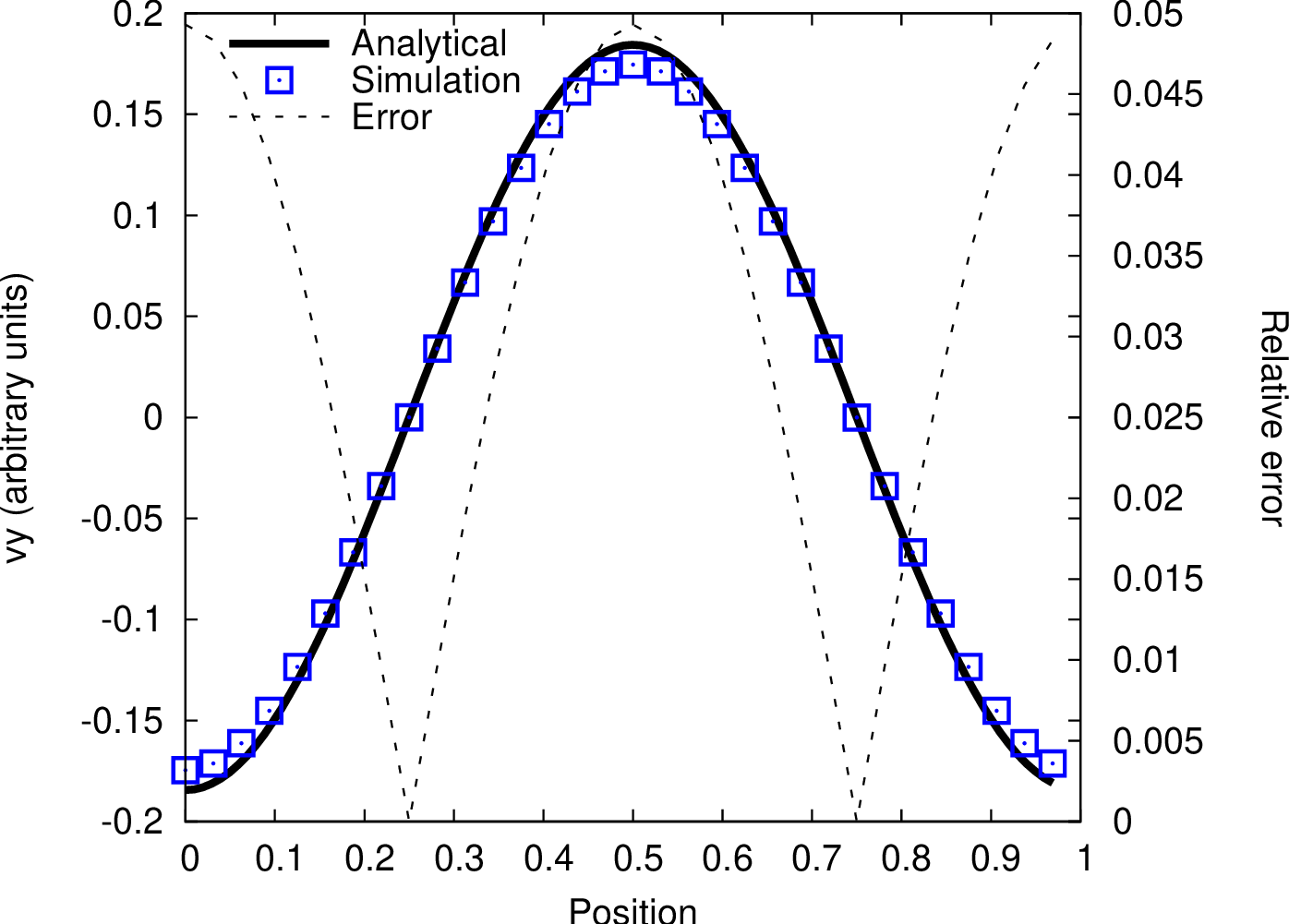}
\\
\includegraphics[width=0.33\textwidth,angle=270]{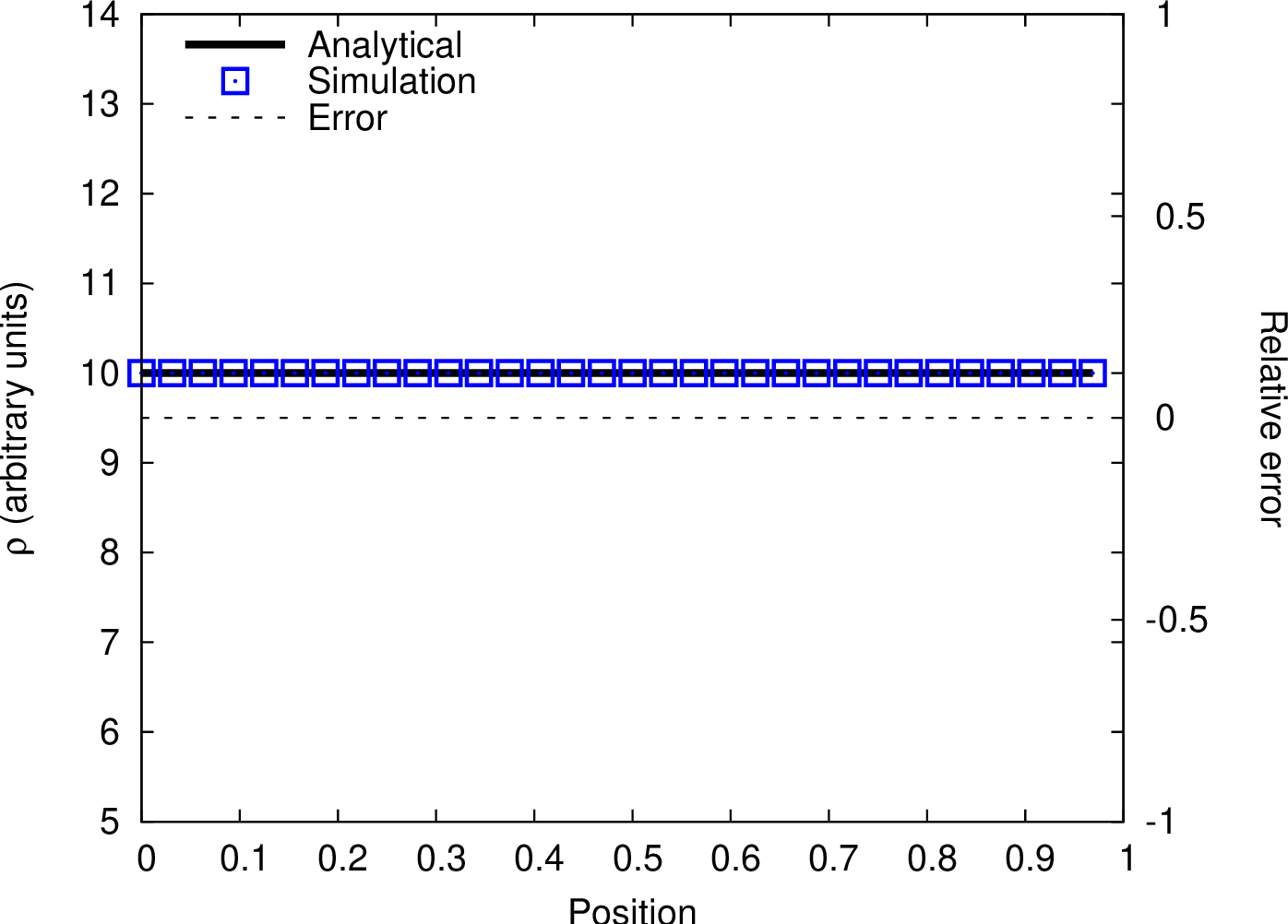}
\includegraphics[width=0.33\textwidth,angle=270]{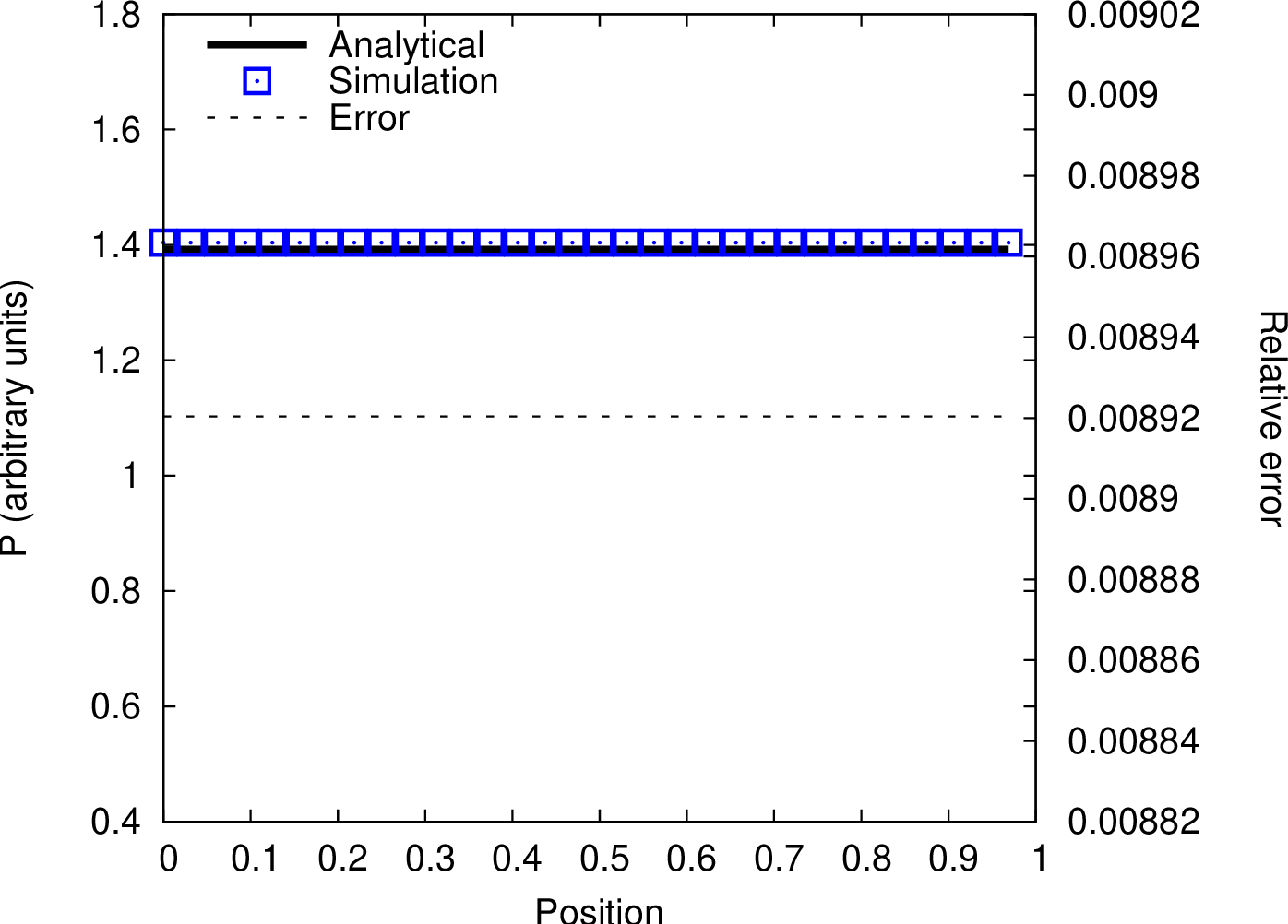}
\caption{Alfv\'en standing waves after four periods and a half, for $\eta_{md}=5\times10^{-3}$. Same caption as in the previous figures.}
\label{figalfvstatioDM}
\end{center}
\end{figure*}

As for the ambipolar diffusion, we start the simulation from an initial perturbed state obtained by adding two propagating waves with opposite values of $s_i$ and the same value of $s_r$, and let the system evolve. 
The evolution of the pressure is (in real notation): 
\begin{align}
P =& P_{init} + (\gamma - 1)\eta_{\Omega} \, k^2 \, \delta b_x^2 \Big[ \frac{e^{2s_r t}-1}{ s_r}   + e^{2s_r t} ( \frac{s_r \cos(2 s_i t) + s_i \sin(2s_i t) }{ |s|^2} ) - \frac{s_r}{|s|^2} \Big]. 
\end{align}
Figure~\ref{figalfvstatioDM} shows a snapshot of the evolution of $\delta {{\bf b}}_x$, $\delta {{\bf u}}_x$, $\delta {{\bf b}}_y$, $\delta {{\bf u}}_y$, $\rho$ and $P$ along $z$ after three wave periods ($t= \frac{3 \times 2.\pi}{s_i}$ ), for $\eta_{\Omega}=5\times10^{-3}$. As seen, once again, the agreement between the numerical and the analytical solution is very good, of the order of or better than a few percents. 
%The density and the pressure are constant at XX \%.

\begin{figure}
\begin{center}
\includegraphics[width=0.57\textwidth,angle=270]{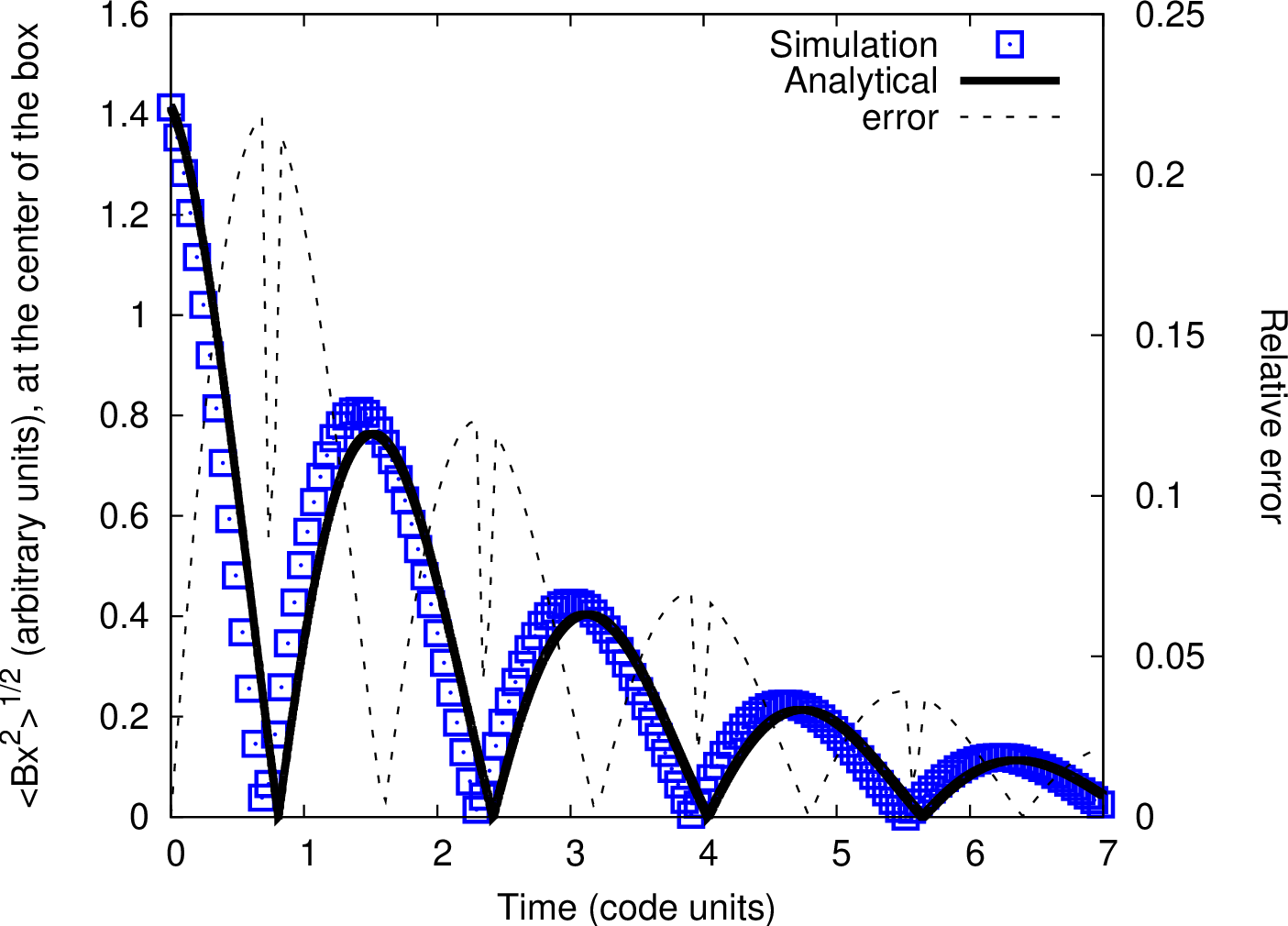}
\caption{Time evolution (expressed in units of the period, $\frac{ 2.\pi}{s_i}$) of $\sqrt{<Bx^2>}$, the root-mean-square of the magnetic field in the $x$ direction, for Alfv\'en standing waves. In this case, $\eta_{\Omega}=2\times10^{-2}$.}
\label{figalfvstatioglobDM}
\end{center}
\end{figure}

\subsubsection{Convergence order}

We tested the evolution of the precision of the implementation of Ohmic diffusion by examining the evolution of the error with the level of refinement, {\it i.e} with the mesh size $\Delta x$ for Alfv\'en standing waves and the Barenblatt test. The error $\epsilon$ is defined here as the maximum difference between the analytical values and the numerical solution, corrected by the damping factor for Alfv\'en waves, and the error at the center of the box for the Barenblatt test. The error as function of cell size follows a power-law, at least in the range studied here (up to 10 periods of the wave). For the standing waves we find:
\begin{eqnarray}
\epsilon \varpropto \Delta x^{3}.
\end{eqnarray}
For the Alfv\'en propagating waves
\begin{eqnarray}
\epsilon \varpropto \Delta x^{2}.
\end{eqnarray}
For the Barenblatt test
\begin{eqnarray}
\epsilon \varpropto \Delta x^{2}.
\end{eqnarray}

\begin{center}
\begin{figure*}
\includegraphics[width=0.35\textwidth,angle=270]{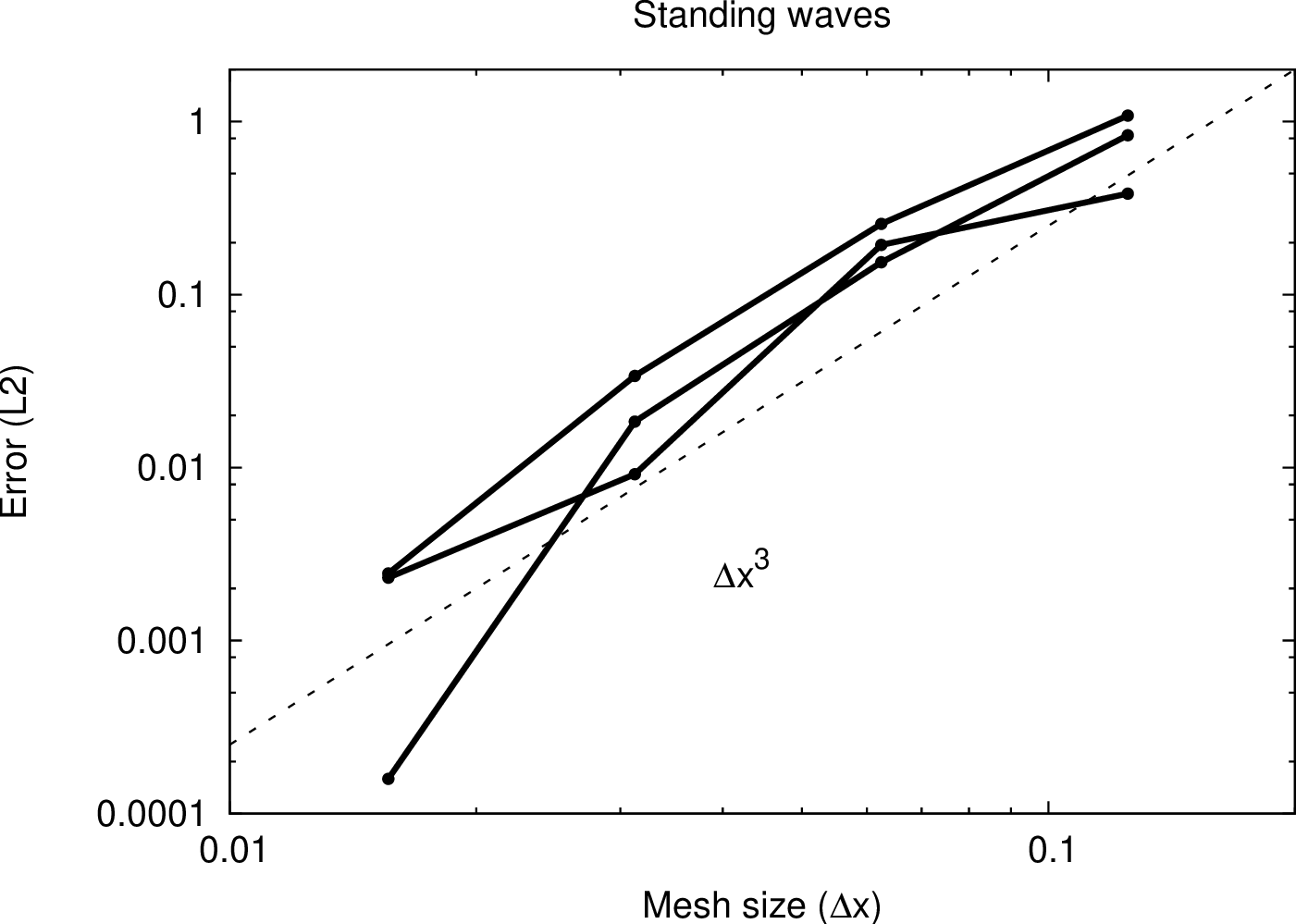}
\includegraphics[width=0.35\textwidth,angle=270]{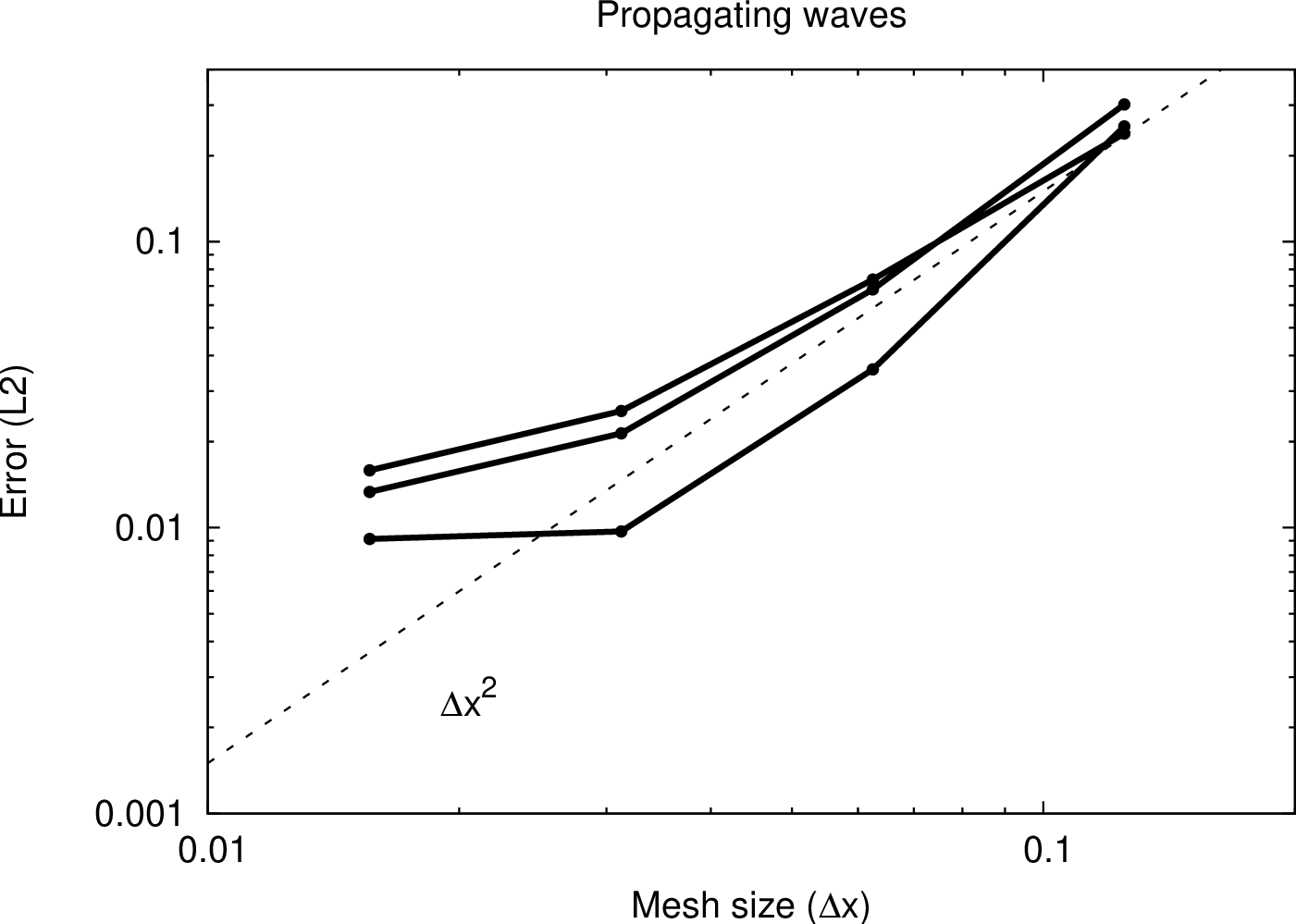}
\caption{Evolution of the error $\epsilon = \sqrt{\sum_{i=1}^{N} \, \frac{(By_{\textrm{numerical}}-By_{\textrm{analytical}})^2}{N}}$ with the mesh size $\Delta x$ for Alfv\'en standing waves (left plot) and Alfv\'en propagating waves (right plot) at different times. The dashed lines correspond to two slopes: $\epsilon \propto \Delta x^{3}$ for the standing waves and $\epsilon \propto \Delta x^{2}$ for the propagating waves.}
\label{errorstandingohmic}
\end{figure*}
\end{center}

A log-log plot of the error as a function of cell size $\Delta x$ for different times is given on Figure~\ref{errorstandingohmic} for Alf\'en standing waves, and on Figure~\ref{errorohmicbaren} for the barenblatt test. The behavior of the error for the propagating waves differs if the mesh size is coarse or fairly refined ( $\epsilon \varpropto \Delta x^{1}$ or $\epsilon \varpropto \Delta x^{2}$ respectively). For the Barenblatt test, the error scales as $\sim \Delta x^{2}$. \label{rev7}\rev{In the case of Ohmic diffusion, Equation~(\ref{eqn09}) reduces exactly to the Heat equation, whereas in the case of ambipolar diffusion, Equation~(\ref{eqn10}) reduces to a non-linear diffusion equation. The error in the two cases scales as $\sim \Delta x^{2}$.}

\subsubsection{Estimate of the numerical drift coefficient of Ohmic diffusion}

As seen in Section~\ref{alfvenpropaDM}, the dissipation of Alfv\'en waves is slightly larger than expected according to the analytical values. 
The spurious dissipation due to the numerical scheme can be  estimated as: 
\begin{eqnarray}
\eta_{mes} = \eta_{\Omega} + \eta_{num},
\end{eqnarray}
where $\eta_{mes}$ is the value  measured in the numerical simulation, with $ \eta_{mes}= - \frac{2 s_{r,num}}{k^2}$, and $\eta_{num}$ is the drift contribution due to numerical dissipation. Another way to proceed is to set $ \eta_{\Omega}= 0 $, to examine how the Alfv\'en waves dissipate, and then to  estimate $\eta_{num}$ as $ \eta_{num}= - \frac{2 s_{r,num}}{k^2 }$. Both methods give about the same value for $\eta_{num}$. For a level of AMR refinement of 4, we get $\eta_{num}= 1. \times 10^{-3}$; for 5,  $\eta_{num} = 1. \times 10^{-4}$ and for 6,  $\eta_{num} = 1.1 \times 10^{-5}$, to be compared with $\eta_{\Omega}=0.005$ or 0.02 for the present simulations. As expected, the better the resolution, the smaller the numerical diffusion.

\section{Conclusion}
\label{concl}

In this paper we have described a numerical method to implement the treatment of the two important terms of non-ideal MHD, namely ambipolar diffusion and Ohmic dissipation, into the multi-dimensional AMR code {\ttfamily RAMSES}. For ambipolar diffusion, we have used a single fluid approach, which is valid when the Lorentz force and the neutral-ion drag force are comparable, corresponding to a domain of strong coupling between the fluid and the field lines. The situations where such an approximation can be made are numerous, \m{of which} cloud collapse or certain protoplanetary disks are two typical examples. The accuracy of our numerical resolution of the MHD equations \m{was} examined by performing a diversity of tests, for which either analytical or benchmark solutions exist. For both ambipolar and Ohmic diffusion, we first explored the purely magnetic case, without any coupling to the hydrodynamics. For ambipolar diffusion, this \m{was} done by comparing the evolution of a Dirac pulse to the solution provided by Barenblatt while for Ohmic diffusion, the solution is confronted to the well known heat diffusion equation. In a second step, we studied the full MHD case \m{(coupling the fluid to the magnetic field)} by considering first an oblique shock, and then the behavior of propagating and standing Alfv\'en waves. For all these tests \m{the solutions obtained with our method show excellent agreement with the analytical predictions, typically} within a few tenths of a percent on average, \m{showcasing} the validity and the robustness of our method. We have also carefully analyzed the main source of numerical error using the Modified Equation framework. \m{In order to} estimate the spatial resolution that is required to model non-ideal MHD effects reliably. This opens the avenue to a vast domain of astrophysical applications, in particular cloud collapse, pre-stellar core formation and protostellar disks where ambipolar and Ohmic diffusion processes \m{are believed to} play a dominant role. Such astrophysical applications of the non-ideal MHD equations with {\ttfamily RAMSES} will be explored in forthcoming papers.

\bigskip
\section*{Acknowledgement}
The research leading to these results has received funding from the European Research Council under the European Community's Seventh Framework Programme (FP7/2007-2013 Grant Agreement no. 247060)

\bigskip
\appendix

\section{The Barenblatt-Pattle solution\protect\label{annexebaren}}

Following \citet{refbaren2}, the solution of Equation~(\ref{bybaren}) in general form ($\frac{\partial B_y}{\partial t} = {\nabla} . \Big( B_y^{\beta} {\nabla} B_y \Big)$), where $\beta$ depends on the problem, is:
\begin{equation}
B_y = \left\{ \begin{split}
& A t^{\alpha} \left[ 1 - (\frac{r}{\eta_0 t^{\delta}})^2 \right]^{{\beta}^{-1}} & \quad & \textrm{if} \quad r \leq \eta_0 t^{\delta}\\
& 0 & \quad & \textrm{if} \quad r > \eta_0 t^{\delta} 
\end{split} \right.
\end{equation}
With $\mu$ the dimensionality of the problem, the various constants are defined as follow:
\begin{align}
\alpha &= \frac{-\mu}{2+\mu \beta} \\
\delta &= \frac{1}{2 + \mu \beta} \\
A &= \left( \frac{\delta \beta \eta_0^2}{2} \right)^{\frac{1}{2}} \\
\int_{x_1}^{x2} B_{y0}(\mathbf{x}) \, d\mathbf{x} &= \eta_0^{\mu +2/\beta}(\frac{1}{2} \delta \beta)^{{\beta}^{-1}} \frac{\Gamma(\frac{1}{2}\mu)\Gamma(1/\beta+1)}{\Gamma(1/\beta+1+\frac{1}{2}\mu)}
\end{align}

\section{Semi-analytical solution for the isothermal C-shock\protect\label{maclo}}

Following \citet{McLow}, in the isothermal case with a constant ion density, we reduce the set of MHD equations to:
\begin{align}
\rho v_{x}^2 + P + \frac{B_y^2}{2} =& C_1 \\
\rho v_{x} v_y - B_y B_x =& C_2 \\
b^2 - b_0^2 =& 2A^2(D-1)(D^{-1}-M^{-2}) \\
\big(D^{-2}-\mathcal{M}^{-2}\big) L \frac{dD}{dx} =& \frac{b}{A}  (b^2+\cos{\theta})^{-1} \nonumber \\
                                                    \times & \Big[ b - D \big( \frac{b-b_0}{A^2} \cos{\theta}^2 + \sin{\theta} \big) \Big] 
\end{align}
with $C_1$ and $C_2$ derived from the initial state, $A = \frac{v}{v_A}$ the Alfv\'en Mach number, and $M = \frac{v}{c_s}$ the Mach number; $\theta = 45^{\circ}$ is the angle between the magnetic field and the velocity field; and $D=\frac{\rho}{\rho_0}$ and $b = \frac{B_y}{B_{0}}$ are the dimensionless density and magnetic field.

\section{Semi-analytical solution for the non-isothermal C-shock\protect\label{duffpud}}

Following \citet{DuffinPudritz} and \citet{Wardle1991}, and reminding that the set of equations is not exactly the same as ours, we solve the set of equations: 
\begin{align}
\frac{db}{dx} &= \frac{\gamma_{AD} \rho_{i0} A^2 r}{v_s b} \\
\left( \frac{1-\gamma r_n p}{(\gamma-1)r_n} \right) \frac{dp}{dx} &= \frac{\gamma_{AD} \rho_{i0} r}{v_s} \left( \frac{1}{r_n} \frac{\gamma}{\gamma-1}p-\frac{S_n+\sin{\theta}}{b} \right) \\
S_n &= \frac{b-b_0}{A^2} \cos^2{\theta} \\
r_n &= \frac{1}{1-(p-p_0)-(\frac{b^2-b_0^2}{2 A^2})} \\
r_i &= r_n \left( \frac{b^2+\cos^2{\theta}}{b r_n (S_n+\sin{\theta})+\cos^2{\theta}}  \right) \\
r &= 1-\frac{r_i}{r_n}
\end{align}
where the dimensionless quantities are $p=\frac{P_n}{\rho_{n0} v_s^2}$, $b=\frac{B_y}{B_0}$, the velocities $v_{nx} = \frac{v_s}{r_n}$, $v_{ny}=\frac{S_n B_0 v_s}{B_x} = \frac{S_n v_s}{b_0}$. $p_0$ and $b_0$ are the initial dimensionless pressure and magnetic fields, $\theta = 45^{\circ}$ is the angle between the pre-shock velocity and the magnetic field, and $A=\frac{v_s}{v_A}$ the Alfv\'en Mach number.

 \bibliographystyle{apj}
% \bibliographystyle{elsarticle-harv}
%\bibliographystyle{plain}
%\bibliographystyle{elsarticle-num-names}
%\bibliographystyle{aa}

%\bibliography{MaBiblio}

   \end{document}